\documentclass{aa}  

\usepackage{txfonts,textcomp}

\usepackage{graphicx}
\usepackage{orcidlink}
\usepackage{adjustbox}
\usepackage{natbib}
\usepackage{booktabs} 
\usepackage{placeins}
\usepackage{cuted}
\usepackage{float}
\usepackage{lastpage}
\usepackage{amsmath}	
\usepackage{xcolor}
\usepackage{lscape}

\newcommand{\HI}{H\,{\sc i}}
\newcommand{\HII}{H\,{\sc ii}}

\newcommand{\Ha}{H$\alpha$}
\newcommand{\Msun}{~M$_{\odot}$}

\newcommand{\Lsun}{~L$_{\odot}$}

\newcommand{\kms}{~km\,s$^{-1}$}

\newcommand{\vhel}{$v_{\rm hel}$}

\begin{document} 

\title{MeerKAT radio continuum imaging of nearby star-forming spirals in the NGC~6221, NGC~3256/3263, and NGC~2434 galaxy groups}

\author{J. Saponara\orcidlink{0000-0003-4562-7694}\inst{1} \and B. S. Koribalski\inst{2,3}, J. English\inst{4} \and P. K. Humire\orcidlink{0000-0003-3537-4849} \inst{5}}

\institute{
Instituto Argentino de Radioastronom\'ia, CONICET-CICPBA-UNLP, CC5 (1897) Villa Elisa, Prov. de Buenos Aires, Argentina.\\ \email{jsaponara@iar-conicet.gov.ar}
\and
Australia Telescope National Facility (ATNF), CSIRO, Space and Astronomy, P.O. Box 76, Epping, NSW 1710, Australia.
\and
School of Science, Western Sydney University, Locked Bag 1797, Penrith, NSW 2751, Australia.
\and
Department of Physics and Astronomy, University of Manitoba, Winnipeg, Manitoba, Canada R3T 2N2, Canada.
\and
Departamento de Astronomia, Instituto de Astronomia, Geofísica e Ciências Atmosféricas da USP, Cidade Universitária, 05508-090 São Paulo, SP, Brazil
}
   
\date{Accepted XXX. Received YYY; in original form ZZZ}

\abstract
{We present high-resolution MeerKAT 1.3~GHz radio continuum images of star-forming spirals in the nearby galaxy groups around NGC~6221, NGC~3256/3263, and NGC~2434. This sample spans the evolutionary timeline for galaxy groups, encompassing early, intermediate, and late stages, respectively. The NGC~6221 group contains an interacting galaxy pair with tidal debris, along with at least three dwarf galaxies. In contrast, the NGC~3256/3263 group represents a loose group consisting of several spiral as well as dwarf galaxies, while
a massive elliptical galaxy dominates the NGC~2434 group.}
{We study the star formation activity in all detected galaxies as it is one of the dominant physical processes in their formation and evolution, seeking evidence of environmental impact.}
{We use MeerKAT radio continuum data and archival WISE infrared data to locate and measure the star formation rate in all group members. In particular, we used polycyclic aromatic hydrocarbons (PAH) as tracers of gas heated due to star formation activity. Furthermore, we create in-band spectral index maps, providing insights into the underlying physical processes associated with the detected star-forming regions. For some galaxies, we also determine key stellar properties like age and mass through SED fittings.}
{We found that galaxies are distributed differently in the WISE colour-colour diagram depending on their evolutionary group stage, as expected. Except for ESO\,059-G012, the galaxies in our sample follow the radio-W3PAH correlation. A possible scenario that explains the ESO\,059-G012 result is that the galaxy has already consumed the gas.
We also found evidence that the interaction among the spiral galaxies NGC~3263, NGC~3256B, and NGC~3256C is causing the Vela Cloud complex and that the galaxies NGC~6221 and NGC~3256 might host a low-luminosity AGN, as was previously proposed in the literature.}
{}
\keywords{galaxies: groups: individual: NGC~2434, NGC~3256/3263 NGC~6221; galaxies: star formation; galaxies: evolution; radio continuum: galaxies; Astrophysics - Astrophysics of Galaxies}

\titlerunning{MeerKAT continuum imaging of star-forming spirals in nearby galaxy groups}
\authorrunning{J. Saponara et al.}
\maketitle
  
%


\section{Introduction}

The radio continuum emission detected in galaxies combines thermal and non-thermal emission. Both processes are typically associated with massive star formation. Star formation is the main internal physical process in the formation and evolution of galaxies. At GHz frequencies, the radio continuum emission is optically thin and is a tracer of recent star formation, unbiased by dust within galaxies \citep{condon-1992,Davies-2017}. 
Thermal emission is produced by free-free emission from the ionised gas, while the non-thermal component is produced by synchrotron radiation caused by cosmic-ray electrons with GeV energies spiralling in the magnetic field of a galaxy. These electrons are primarily accelerated in the star-forming region's shock fronts of supernova explosions or stellar winds.
Most optical or ultraviolet indicators of star formation rate (SFR) in galaxies are either indirect tracers of young stars or affected by dust extinction.
The ultraviolet light from young, fairly massive stars ($>$5\Msun) also ionises the surrounding hydrogen gas cloud, producing thermal radio emission and heating the nearby dust, which re-radiates in the infrared regime.
Thus, assuming that infrared and radio emissions are both connected to massive stars and, therefore, to star formation, the above characteristics mentioned underlie the basis for the infrared-radio correlation \citep{Helou-1985,condon-1992,Yun-2001,Bell_2003}. Different studies were conducted to understand this correlation better, from a detailed analysis of individual galaxies \citep[e.g. NGC~6946,][]{Tabatabaei-2013}, to statistical analysis of many galaxies \citep{Shao-2018,Grundy-2023} and also considering galaxies at different redshift \citep{Seymour_2009,Basu_2015,Delhaize_2017,Yoon_2024}. Sometimes, it is observed that the infrared-radio relationship is not tight when considering interacting or merging galaxies. For example, an excess of radio emission was found in the case of the so-called Taffy system, a pair of interacting galaxies with a strong synchrotron-emitting gas bridge between them \citep{Condon-2002}. \cite{Murphy_2013} suggests that the excess in non-thermal emission is probably due to particle acceleration in large-scale shocks in bridges between interacting galaxies \citep{Lisenfeld-2000}. Moreover, \cite{Donevski_2015} studies the possibility of such departure in the relation not only due to the bridges and in ‘Taffy’ systems but also from cosmic rays accelerated in tidal shocks in the galaxies themselves. Then, deviations from the expected infrared-radio relation can show how galaxy-galaxy interactions or interactions between galaxies and the intra-cluster or group medium affect the gaseous and relativistic phases of their interstellar medium \citep{Condon-2002,Murphy_2009, Murphy_2013,Donevski_2015}. Several authors have investigated the non-linearity of the infrared–radio correlation, exploring its dependence on radio and infrared luminosities, stellar mass, and redshift. \citet{Molnar-2021} conducted a study focusing on galaxies at low redshift ($z < 0.2$), distinguishing between star-forming galaxies (SFGs) and active galactic nuclei (AGN). They found that galaxies with higher radio luminosities ($\log \, L_{1.4\,\rm{Ghz}} \, [\mathrm{W\,Hz}^{-1}] \geq 22.5$) exhibit relatively stronger radio emission compared to the infrared. \citet{Delvecchio-2021} examined potential dependencies of the infrared-radio correlation on stellar mass and redshift. Their results show that for a given star formation rate, more massive galaxies emit more strongly in the radio than in the infrared.

We obtained single-pointing MeerKAT 1.3~GHz observations of three nearby galaxy groups (around the galaxy NGC~6221, galaxies NGC~3256 and NGC~3263, and galaxy NGC~2434), known to contain widespread tidal debris and sampling the evolutionary timeline of groups \citep[e.g.][]{Verdes-Montenegro_2001}. In the first group, the pair-wise interactions between the two dominant spirals, NGC~6215 and NGC~6221, resulted in an extended \HI\ bridge \citep{Koribalski-2004}. In contrast, the second group is a loose association of galaxies around NGC~3256 and NGC~3263, where members pre-process via pair-wise interactions as they begin to fall towards the group’s centre of mass. Among the widespread tidal debris is the large Vela \HI\ cloud \citep{English-2010}.
The third group, named after the central elliptical galaxy NGC~2434, represents the end of the evolutionary sequence for groups; the group members appear to be in the process of accreting onto this central object \citep{Ryder-2001}.

We aim to map the star formation rate in the group members, described in Appendix~\ref{appen:galaxy_groups}, using our MeerKAT radio continuum data, unobscured by dust, along with supplemental high-angular resolution WISE data, and to find evidence of environmental impact on star formation activity. The following subsections introduce the galaxy groups, followed by a description of the MeerKAT observations and ancillary data in Section~2. The results are presented in Section~3, followed by our discussion in Section~4. A summary is provided in Section~5.

\section{Observations and data processing}
\subsection{MeerKAT observations and data processing}\label{sec:obs}

MeerKAT is a powerful radio interferometer located in the Karoo desert of South Africa, consisting of 64 dishes with baselines out to 8~km \citep{Jonas-2009,Jonas-2016,Mauch-2020}. The parabolic 13.5~m diameter antennas have offset Gregorian receivers. Of the 64 antennas, 48 are located in the inner core (within a 1~km radius), providing the shortest baseline of 29~m. Spectral line and radio continuum observations of three nearby galaxy groups (project SCI-20210212-BK-01) were carried out during 2021 and 2022 (see Table~\ref{tab:obs}). Here we present the radio continuum data for which we used a centre frequency of 1.3~GHz and a bandwidth of 856~MHz. The latter was divided into 10 channels for the in-band spectral analysis of the radio continuum emission. The resulting frequency range is 856 to 1712~MHz. The primary calibrators were PKS\,1934--638 and PKS\,0408--658 with total 1.3~GHz intensities of 15.0~Jy and 15.7~Jy, respectively. The target field and the secondary calibrator were observed alternately for 36~min and 2~min, respectively. 

The MeerKAT radio continuum data processing was done by SARAO using the continuum and \HI\ pipeline CARACal\footnote{https://caracal.readthedocs.io/en/latest/}, which has been extensively tested and optimised for MeerKAT data \citep{Jozsa-2020ASPC,Jozsa-2020ascl}. For the NGC~3263 field, we used the combined data from two $\sim$5h observations, described in \citet{Koribalski2024-MeerKAT}, along with the processing procedure.  The achieved angular resolutions and rms sensitivities are listed in Table~\ref{tab:obs} together with the observing parameters. The data analysis was done using {\sc MIRIAD} software tasks. 

In Table~\ref {tab:obs} are listed the MeerKAT observing parameters, and in Figures~\ref{fig:6221} and \ref{fig:continuo} are displayed the high-resolution radio continuum images of the remaining galaxies detected within the three galaxy groups.

\subsection{WISE ancillary data}
\label{sec:infrared}

The three galaxy groups were observed with the 
Wide-field Infrared Survey Explorer \citep[WISE,][]{Wright-2010}. During this mission, the sky was mapped at 3.4, 4.6, 12, and 22 $\mu$m (W1, W2, W3, and W4) with angular resolutions of 6.1\arcsec, 6.4\arcsec, 6.5\arcsec, and 12.0\arcsec, respectively. We gathered WISE images from the NASA/IPAC Infrared Science Archive (IRSA). We measured the integrated WISE W1, W2 and W3 band flux densities for our galaxy sample, because the high angular resolution makes them useful for analysing resolved properties. The procedure was performed using {\sc Photutils} \citep{Bradley-2016,Bradley-2020}, an Astropy-affiliated photometry package written in Python. The tool has functions such as detecting astronomical target sources, evaluating the backgrounds of astronomical images and performing aperture photometry. We explain the procedure in Section~\ref{sec:infrared-photutils}.

\section{Results}

\subsection{Radio continuum}

\begin{figure*}
\centering
\includegraphics[width=0.3\textwidth]{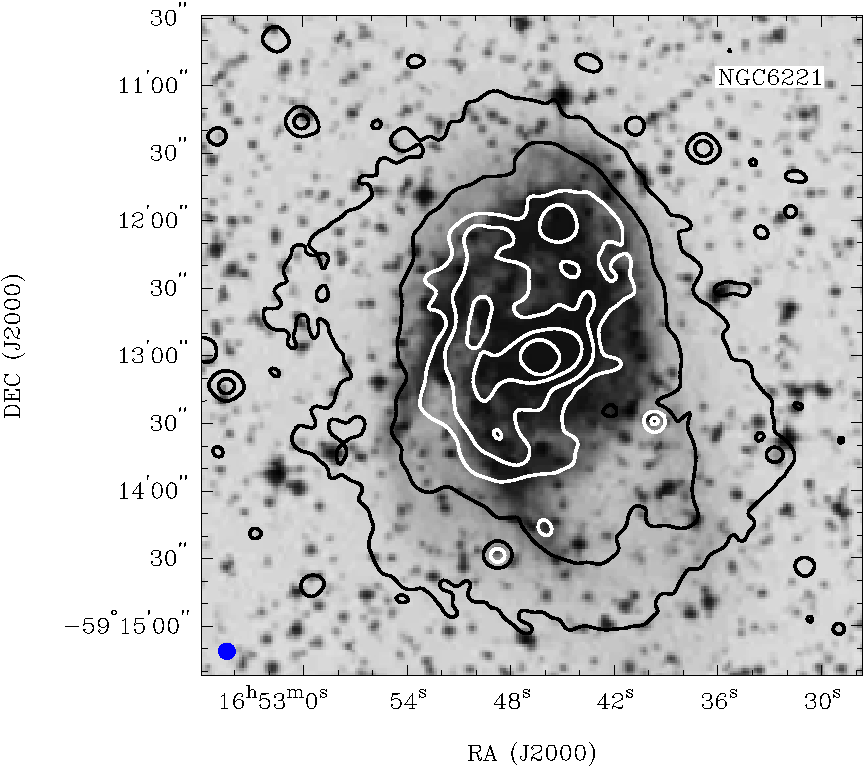}
\includegraphics[width=0.305\textwidth]{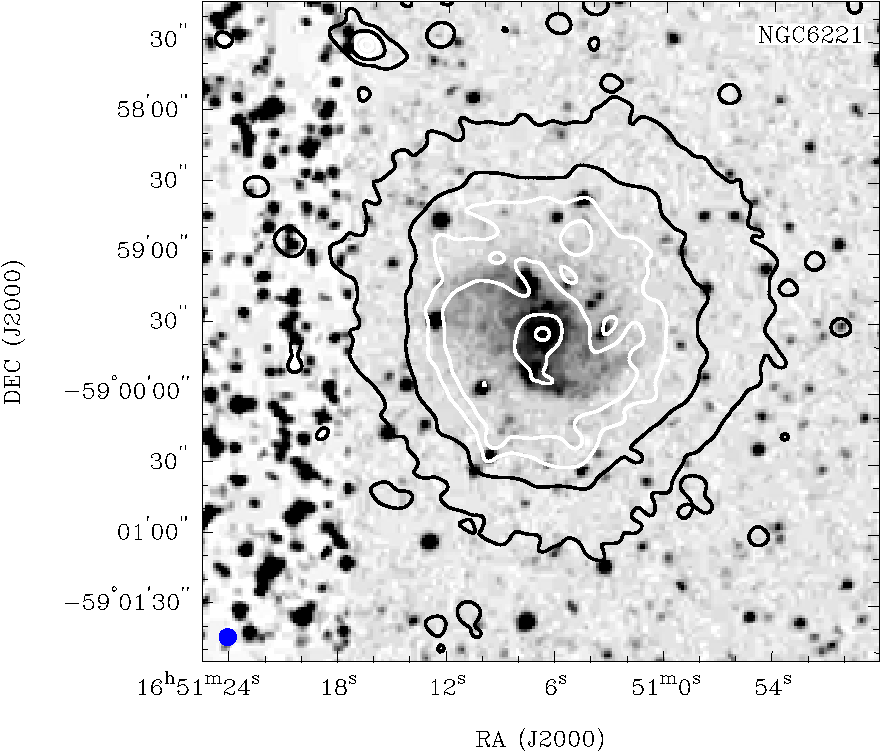}
\includegraphics[width=0.32\textwidth]{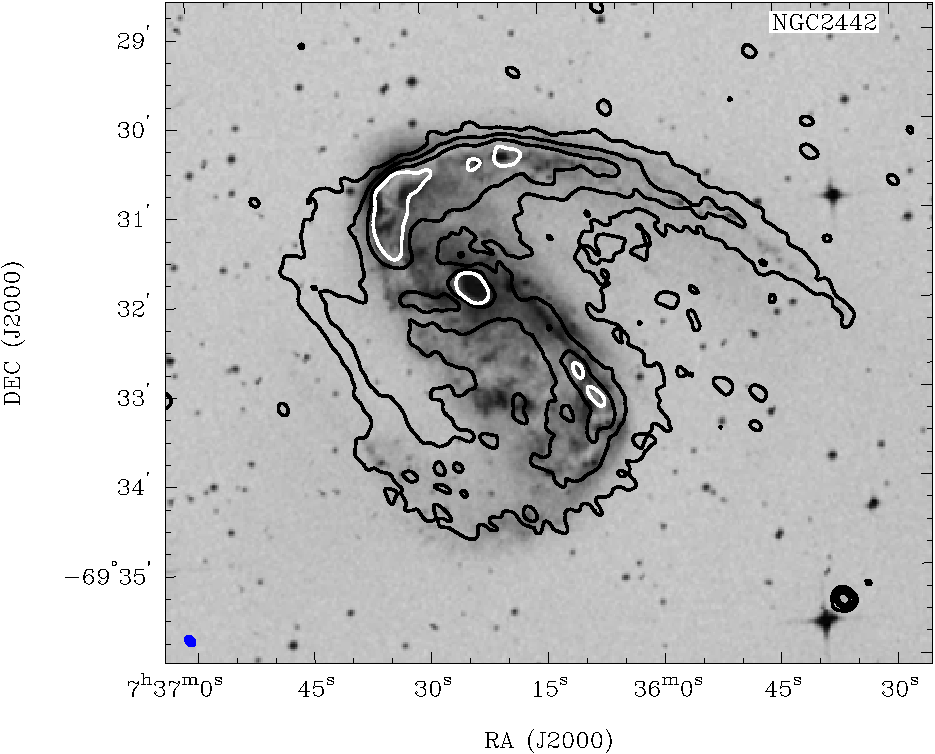}\\
\includegraphics[width=0.44\textwidth]{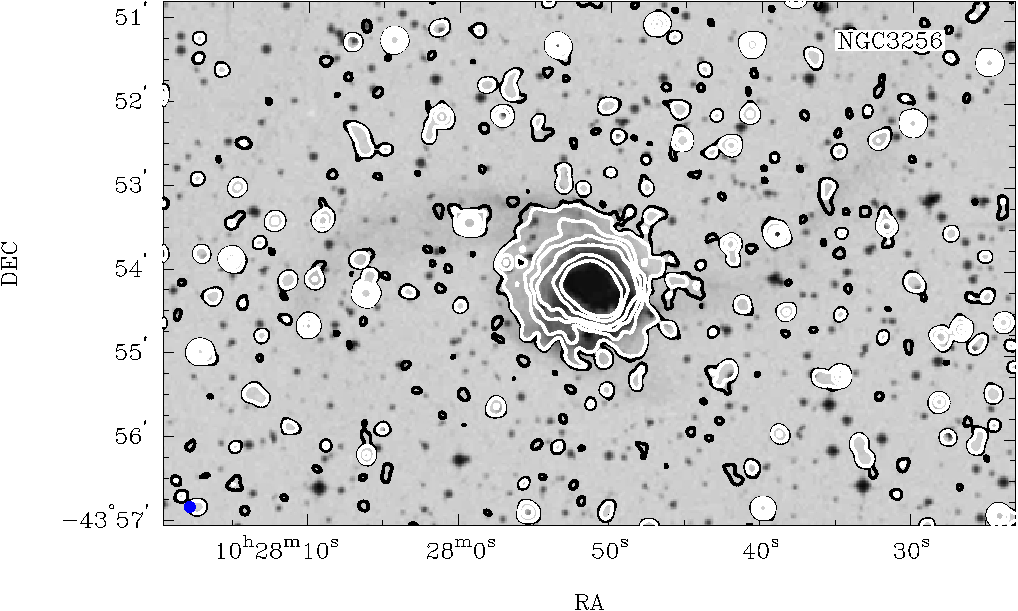}
\includegraphics[width=0.54\textwidth]{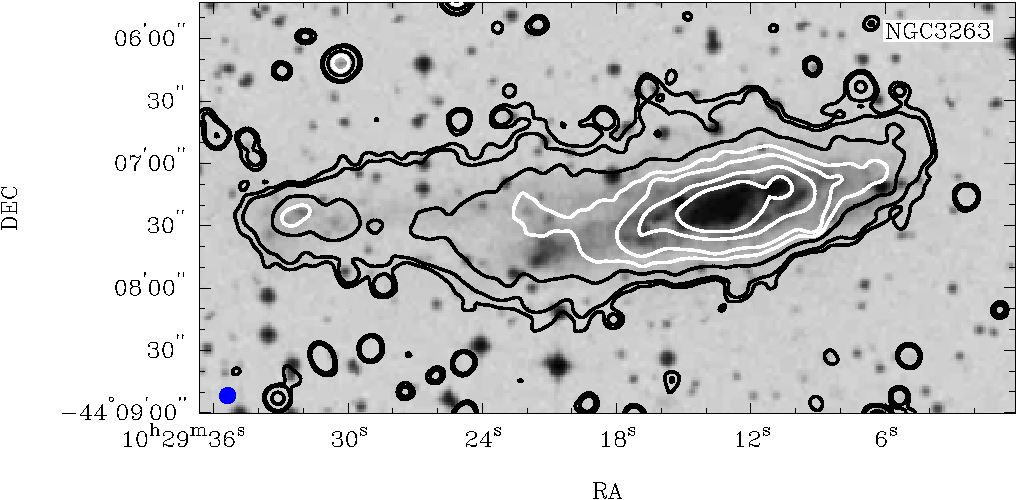}\\
\caption{ High-resolution MeerKAT 20-cm radio continuum emission of five galaxies overlaid onto DSS optical images.
 First row: NGC~6221, NGC~6215 and NGC~2442. The radio continuum contour levels of NGC~6221 and NGC~6215 are 0.03, 0.2, 0.5, 1, 3 and 8~mJy\,beam$^{-1}$, while for NGC~2442, the levels are 0.027, 0.09, and 0.27~mJy\,beam$^{-1}$. Second row: NGC~3263 and NGC~3256. The radio continuum contour levels are 0.027, 0.045, 0.18, 0.45, 0.9, 2.7 and 7.2~mJy\,beam$^{-1}$. Contours are shown in black or white for display purposes. The synthesised beam is displayed at the bottom-left corner of each panel.}
\label{fig:6221}
\end{figure*}

We detect radio continuum emission from several spiral galaxies in all three groups and measure their flux densities using standard procedures (see Table~\ref{tab:prop}, and Figs.~\ref{fig:6221} and ~\ref{fig:continuo}).

In the NGC~6221 group, we detect radio continuum emission from the galaxies NGC~6221 and NGC~6215; see Table~\ref{tab:prop}. No radio continuum emission was detected from the three gas-rich dwarf galaxies catalogued by \cite{Koribalski-2004}, nor from the \HI\ bridge nor the \HI\ clumps. Considering a point-source distribution, we estimate an upper limit to their flux densities of $\sim$~35$\mu$Jy (5$\sigma$). Furthermore, three background galaxies were detected in this field: ESO\,138-G001, ESO\,138-G004, and ESO\,138-G005, see Fig.~\ref{fig:continuo}. Fig.~\ref{fig:6221} shows the high-resolution radio continuum images of the NGC~6221/6215 galaxy pair.

In the NGC~3256/3263 group, we detect radio continuum emission from eight out of ten galaxies proposed as members by \cite{kourkchi-2017}, see Table~\ref{tab:prop}. No continuum emission is detected towards the Vela Cloud and any other \HI\ debris within the group. We consider the galaxies ESO\,263-44 and ESO\,263-46 as group members, despite the fact that they are usually not included in previous studies, because their systemic velocity agrees with a moderate group velocity dispersion (< 300\kms). 
Furthermore, we detect three background galaxies: ESO\,263-45, LEDA\,3097821, and LEDA\,087403. In Fig.~\ref{fig:6221}, we present the high-resolution radio continuum images of NGC~3256 and NGC~3263. The remaining galaxies, along with the background galaxies, are displayed in  Fig.~\ref{fig:continuo}.

In the NGC~2434 galaxy group, we detect four out of twelve galaxies, proposed as members by \cite{kourkchi-2017}. The group members NGC~2397 (detected) and NGC~2397B lie in the outskirts of the primary beam; see Table~\ref{tab:prop} for further information. 
No continuum emission is detected towards the \HI\ debris described in \cite{Koribalski-2004}. The high-resolution radio continuum image of NGC~2442 is depicted in Fig.~\ref{fig:6221}, while the continuum of the other three detected galaxies can be seen in Fig.~\ref{fig:continuo}.

\begin{table*}
\centering
\caption{Radio continuum and infrared properties of the detected galaxies.}
\setlength{\tabcolsep}{2.5pt} 
\begin{adjustbox}{max width=1.0\textwidth}
\begin{tabular}{@{}lcccccccccc@{}}
\hline\hline
Galaxy & RA & DEC & Morphology & MS & $S_{\rm 20cm}$ & $F_{\rm W3PAH}$  & $q_{\rm W3PAH}$ & $log(F_{\rm TIR})$ & $q_{\rm TIR}$ & $\alpha^{1021}_{1345}$ \\
 &  (h m s) & ($^{\circ}\,'\,''$) &  &  &  (mJy) &  (mJy) & & (W m$^2$) &  &   \\
(1) & \multicolumn{2}{c}{(2)} & (3) & (4) &  (5)&  (6) & (7) & (8) & (9) & (10)  \\
\midrule
NGC~2397 & 07:21:19.9 & $-$69:00:5.30 & SB(s)b & 2 & 50$\pm$2 & 482$\pm$23 & 2.8 & $-$9.3 & 2.4 &... \\
NGC~2442 & 07:36:23.8 & $-$69:31:51.0 & SAB(s)bc pec & 2 & 327$\pm$11 & 473$\pm$23 & 2.0 & $-$9.3 & 1.6 & ...\\
ESO\,059-G012 & 07:38:32.0 & $-$68:46:14.4 & Sa & 0 & 0.2$\pm$0.1 & 17$\pm$1 & 3.7 & $-$10.6 & 3.5 & ...\\
NGC~2434 & 07:34:51.6 & $-$69:17:02.9 & S0 & 0 & 2.0$\pm$0.1 & ...& ...&... &... &... \\
\midrule
ESO\,263-G033 & 10:24:47.46 & $-$43:57:51.60 & (R)SA0(s)? & & 0.3$\pm$0.1 & 0.5$\pm$0.3 & 2.0 & ...& ...& ...\\
NGC~3256 & 10:27:51.2 & $-$43:54:13.4 & Sb(s) pec & 3 & 476$\pm$1 & 2311$\pm$106 & 2.5 & $-$8.8 & 1.9 & $-0.9\pm0.01$ \\
PGC~087403 & 10:28:25.8 & $-$44:16:17.0 & dIrr & 2 & 0.5$\pm$0.1 &... &... &... & ...&... \\
*LEDA~3097821 & 10:28:59.6 & $-$44:41:12.0 &... & ...& 0.4$\pm$0.1 & ...& ...&... &... & ...\\
NGC~3256B & 10:29:01.0 & $-$44:24:10.4 & SB(s)bc & 2 & 38$\pm$1 & 142$\pm$7 & 2.4 & $-$9.8 & 2.0 & $-0.8\pm0.2$ \\
NGC~3261 & 10:29:01.4 & $-$44:39:24.6 & SB(rs)b & 2 & 65$\pm$1 & 145$\pm$9 & 2.2 & ...& ...&... \\
NGC~3256C & 10:29:05.7 & $-$43:50:57.4 & SB(rs)d & 2 & 23$\pm$1 & 111$\pm$5 & 2.5 & $-$9.9 & 2.1 & $-1.5\pm0.4$ \\
NGC~3263 & 10:29:13.3 & $-$44:07:22.5 & SB(rs)cd & 2.5 & 148$\pm$1 & 450$\pm$22 & 2.3 & $-$9.3 & 1.8 & $-1.2\pm0.1$ \\
ESO\,263-G044 & 10:29:29.5 & $-$44:16:06.0 & ...& 2 & 2.3$\pm$0.1 & 3.0$\pm$0.2 & 2.0 & $-$11.2 & 1.8 &... \\
NGC~3256A & 10:25:51.0 & $-$43:44:53.0 & dIrr & 2 & 0.2$\pm$0.1 & 1.2$\pm$0.2 & 2.6 & $-$11.2 & 2.4 &... \\
*ESO\,263-G045 & 10:29:34.5 & $-$44:22:38.0 &... & ...& 1.4$\pm$0.2 &... &... &... &... & ...\\
*LEDA~537146 & 10:30:04.9 & $-$44:27:13.0 & dE? & ...& 1.0$\pm$0.1 & ...& ...& ...& ...& ...\\
ESO\,263-G046 & 10:30:15.4 & $-$44:18:28.0 & SB(s)d & 2.5 & 0.7$\pm$0.1 & 3.9$\pm$0.2 & 2.5 & $-$11.2 & 2.3 &... \\
\midrule
NGC~6215 & 16:51:06.80 & $-$58:59:36.0 & SA(s)c & 2.5 & 293$\pm$1 & 887$\pm$41 & 2.4 & $-$9.0 & 1.9 & $-1.2\pm0.1$ \\
NGC~6221 & 16:52:46.32 & $-$59:13:00.0 & SB(s)c & 2.5 & 353$\pm$1 & 1951$\pm$91 & 2.6 & $-$8.7 & 2.1 & $-1.4\pm0.1$ \\
*ESO\,138-G004 & 16:52:48.40 & $-$58:56:46.1 &... & ...& 0.3$\pm$0.2 & ...&... &... &... &... \\
*ESO\,138-G005 & 16:53:53.37 & $-$58:46:40.8 &... & ...& 4.4$\pm$0.2 & ...& ...& ...& ...&... \\
*ESO\,138-G001 & 16:51:20.26 & $-$59:14:04.2 &... &... & 36$\pm$19 &... &... &... &... &... \\
\bottomrule
\end{tabular}
\end{adjustbox}
\label{tab:prop}
\tablefoot{Derived sample properties. From left to right: (1) Galaxy name, (2) J2000 galaxy coordinates, (3) morphological type from NED, (4) merger stage (MS) based on \citet{Haan-2011}, (5) integrated 20-cm flux density, (6) infrared flux density, (7) $q_{\rm W3PAH}$ parameter, (8) Total infrared flux estimated using \cite{Cluver_2017}, (9) $q_{\rm TIR}$ parameter and (10) in-band spectral index.``*'' indicates detected background galaxies.}
\end{table*}

\subsection{Mid-infrared photometry}
\label{sec:infrared-photutils}

To obtain accurate photometric measurements of the sampled galaxies, we must first locate and catalogue the sources in the images. Then, we need to identify the pixels in the images that belong to each source to create a mask. This mask will help us to exclude the light profile from neighbouring objects. Hence, we use the photometry routine {\sc Photutils} \citep{Bradley-2016,Bradley-2020}. 

We produced the catalogues and segmentation maps for the masks using the {\sc detect$\_$sources} and {\sc SourceCatalog} functions to separate the target galaxy from its neighbours. To do this, we first smoothed the W1 band image with a Gaussian kernel of 3$\times$3 and then segmented this image. We used the W1 band, which covers a larger area, to also choose the segments to use for the W2 and W3 bands in order to avoid missing flux from longer wavelength WISE bands. The subsequent segmentation requires a minimum source size specified by the number of connected pixels, which we set to 10, with values above a threshold. This threshold was computed using Background 2D from {\sc Photutils} and {\sc sigma$\_$clipped$\_$stats} with a 3$\sigma$ cut-off (this resulted in the best deblending). This cut-off, in turn, had been carefully estimated from the noise in the smoothed image. If required, overlapping sources were separated using {\sc Photutils' deblend$\_$sources} function that combines multi-thresholding and watershed segmentation.

\begin{figure*}
\centering
\includegraphics[width=0.9\textwidth]{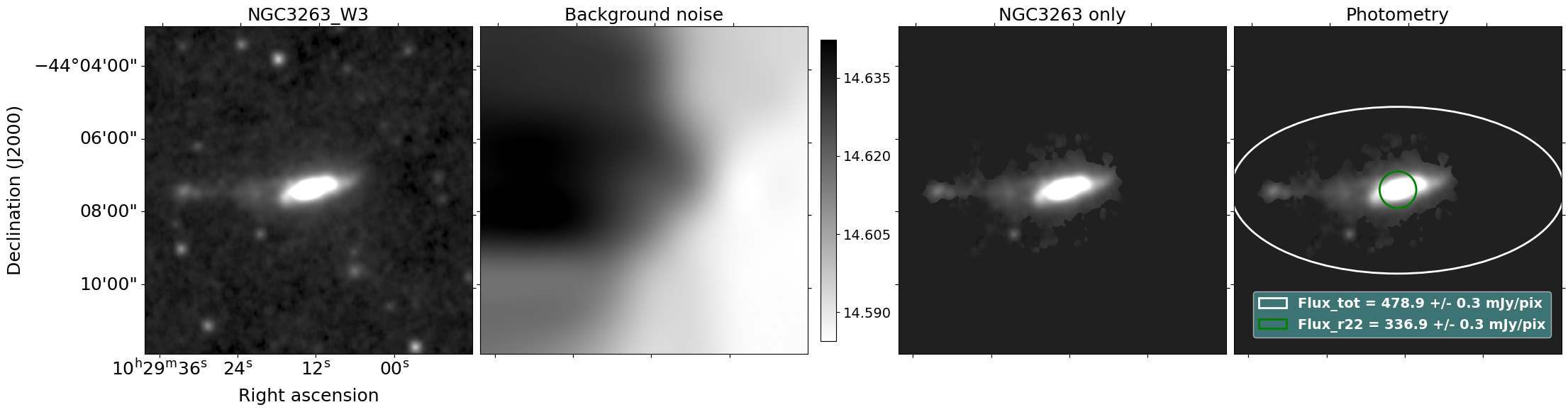}
\caption{Example of source identification and photometry with {\sc Photutils} on the WISE W3 image of NGC~3263. From left to right: original W3 image, background model, background-subtracted galaxy image, and final photometry. The white ellipse indicates the aperture used to measure the total flux density. The green circle corresponds to the 22\arcsec\ isophote employed for comparison with catalogue values (see text).}
\label{fig:ngc3263_wise}
\end{figure*}

Subsequently, the photometry was performed using the {\sc Photutils} function {\sc aperture$\_$photometry}. We used circular apertures centred at the galaxy's centre, except for NGC~3263, where we used an elliptical aperture that properly adjusted to the galaxy's shape. The axes were selected big enough to contain the entire source, primarily covering the same size observed in the radio images. We were unable to perform the photometry for the background galaxy PGC\,087403, especially at the W3-band, because we could not resolve the galaxy from a source close to it.
In Fig.~\ref{fig:ngc3263_wise}, we show, as an example, the procedure applied for the galaxy NGC~3263. The four panels display the W3 band image, the 2D background estimation, the galaxy only and the photometry. The results for the entire sample of galaxies are shown in Appendix~\ref{sec:WISE}.

The photometric units were converted using the functions provided by {\sc Photutils}. Uncertainties in the flux density are measured using {\sc ApertureStats} and added in quadrature to the flux calibration errors of 2.4$\%$, 2.8$\%$ and 4.5$\%$ in W1, W2 and W3, respectively \citep{Wright-2010}. The infrared fluxes are listed in Table~\ref{tab:infrared_fluxes}. To check the photometry procedure, we also measured the fluxes within the 22\arcsec\ isophote and compared these values with the catalogued values. We found that for W1, W2 and W3, all but one (NGC~3261) agree within 10\% of the fluxes. The NGC~3261 flux in the W3 band is offset by 12\%. 

\begin{table} 
\centering 
\caption{Integrated WISE infrared flux densities.}
\begin{tabular}{lccc}
\hline\hline
Galaxy       &  W1 &  W2  & W3       \\
             & (mJy) & (mJy)  & (mJy)    \\
\midrule      
NGC~2397  &190$\pm$0.4 &119$\pm$0.3&       531.1$\pm$1.1\\
NGC~2442  &232.4$\pm$0.6 & 143.3$\pm$0.4&     509.9$\pm$1.0 \\
ESO\,059-G012  &  24.1$\pm$0.1 & 14.0$\pm$0.1 &  20.8$\pm$0.1\\
NGC~2434  & 141.7$\pm$0.6 &78.0$\pm$0.3&    20.6$\pm$0.1\\
\midrule           
NGC~3256  &  269.1$\pm$1.2 &249.6$\pm$1.2 &  2354.2$\pm$9.4\\
NGC~3256B  &  45.6$\pm$0.2&   30.5$\pm$0.1&  150.7$\pm$0.6\\
NGC~3261  &   174.8$\pm$0.6 &  164.1$\pm$0.3&   173.3$\pm$0.2\\
NGC~3262 &31.0$\pm$0.1 &17.2$\pm$0.1 & 4.9$\pm$0.1 \\
NGC~3256C  &  32.0$\pm$0.1   &20.4$\pm$0.1   &116.1$\pm$0.3\\
NGC~3263  &  182$\pm$0.6  & 111.7$\pm$0.6  & 478.9$\pm$0.6\\
ESO\,263-G044  & 2.8$\pm$0.1       &1.7$\pm$0.1     &3.9$\pm$0.1\\
NGC~3256A  &  3.5$\pm$0.1        &2.0$\pm$0.1    &1.8$\pm$0.1\\
ESO\,263-G046  & 3.2 $\pm$0.1     &1.8$\pm$0.1& 4.5$\pm$0.1\\
\midrule
NGC~6215  &   174.2$\pm$0.5&  112.3$\pm$0.3   &914.6$\pm$2.3\\
NGC~6221  &  475.8$\pm$1.0  &333.6$\pm$0.7    &2026.5$\pm$4.1\\
 \bottomrule
\end{tabular}
\label{tab:infrared_fluxes}
\end{table}

\subsection{WISE colour-colour diagram}
\label{appen:color-color}

\begin{figure}
\centering
\includegraphics[width=0.85\columnwidth]{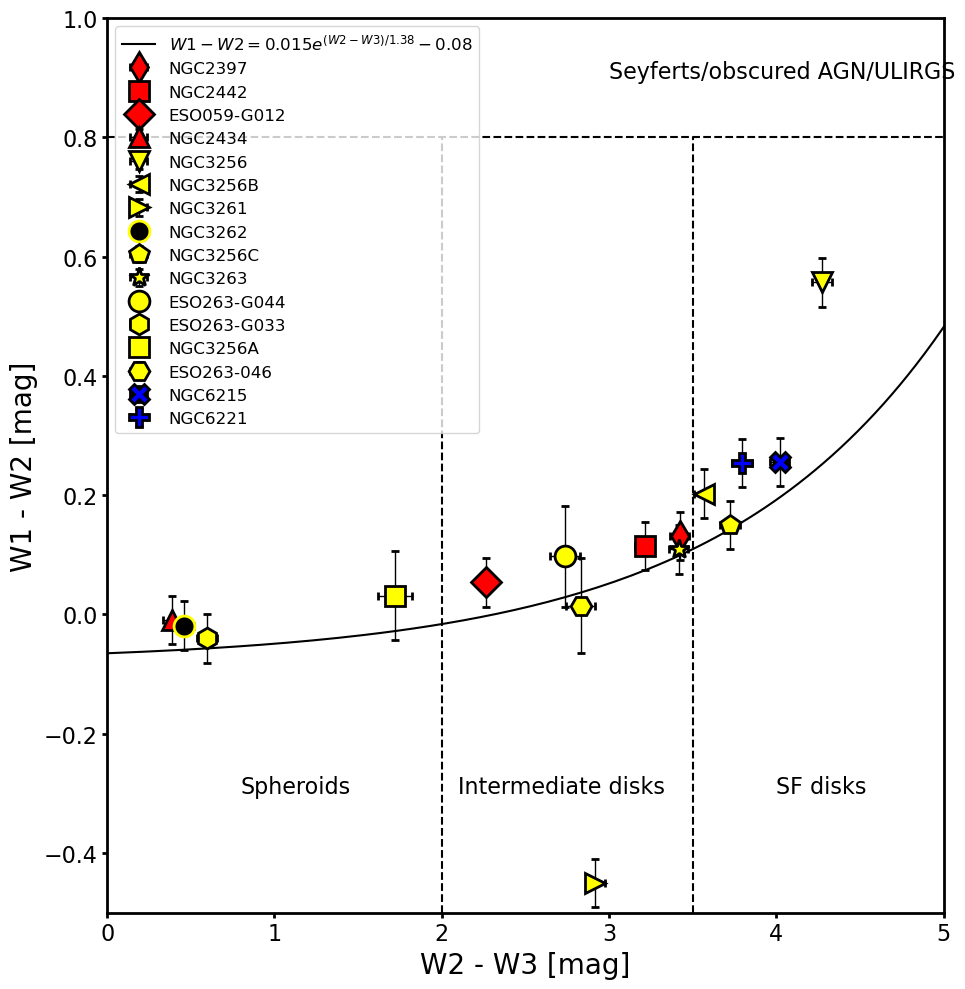}
\caption{WISE colour-colour diagram following \citep{Cluver_2017}. Early-type galaxies with low star formation rates are located at the bottom left ($W2 - W3 < 2$), and star-forming disks are expected to be on the right side ($W2 - W3 > 3.5$). Intermediate disks are likely to be found between these regions. Above $W1 - W2  = 0.8$, heating from dusty active galactic nuclei (AGN) primarily influences the mid-infrared emission. Error bars are displayed for each object unless smaller than the symbol. The black line traces the ``star formation sequence'' defined by \citet{Jarrett_2019}. The galaxy represented with a black dot is not detected in our radio continuum images.}
   \label{fig:class}
\end{figure}

The WISE colour-colour diagram presented in \citet{Wright-2010} has been explored as a diagnostic tool for large samples of galaxies \citep[e.g.][]{Cluver_2017}, large-sized galaxies \citep{Jarrett_2019}, and pairs of galaxies \citep{bok-2020}. The $W2 - W3$ axis is commonly used to indicate the dust content and so roughly follows the galaxy type. For example, low-star-forming, dust-free early-type galaxies have bluer $IR$ colours and are expected to lie in the bottom left of the colour-colour diagram ($W2 - W3 < 2$), while star formation dominated, dusty disk galaxies have redder IR colours ($W2 - W3 > 3.5$) and occupy the right side of the diagram. The intermediate disk spiral galaxies are between these extremes ($2.0 <W2 - W3< 3.5$). The $W1 - W2$ colour traces the amount of hot dust. Above $W1 -  W2 = 0.8$, are expected to lie the active galactic nuclei,  assuming that the mid-infrared emission primarily comes from the dust heated by AGN. The WISE colour-colour diagram of our galaxy sample is shown in Fig.~\ref{fig:class}. In the same plot, we depict the ``star formation sequence'' curve fit by \cite{Jarrett_2019} to 100 large galaxies. Departures from this sequence may reveal nuclear, starburst, and interacting/merging events. 
Thus, the WISE colour-colour diagram suggests that most galaxies in our sample are unlikely to harbour a powerful AGN. The highest $W1 - W2$ value in our sample is associated with NGC~3256, for which there is evidence of an obscured, low luminosity AGN \citep{Ohyama_2015}.

\subsection{Infrared-radio correlation}

The infrared radio correlation is one of the most intriguing relations in astronomy. It is widely studied by many authors, empirically in local universe \citep{Helou-1985,condon-1992,Yun-2001,Bell_2003}, at different redshift \citep{Seymour_2009,Basu_2015,Delhaize_2017, Yoon_2024,Molnar-2021} and from a detailed analysis of certain galaxies \citep[e.g. NGC~6946,][]{Tabatabaei-2013} to statistical analysis of many galaxies local galaxies \citep{Shao-2018,Grundy-2023}.
 
\begin{figure}
\centering
    \includegraphics[width=0.9\columnwidth]{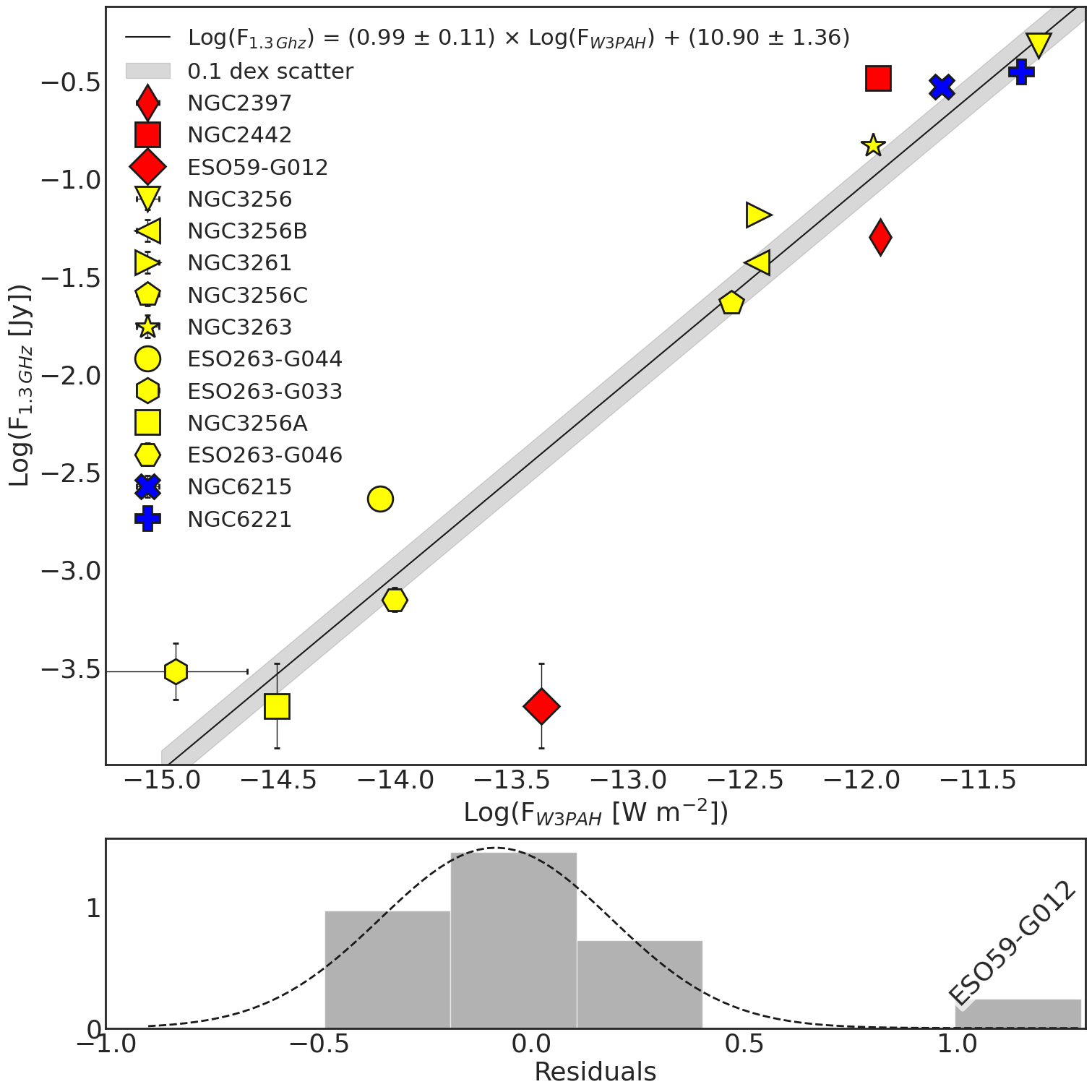}

\caption{Infrared-radio correlation using only the PAH component and radio continuum fluxes. Top panel: $\log(F_{\rm W3PAH})$ vs.\ $\log(F_{\rm 1.3GHz})$ for galaxies with enough data (see text), following the legend.  
Error bars are shown unless smaller than the symbol. A linear fit, $\log(y) = m \cdot \log(x) + b$, is plotted as a black line, with a shaded region indicating a 0.1~dex scatter. Bottom panel: histogram of the squared differences between observed and fitted values. }
\label{fig:FIRC}
\end{figure}

Our work studies the infrared-radio relation, focused on the infrared WISE W3 band. It is an excellent tracer of ISM emissions, with contributions from polycyclic aromatic hydrocarbons (PAH) in the photodissociation regions. 
However, the WISE W3 band is contaminated by evolved stellar populations. Thus, \cite{Cluver_2017} proposed a method to remove this component from the evolved stars using the W1 band emission. The method consists of subtracting 15.8\% of the W1 band flux from the W3 band flux for both the global and pixel-mapped flux densities. We denote these modified W3 band flux densities and any other measurements derived from them using the W3PAH subscript. Two galaxies have been excluded from the calculations: NGC~2434 due to its negative $F_{\rm W3PAH}$ value, which suggests that the emission is dominated by the evolved stellar population with no significant PAH contribution, and NGC~3262 because it was not detected in our radio continuum images. The W3PAH flux for each galaxy is listed in Table~\ref{tab:prop}. In Fig.~\ref{fig:FIRC}, it is shown that the W3PAH flux density correlates with 1.4~GHz radio continuum flux density. Therefore, we inferred the best-fit model:

\begin{equation}
log \left( \frac{F_{\rm 1.3GHz}}{\rm Jy} \right) = m \cdot log \left( \frac{F_{\rm W3PAH}}{\rm W \, m^{-2}} \right) + b .
\end{equation}

We use the least squares method, which minimises the squared orthogonal distances to the modelled relation. We measure the dispersion as the standard deviation of the data's orthogonal offset distribution relative to the best-fit model. We find $m=0.99\pm 0.11$, $b=10.9\pm1.4$ and the dispersion is $\sigma = 0.3$. The histogram shows the squares of the differences between the observed and fitted values. The galaxy ESO\,059-G012 lies outside the three-sigma range (Fig.~\ref{fig:FIRC}). To measure the W3PAH-radio relation, we computed the more commonly used parameter defined as, 
\begin{equation}
    q_{\rm W3PAH} = log \left( \frac{F_{\rm W3PAH}}{3.75 \times 10^{12}~\rm W~m^{-2}} \right) - log \left( \frac{F_{\rm 1.4GHz}}{10^{26} ~\rm Jy} \right),
    \label{eq:W3pah-radio}
\end{equation}
\noindent
where $F_{\rm 1.3GHz}$ is the radio continuum flux and $F_{\lambda}$ depends the infrared data; see \citep{Yun-2001} for further details. We find a mean value of $ q_{\rm W3PAH} = 2.5 \pm 0.1$. Fig.~\ref{fig:qlum} shows this $q_{\rm W3PAH}$ value as a function of equivalent W3PAH and radio continuum luminosity. The galaxy ESO\,059-G012 has three times larger infrared flux density than expected using \cite{Yun-2001} empirical relation. In Section~\ref{sec:discussion}, we will discuss the physical properties inferred from these results.

We estimate the total infrared (TIR) luminosity implementing the relationship between W3PAH and TIR found in \cite{Cluver_2017}:

\begin{equation}
    log \left( \frac{L_{\rm TIR}}{L_{\odot}} \right) = (0.889\pm0.018) log \left( \frac{\nu L_{\rm W3PAH}}{L_{\odot}} \right) + (2.21\pm 0.15),
    \label{eq:TIR}
\end{equation}

\noindent
where $\nu L_{\rm W3PAH}$ is the monochromatic W3PAH luminosity. The infrared-radio relation remains linear. The fitted values are $m_{TIR}=1.1\pm0.12$ and $b_{TIR}=9.27\pm1.24$ and the dispersion is $\sigma_{\rm TIR} = 0.5$. The $q_{\rm TIR}$ values are slightly lower than $q_{\rm W3PAH} $ values. We find a mean value of $ q_{\rm TIR} = 2.1 \pm 0.5$. Fig.~\ref{fig:qlum} shows the $q_{\rm TIR}$ value as a function of total infrared luminosity.

\subsection{SED fitting}
\label{sec:SED-fitting}
We performed a photometric analysis in order to construct spectral energy distributions (SEDs) for a subset of the galaxies in our sample. Using the photutils package, we remeasured fluxes in all four WISE bands and the radio continuum image. The Kron isophote was determined in the WISE W1 band and applied consistently across the other WISE bands and the radio image to ensure matched apertures. We were able to obtain reliable WISE photometry in all four bands only for the following galaxies: ESO59-G012, NGC~
2434, NGC~3256, NGC~3256B, NGC~3256C, NGC~3263, NGC~6215, and NGC~6221. We performed the SED fitting using WISE and radio continuum data only. For NGC~2434, we used catalogued photometric data\footnote{\url{http://vizier.cds.unistra.fr/vizier/sed/?submitSimbad=Photometry&-c=07+34+51.1504542360-69+17+02.996020032&-c.r=5&-c.u=arcsec&show_settings=1}}.

This modelling was conducted using GalaPy \citep{Ronconi2024}, an open-source software package implemented in Python/C$^{++}$ capable of modelling emission from the X-ray to the radio regimes. The underlying stellar populations were assumed to follow a Chabrier initial mass function \citep{Chabrier2003}, within a $\Lambda$CDM cosmological framework \citep{Planck2020}. The main results are listed in Table~\ref{tab:SED-results}, along with a refined SFR estimation, the stellar age and stellar mass.  A complete description of the process is presented in section~\ref{appen.SED_modeling}.

\begin{figure}
\centering
   \includegraphics[width=0.8\columnwidth]{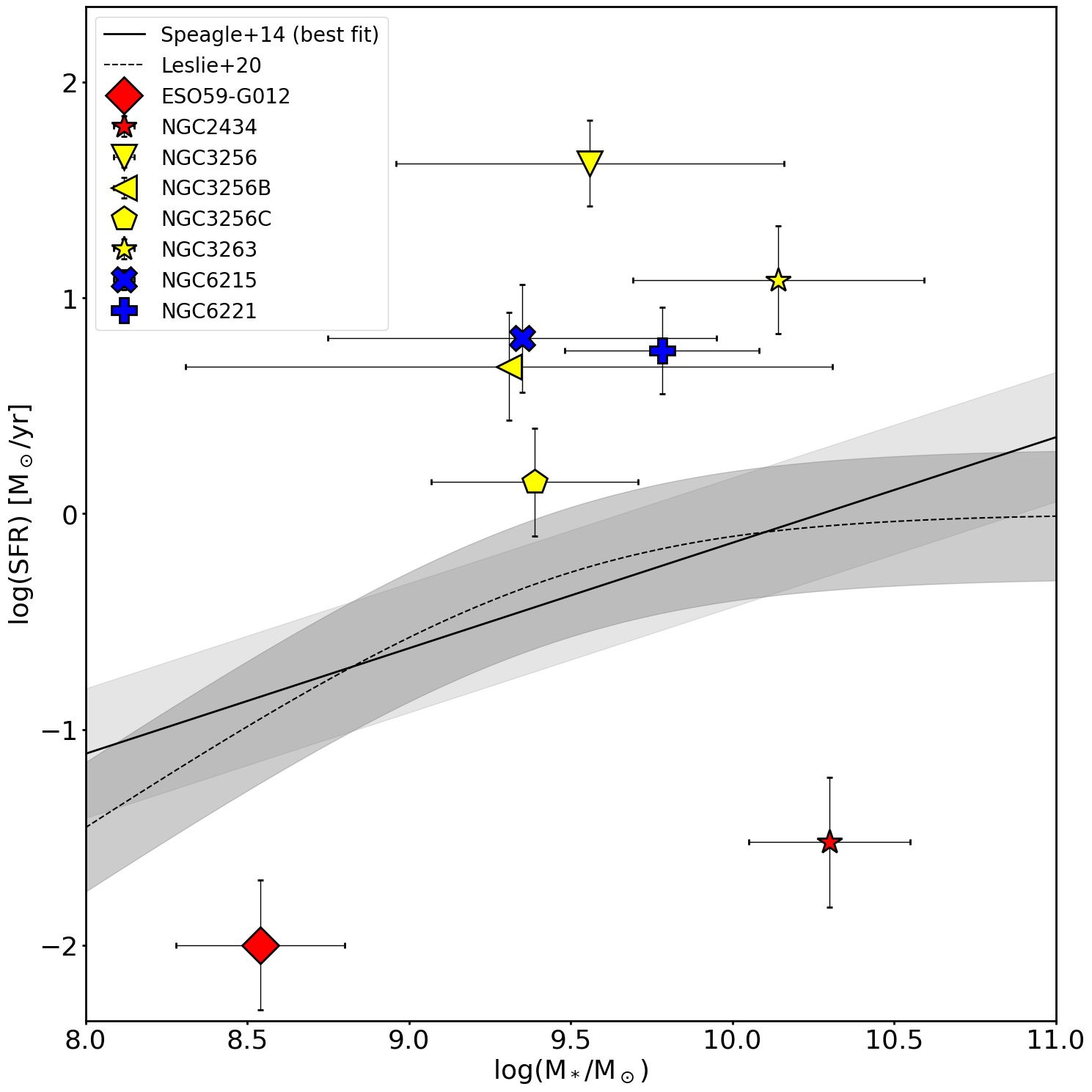}\\
   \caption{Star formation rate as a function of stellar mass, both quantities derived from the SED fitting analysis. Main sequence relation from \citet{Leslie-2020} (dashed line). Best fit relation from \citet{Speagle-2014} (solid black line). The shaded region indicates the $0.3$~dex scatter \citet{Leslie-2020, Speagle-2014}.}
   \label{fig:MS}
\end{figure}

To place our galaxies in the context of the star-forming main sequence, we compare their stellar masses and star formation rates with the parametrisations of \cite{Speagle-2014} and \cite{Leslie-2020}, as shown in Fig.~\ref{fig:MS}. The former provides a redshift-dependent relation widely used across cosmic time, while the latter offers an updated characterisation based on a consistent analysis of nearby galaxies using multiple SFR tracers.

\subsection{Star-formation properties}

We estimate the SFR of the galaxies using the 20~cm radio continuum flux density and the W3PAH monochromatic luminosity. SFR based on the 20~cm radio continuum data was estimated using the formula by \cite{Condon-2002}. The recent high-mass SFR ($>$5\Msun) is given by:

\begin{equation}
  SFR~(\rm M_{\odot}\,yr^{-1} ) = 0.03 \cdot D^2 \cdot S_{20},
  \label{eq:SFR_cont}
\end{equation}
\noindent
where $D$ is the distance in Mpc and $S_{20}$ the 20~cm radio continuum flux density in Jy. To estimate the SFR of all stars ($>$0.1\Msun), we multiply the previous expression by 4.76; see \cite{Condon-2002}. The estimation of the SFR using W3PAH contribution is calculated using the following formula:

\begin{equation}
    \rm log \left(  \frac{SFR_{\rm W3PAH}} {\rm M_{\odot}\,yr^{-1}}    \right) = 1.13 \, \rm log \left( \frac{\nu~L_{\rm W3PAH}}{L_{\odot}} \right) - 10.24.
\end{equation}

Here, $\nu$ is the WISE W3 band central frequency (2.498 $\times$ 10$^{13}$~Hz) and L$_{\odot} = 3.828 \times 10^{26}$~W \cite[see][]{Cluver_2014}. Because the initial mass functions (IMFs) adopted in \citet{Cluver_2014} and \citet{Condon-2002} differ, Kroupa IMF \citep{Kroupa-2002} in the former and Salpeter IMF \citep{Salpeter-1955} in the latter, we apply a correction to ensure consistency. Specifically, to SFRs derived assuming a Kroupa IMF to those corresponding to a Salpeter IMF, we divide by a factor of 0.67 \citep{Madau-2014}.
The results are listed in Table~\ref{tab:SFR}. Despite the evolutionary stage of the host group, most of the galaxies in our sample show a linear trend between their SFRs as determined by their W3PAH and radio continuum flux density; see Fig.~\ref{fig:SFR}. In particular, the linear relation is close to 1:1 if considering the recent high-mass SFR above 5\Msun, only one galaxy lies outside the 10\% scatter: ESO\,059-G012. In contrast, when considering the star formation rate of stars with solar masses above 0.1, two galaxies are located outside the $10\%$ scatter of the 1:1 linear relation. These galaxies are ESO\,263-G033 and ESO\,263-G044.

\begin{table} 
\centering 
\caption{Derived SFR properties } 
\begin{tabular}{l@{~~}c@{~~}c@{~~}c@{~~}c@{~~}} 
\hline\hline
Galaxy       & SFR$_{\rm 20}$	 & SFR$_{\rm 20}$  & SFR$_{\rm W3PAH}$\tablefootmark{a}\\
             &  ($>$5\Msun\,yr$^{-1}$) &   ($>$0.1\Msun\,yr$^{-1}$) &  (\Msun\,yr$^{-1}$)  \\
\hline 
NGC~2397 &   0.4       &  2.0    &  1.4    \\
NGC~2442 &   2.8       &  13.4   &  1.4    \\
ESO\,059-G012 &   0.001      &  0.008    &  0.03   \\
NGC~2434 &   0.056      &  0.26    &  ...   \\
\midrule
ESO\,263-G033 &   0.01       &  0.06    &  0.003     \\       
NGC~3256 &   20.6       &  98.15    &  50   \\
PGC~087403 & 0.1 & 0.02 &...\\
LEDA\,087403 &   0.02       &  0.10     &   ...       \\
LEDA\,3097821 &   0.01       &  0.08     &   ...        \\
NGC~3256B &   1.6        &  7.62     &  2.1   \\
NGC~3261 &   2.81       &  13.40    &  2.2    \\
NGC~3256C &   0.99       &  4.74     &  1.6    \\
NGC~3263 &   6.41       &  30.5     &  7.9   \\
ESO\,263-G044 &   0.09       &  0.43     &  0.03     \\
NGC~3256A &   0.008     &  0.04    & 0.01 \\
LEDA\,533801  &   0.04  &  0.19    & ...     \\
ESO\,263-G046 &   0.03  &  0.14    & 0.03\\
\midrule
NGC~6215 &   2.8       &  13.5    & 3.2 \\
NGC~6221 &   3.4       &  16.3    & 7.7 \\
\hline
\end{tabular}
\tablefoot{References: SFR$_{\rm 20}$ \citep{Condon-2002}, \tablefoottext{a}{SFR$_{\rm W3PAH}$ / 0.67 to convert from Kroupa IMF to Salpeter IMF} \citep{Cluver_2014}}
\label{tab:SFR}
\end{table}

\subsection{Radio spectral index}
A galaxy's spectral index ($\alpha$) provides insight into the different processes occurring within the galaxy. The spectral index is defined by the integrated flux density $S$ at frequency $\nu$; here, we used the convention $S_{\nu} \propto \nu^\alpha$. We measure the spectral indices by applying the in-band method. This method takes advantage of a single observation with a wide bandwidth. Our MeerKAT observations have a bandwidth of 856~MHz. However, to improve the rms of the continuum images to calculate the spectral index map, we combined channels 4 with 5 and channels 8 with 9. Thus, the in-band spectral index was calculated over a frequency range of 324 MHz, from 1021 MHz to 1345 MHz. The final rms in each image is 6.5~$\mu$Jy and 2.1~$\mu$Jy. We applied the primary beam correction before the calculation. Despite the effort to improve the signal-to-noise ratio, we can only obtain confident global spectral index values for the strongest radio-emitting galaxies within the primary beam corrected spatial region of each image; see Table~\ref{tab:prop}.
According to observations \citep{Lisenfeld-2000}, thermal emission has an optically thin spectral index of $\alpha \geq -0.1$, while non-thermal emission can have a spectral index that ranges between approximately $-0.5$ to $-1.1$, where $-0.5$ value corresponds to the non-thermal emission originating in star-forming regions where thermal fraction is high.

\begin{figure}
    \centering
    \includegraphics[width=\columnwidth]{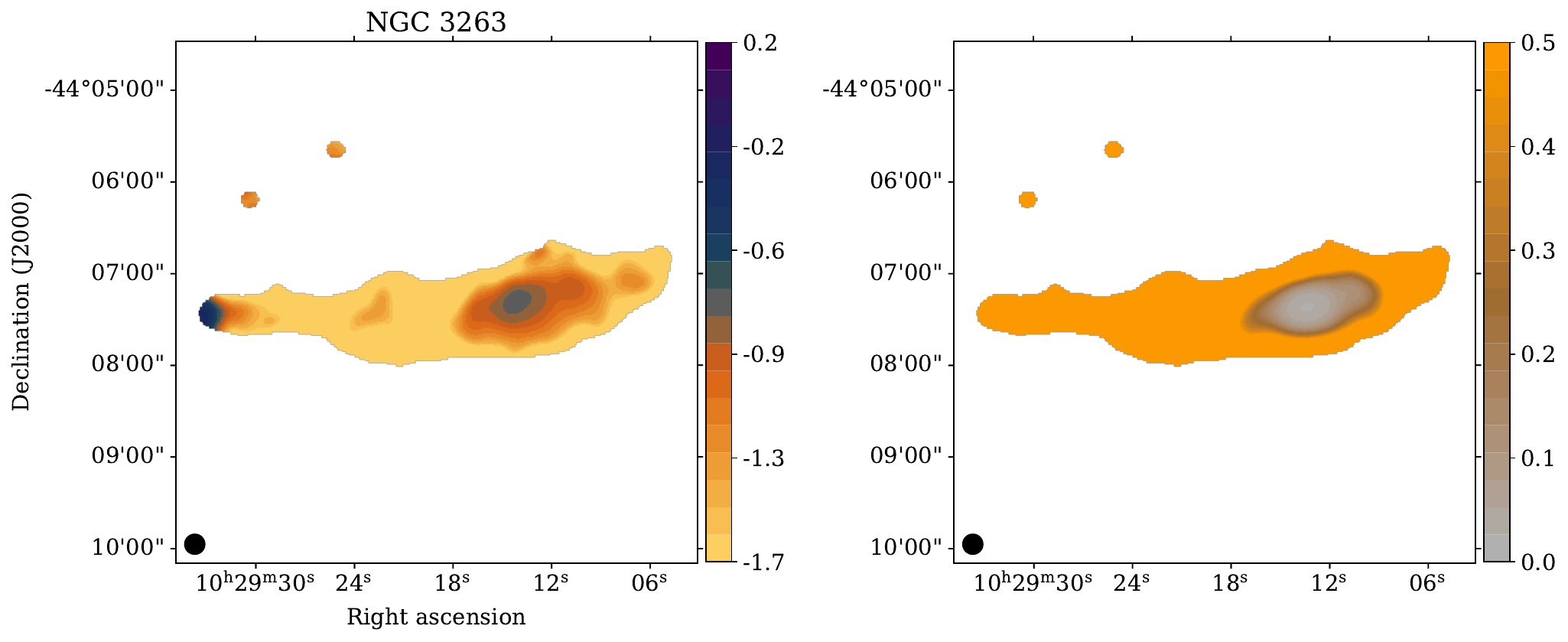}
 \includegraphics[width=\columnwidth]{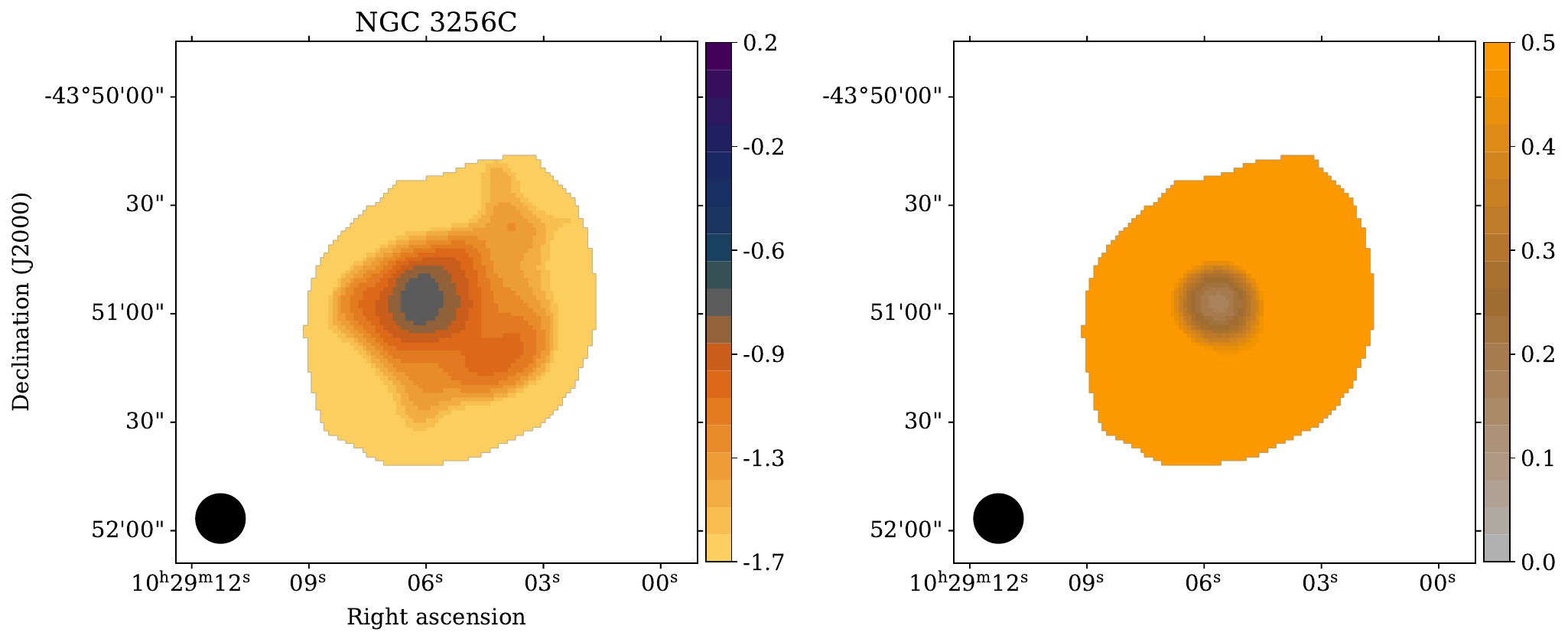}
  \includegraphics[width=\columnwidth]{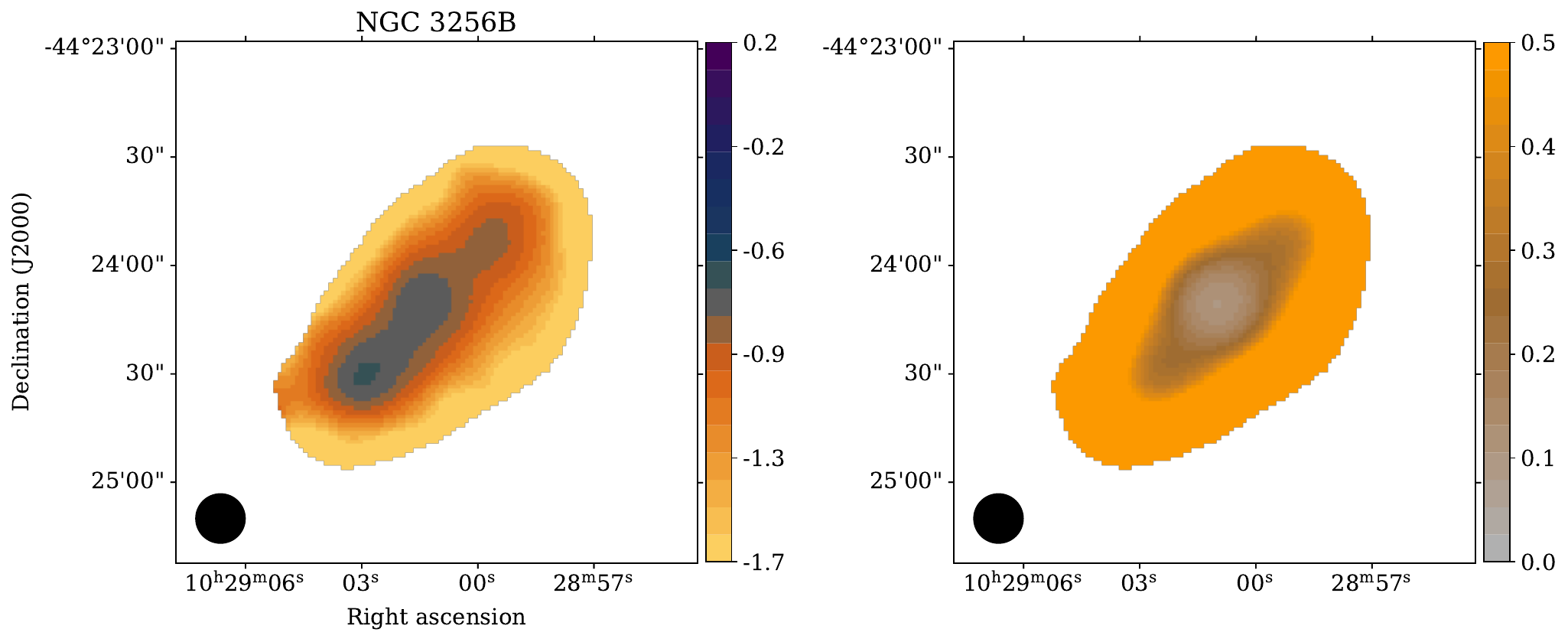}
  \includegraphics[width=\columnwidth]{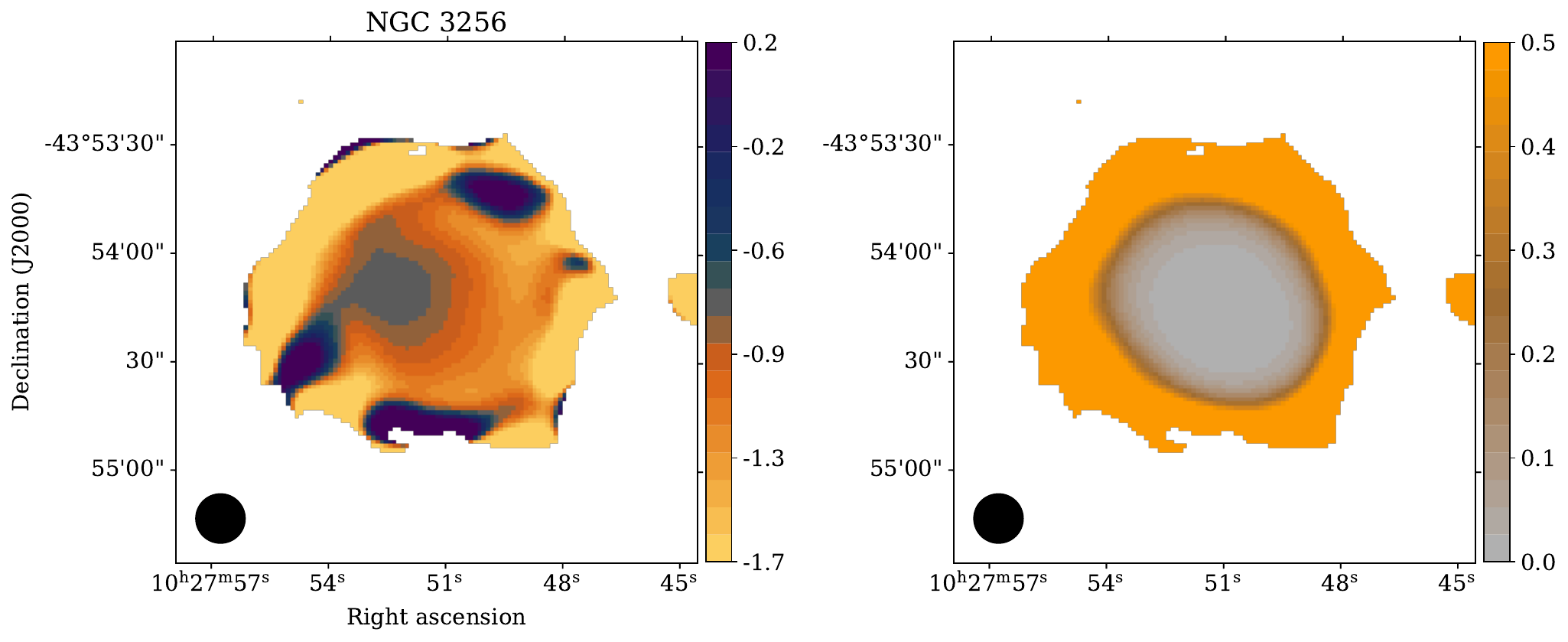}
\includegraphics[width=\columnwidth]{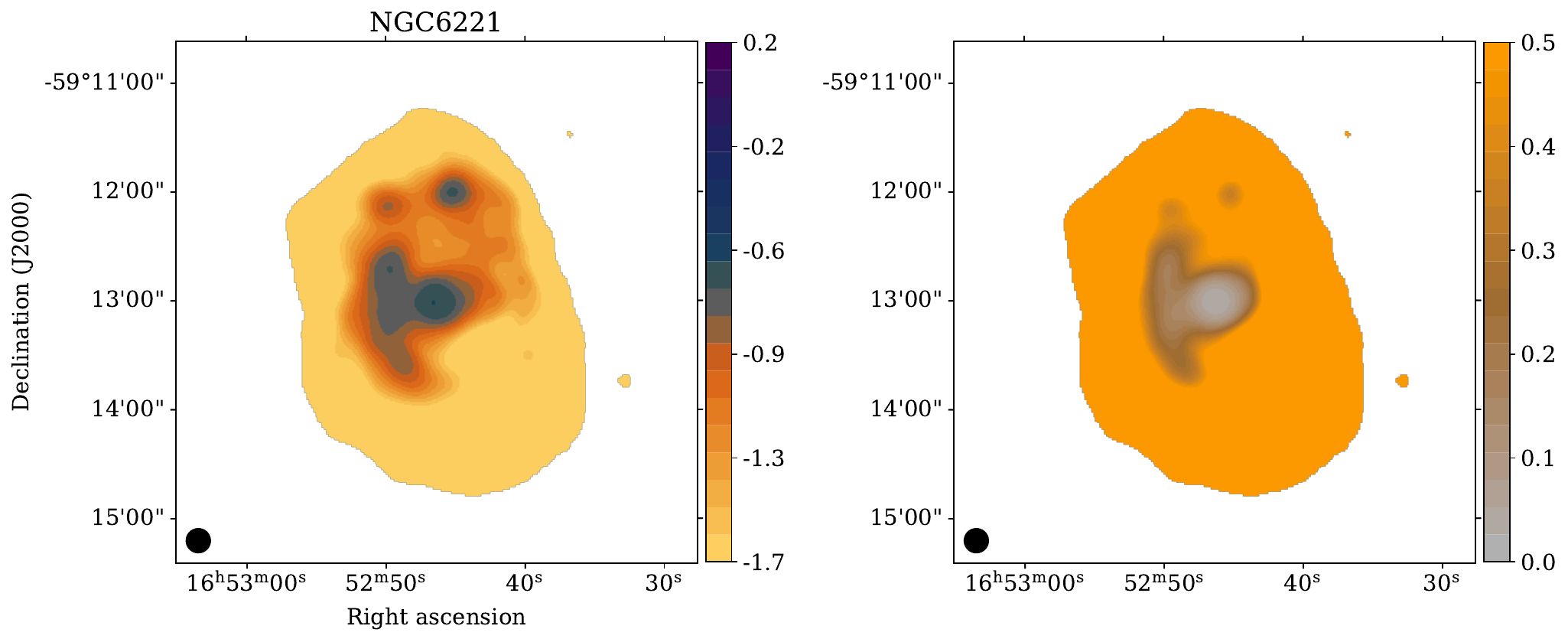}
\includegraphics[width=\columnwidth]{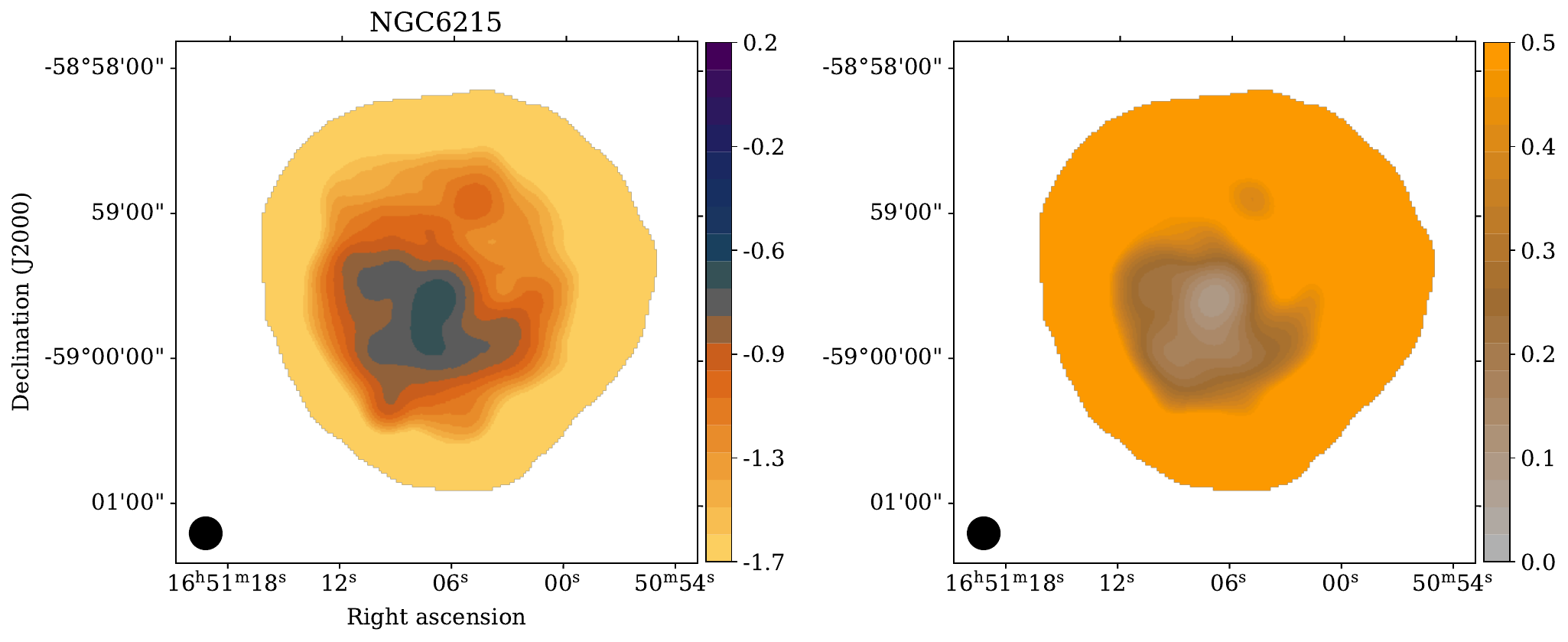}
\caption{The resolved spectral index and error maps of the brightest galaxies in our sample. The dark plum colour pinpoints the flat spectral index associated with thermal emission, while orange marks steep spectral values associated with non-thermal emission \citep{Cosmoscanvas}.}
    \label{fig:spectral_index}
\end{figure}

We created the resolved spectral index map by using the spectral index definition for each pixel with a sigma-to-noise ratio above 5$\sigma$. To prevent bias in $\alpha$, we compared the flux cuts, which were chosen based on sensitivity limits and frequencies. In Fig.~\ref{fig:spectral_index}, we show the spectral index and error maps for the galaxies NGC~3263, NGC~3256C, NGC~3256B, NGC~3256, NGC~6221 and NGC~6215. The galaxy NGC~2442 is close to the edge of the region affected by the primary beam correction, resulting in high error maps. The dark plum colour (more positive than $\alpha=-0.1$) in our maps pinpoints the regions of flat spectral index associated with thermal emission. However, positive values can also be associated with errors in the map. The orange (more negative than $\alpha=-0.8$) through yellow colours mark steep spectral values associated with non-thermal emission \citep{Cosmoscanvas}. 

Our spectral index maps show some structure in the spectral index distribution within the galaxies despite being created using the in-band spectral method. 
Some galaxies have regions with in-band spectral indices that are more positive than -0.2. However, these pixels occur in spatial areas of high error and may be associated with signal-to-noise artefacts -- an example is NGC~3256. Within the spatial regions associated with low uncertainty, the flattest value observed is around $-0.6$, and generally these pixels occur in areas with strong radio continuum emission near the nuclei of the galaxies.  The exception is NGC~3256C, in which the spectral index value $\approx -0.6$ is located in the disk plane but offset from the galaxy's centre. The steepest values observed in all galaxies are around $-1.7$, and these pixels occur towards the peripheral regions of the disk.

In Fig.~\ref{fig:qvssp}, we compare the spectral index value (see Table~\ref{tab:prop}) of the galaxies with the obtained $q_{\rm W3PAH}$ parameter. All the galaxies in the plot show a radio spectral index steeper than typical $\alpha = -0.8$. Also, $q_{\rm W3PAH}$ values are higher than $q = 2.34$ found by \citet{Yun-2001}, except for the galaxy NGC~3263 ($q_{\rm W3PAH} = 2.3$).
\cite{Bressan-2002} revisits the nature of the far infrared-radio correlation and explores a possible relation between spectral index and $q$ parameter in the context of evolutionary models for starburst galaxies, predicting that galaxies follow specific tracks as they age. They noted that more evolved systems, located toward lower $q$ and steeper spectral index, tend to have higher stellar masses. Additionally, \cite{Thomson-2014} found observational evidence supporting this framework in a sample of high-redshift, and highlighted that galaxies with steep radio spectra are often associated with merger or interaction processes.

\subsection{\texorpdfstring{$q_{\rm W3PAH}$}{qW3PAH} maps}

\begin{figure*}
\centering
\includegraphics[width=0.6\columnwidth]{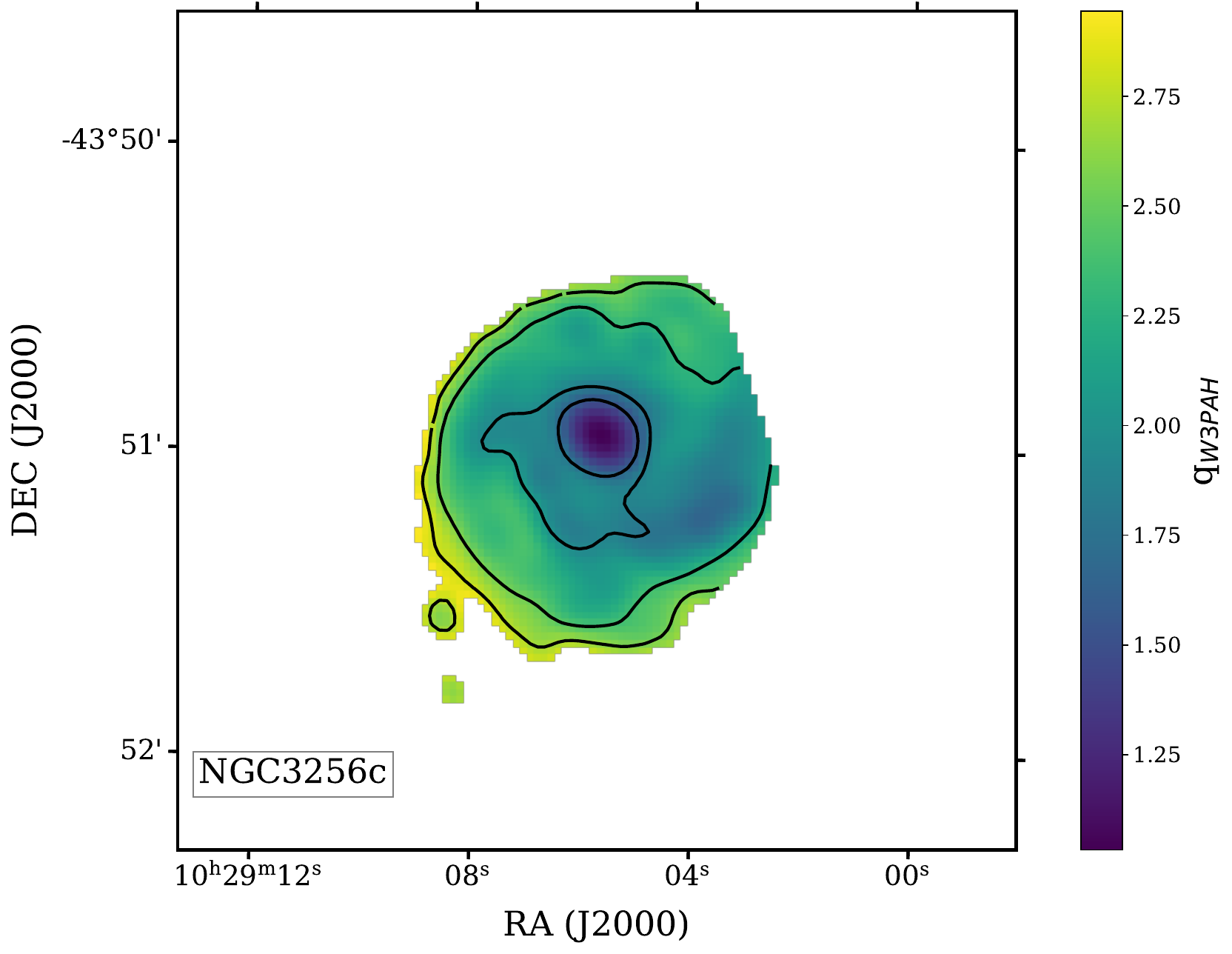}
\includegraphics[width=0.6\columnwidth]{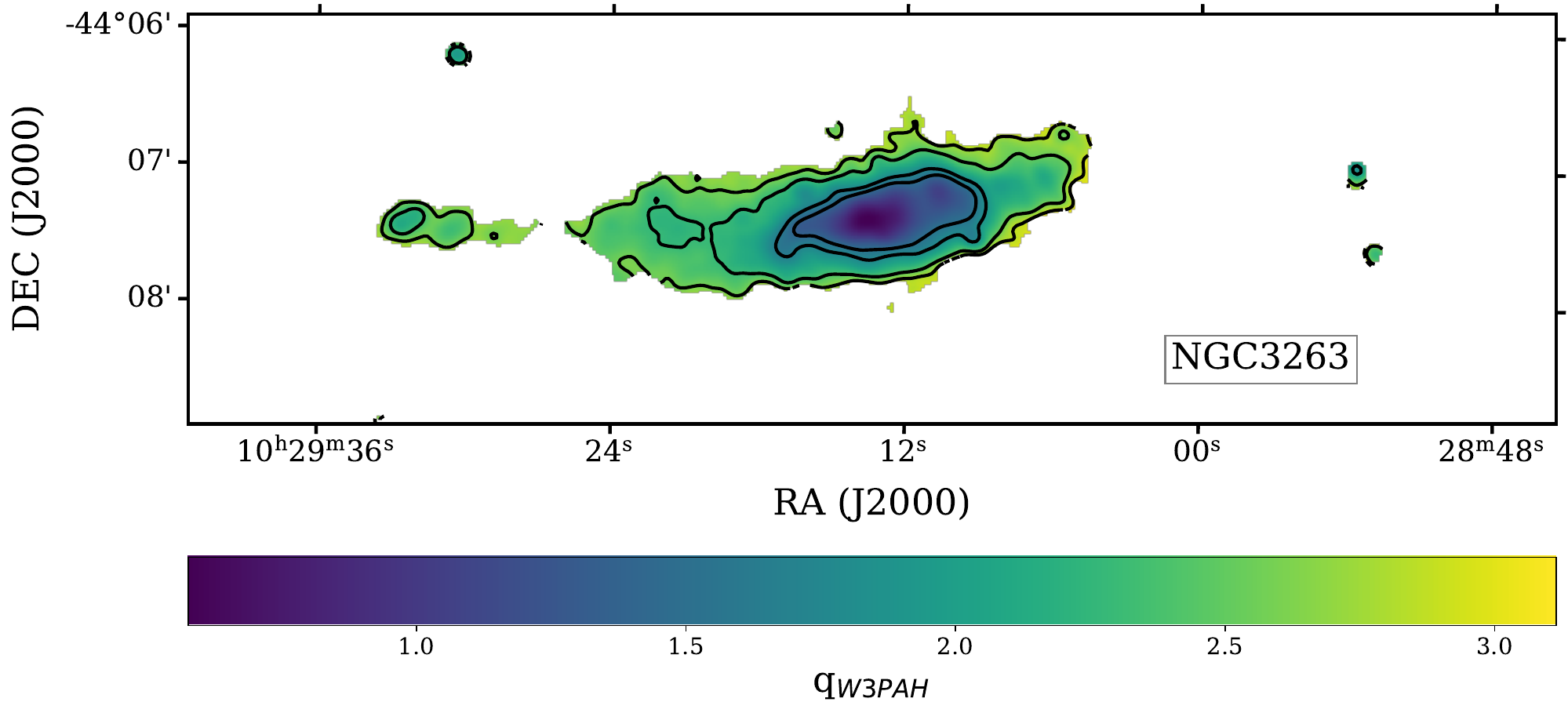}
\includegraphics[width=0.6\columnwidth]{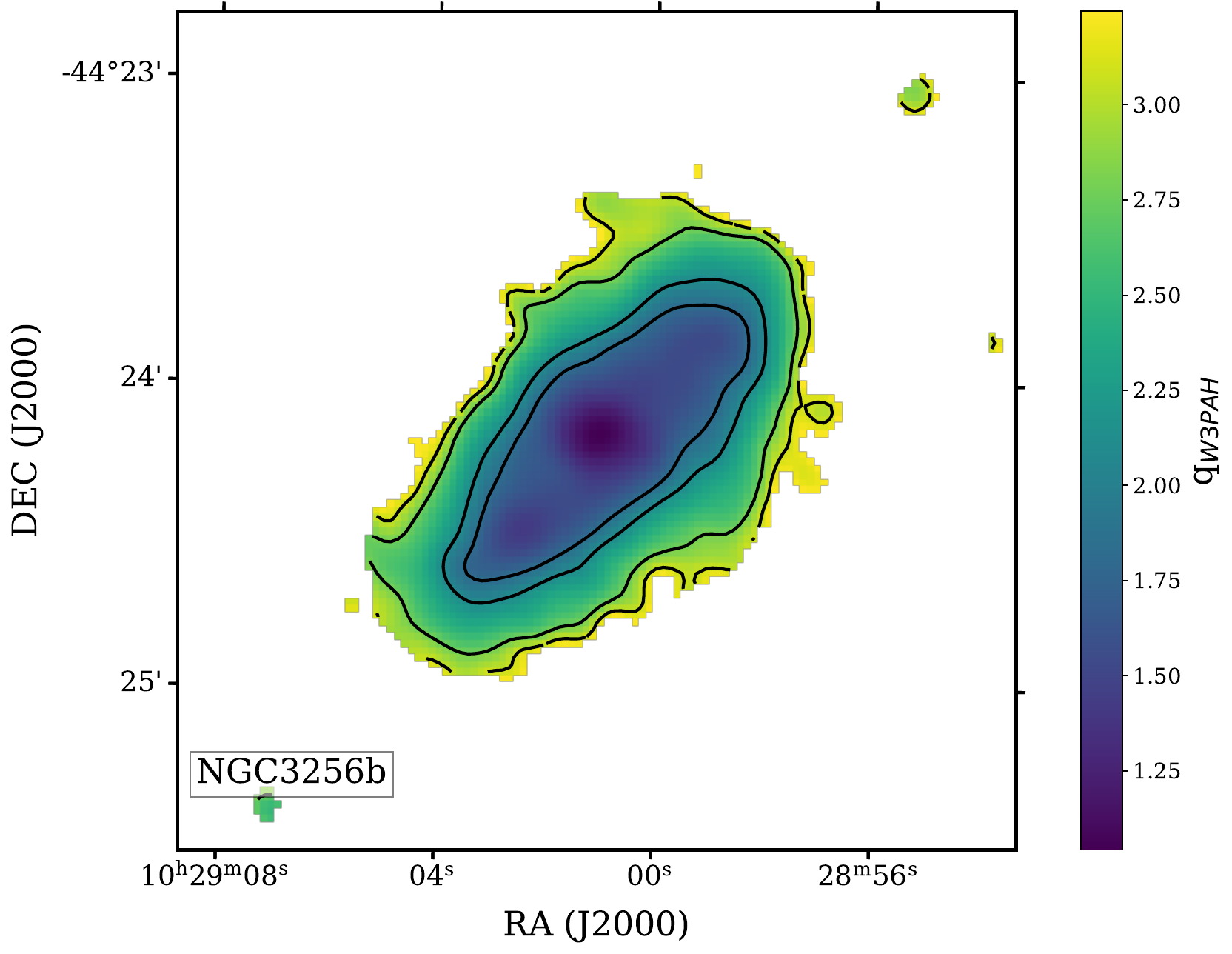}\\
  \includegraphics[width=0.6\columnwidth]{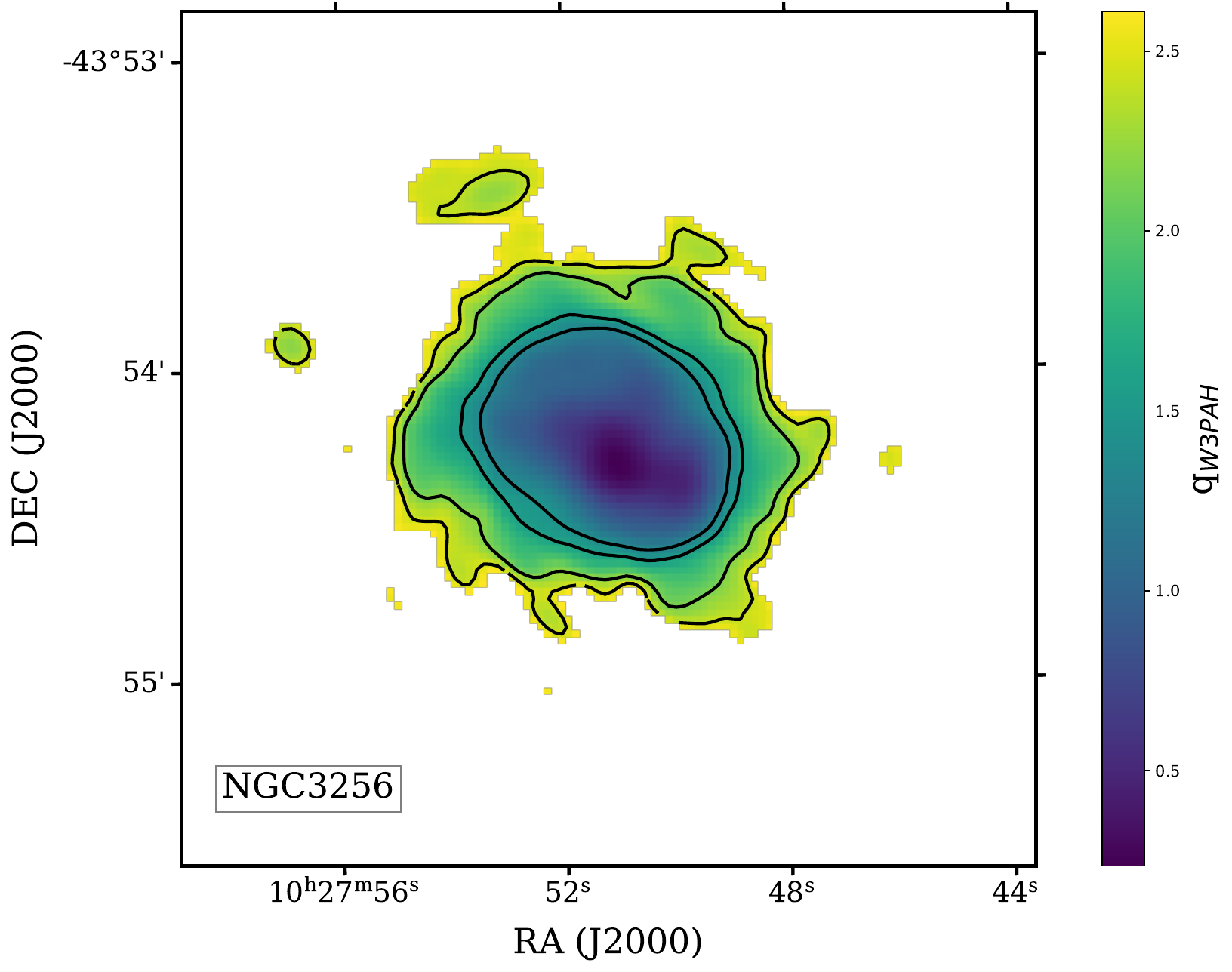}
\includegraphics[width=0.6\columnwidth]{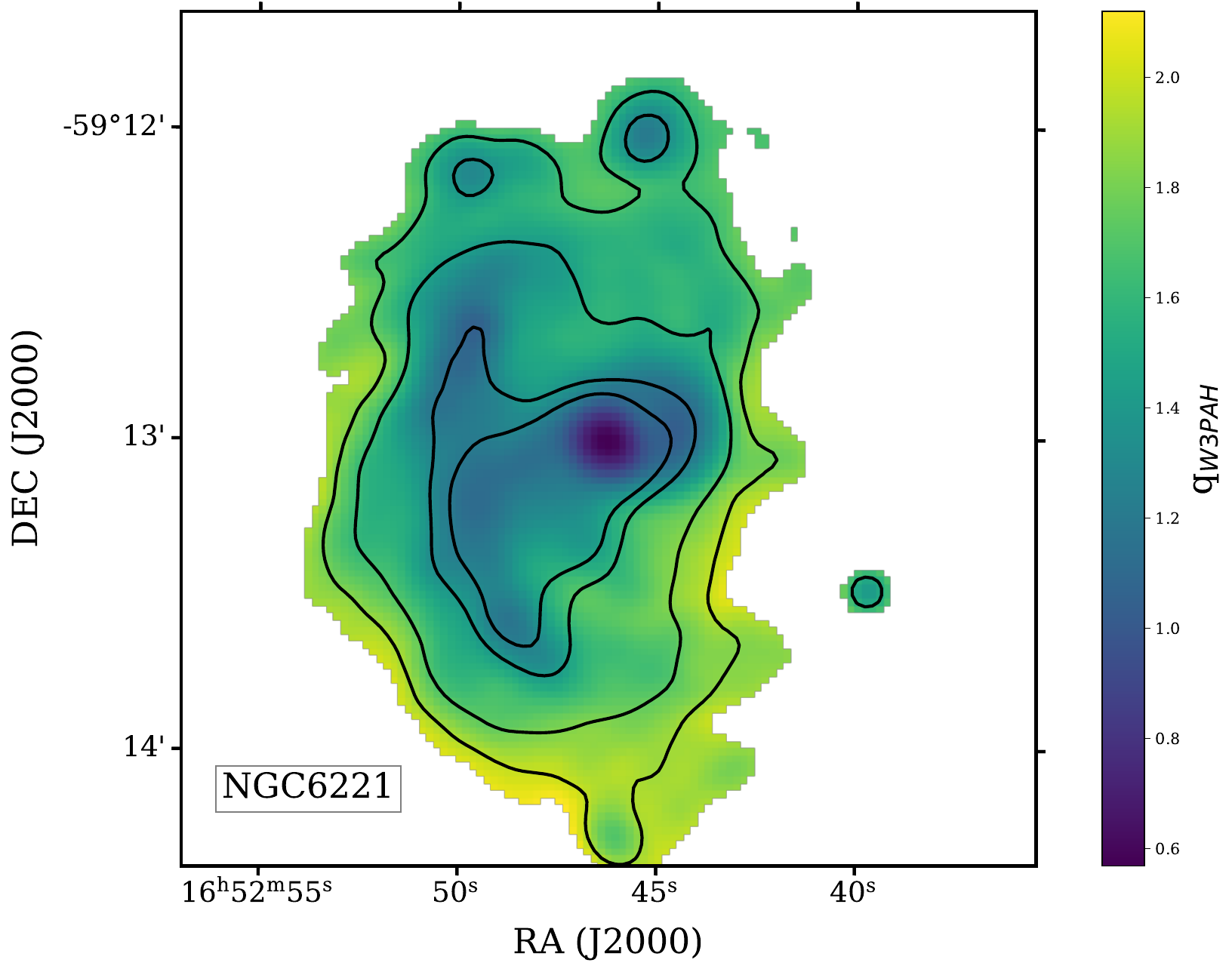}
\includegraphics[width=0.6\columnwidth]{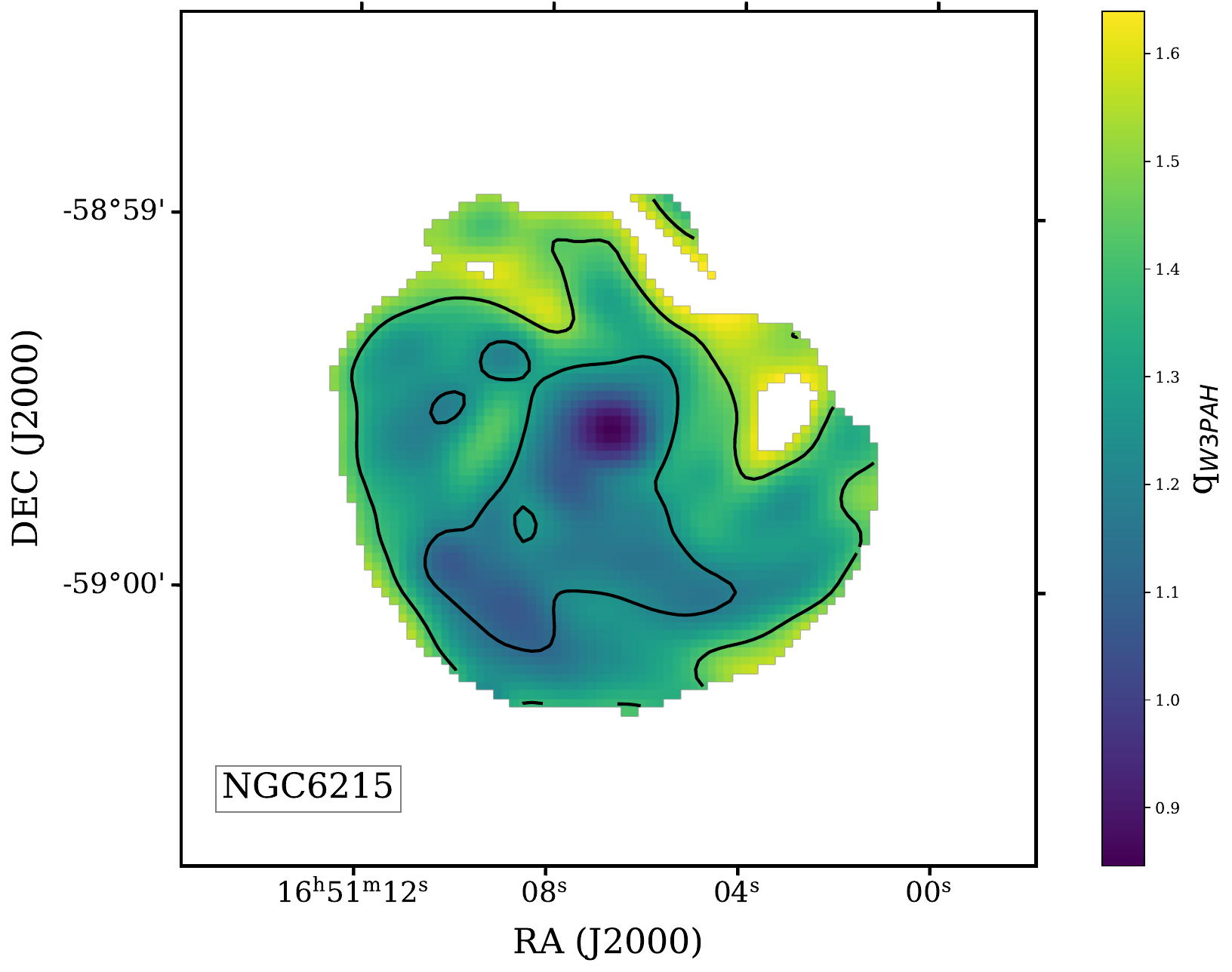}
\caption{NGC~3263, NGC~3256C, NGC~3256B, NGC~3256, NGC~6221, NGC~6215 $q_{\rm W3PAH}$ maps. The colour bar shows the $q_{\rm W3PAH}$ range values.  Black contours correspond to errors 0.005, 0.01, 0.02, 0.05, in parameter determination. High values of $q_{\rm W3PAH}$ indicate radio-deficiency while low values indicate radio-excess.}

\label{fig:qmaps}
\end{figure*}

The $q_{\rm W3PAH}$ maps show the ratio of W3PAH to radio emission across different regions of a galaxy, as indicated in Fig.~\ref{fig:qmaps}. Physically, a high value of $q_{\rm W3PAH}$ indicates strong star formation, where dust heated by young stars dominates the infrared emission and a low value may point to non-thermal processes like synchrotron emission from AGN, where radio emission is stronger. The $q_{\rm W3PAH}$ maps are created using Eq.~\ref{eq:W3pah-radio} on a per-pixel basis. Each pixel at MeerKAT and W3PAH images requires emission above three sigma detections. 

The galaxies have $q_{\rm W3PAH}$ maps with a lower median than their global $q_{\rm W3PAH}$ values. The discrepancy between the global and resolved $q_{\rm W3PAH}$ values likely arises from the spatial scale at which each measurement is performed.
The galaxies NGC~3263, NGC~3256C, NGC~3256B, and NGC~6215 exhibit in their maps values of $q_{\rm W3PAH}$ between $1.0$ and $3.0$, while NGC~3256 and NGC~6221 near the centre of the galaxies show values of $q_{\rm W3PAH}$ less than $0.5$.

\subsection{Degree of interaction}

\cite{Haan-2011} proposed a merger stage classification based on $I$-band images. The numerical classification is defined in the following way: 0 = non-merger, 1 = separate galaxies, but disks symmetric (intact), no tidal tails, and the maximum line-of-sight velocity difference range is from 65 to 160 \kms, 2 = progenitor galaxies distinguishable with disks asymmetric or amorphous and/or tidal tails, 3 = two nuclei in a common envelope, 4 = double nuclei plus tidal tail, 5 = single or obscured nucleus with long prominent tails, 6 =  single or obscured nucleus with disturbed central morphology and short faint tails. Therefore, we classified the individual merger stages of the galaxies by an integer value ranging from 0 to 6, based on the work of \cite{Haan-2011}, but adding a sub-classification in stage 2. We separate it into two groups: the galaxies that show asymmetric disks and the galaxies with tails. The classification is written in Table~\ref{tab:prop}. Most galaxies in our sample can be classified as stage two, except for NGC~3256 (stage 3), NGC~2434, ESO\,059-G033, and ESO\,059-G012 (stage 0).
In Fig.~\ref{fig:qvsMS}, we compare the merger stage classification with $q_{\rm W3PAH}$ value. The galaxy ESO\,059-G012 has an excess in the W3PAH emission and is classified as stage 0. All galaxies in merger stages 2, 2.5, and 3 display $q_{\rm W3PAH}$ values within the excess lines.

\section{Discussion}{\label{sec:discussion}}

\subsection{Global properties}

The upper limit end of the radio luminosity distribution of the galaxy sample is $<$10$^{23}$~W\,Hz$^{-1}$. This is consistent with that of the field galaxies, other groups such as Hickson compact groups \citep{Menon_1995,Omar_2005} and late-type galaxies in nearby clusters \citep{Reddy_2004}. The W3PAH luminosity range also agrees with what is observed in other groups of galaxies, such as Eridanus \citep{Grundy-2023}.

The WISE colour-colour diagram in Fig.~\ref{fig:class} shows that galaxies are distributed differently in the diagram based on their evolutionary group stage. The galaxies representing an intermediate evolutionary group stage (yellow) span along the diagram, from ellipticals to spirals, normal and star-forming. One (NGC~3256) even resides close to the AGN region of the diagram. The most evolved group (red) presents galaxies in the elliptical and spiral regimes, and the early-stage group (blue) presents galaxies in the star-forming region. This result agrees with the group evolution theory. 

It is shown in Fig.~\ref{fig:MS}, galaxies with clear signs of interaction lie above the MS defined by \cite{Leslie-2020} and \cite{Speagle-2014}. However, it is important to note that the low dust content derived from the SED fitting of the galaxies NGC~2434 and ESO59-G012 might result from limitations in the model due to the lack of infrared data. A more complete infrared coverage is needed to assess whether these sources are truly quenched or if their SEDs are simply not well constrained. The low number of photometric measurements also impedes to constrain the impact of AGN activity on the FIR-radio correlation. Therefore, we can only estimate a low AGN contribution based on WISE colour diagnostics, and this should be interpreted with caution.

The majority of the galaxies in our sample follow the radio-W3PAH correlation. Also, the SFR estimated using radio and W3PAH methods agrees, especially considering the star formation of massive stars. \cite{Leroy-2021}, report that W3 emission is tightly correlated with CO(2–1) emission, with a scatter $<0.1$~dex. Since CO traces molecular gas, our result supports the use of W3PAH as a proxy for H$_{2}$ mass and suggests that the radio continuum emission scales nearly linearly with the molecular gas reservoir traced by PAH. Galaxies that deviate from the mean relation may therefore reflect variations in the relative abundance of molecular gas or differences in the dominant radio emission mechanisms.

The only galaxy that deviates beyond $3\sigma$ from the radio-W3PAH relation is ESO\,059-G012 (Fig.~\ref{fig:FIRC}). The galaxy is radio-deficient with a $q_{\rm W3PAH}$ value of 3.7. This is not common, \cite{Yun-2001} found only 9 galaxies in a sample of 1800 galaxies with $q$ values greater than 3.0. ESO\,059-G012 is classified as Sa and is the faintest galaxy detected in our radio continuum images. It is located in the north-east, away from the \HI\ intergalactic gas cloud. We propose that the galaxy has already consumed the gas, and the infrared excess relative to the radio continuum emission is due to the heating of dust from the old stellar population. Moreover, no IR-bump is seen in its SED shape (Fig.~\ref{fig:SEDs}.

We find a mean value of $q_{\rm W3PAH} = 2.5 \pm 0.1$. Even though it is offset from the empirical relation \citep[$q = 2.34$,][]{Yun-2001}, within errors, it agrees with the result presented in \cite{Shao-2018}, considering the entire sample, and \cite{Grundy-2023}, calculated at the same frequency. In our galaxy sample, the $q_{\rm W3PAH}$ parameter remains independent of the radio continuum and W3PAH luminosities. We do not observe a break at low W3PAH luminosities or a decrease of $q_{\rm W3PAH}$ values in terms of an increment in stellar mass. 
Within the uncertainties, the range of in-band spectral indices for our galaxy sample is consistent with those of the 35 edge-on galaxies in the CHANG-ES\footnote{https://projects.canfar.net/changes} survey \citep{Schmidt-2019}.
In our maps of the edge-on galaxies NGC~3263 and NGC~3256B, the spectral index steepens as height away from the mid-plane increases.  The trend remains, although the error in our maps also increases. This is also consistent with findings noticed in CHANG-ES  in-band observed spectral index maps and confirmed in analysis of non-thermal spectral index maps \citep[e.g.][]{Schmidt-2019} and band-to-band non-thermal spectral index maps \cite[e.g.][and references therein]{Stein-2023}. This steepening would be expected from cosmic ray ageing due to synchrotron electron energy loss \citep{2024-Irwin}.

Galaxies NGC~6215,  NGC~6221, NGC~3263, and NGC~3256C have in-band {\em global} spectral index values, listed in Table~\ref{tab:prop}, associated with synchrotron radiation. However, they are steeper than the canonical value of $-0.8$ for non-AGN galaxies by \cite{condon-1992} and more recent work such as \cite{An-2021}, who included MeerKAT data for band-to-band spectral index calculations. However, for their sample of star-forming galaxies, \cite{An-2021} found a scatter in the distribution of $\alpha^{3GHz}_{1.3GHz}$  of 0.35 and their Fig.~2 shows a range as negative as $-1.8$. Thus, within our uncertainties, all our in-band determinations, shown in Fig.~\ref{fig:qvssp}, result in global spectral indices that are consistent with previous results.

We note that the non-thermal emission is generally attributed to cosmic rays from star formation. However, a contribution to the scatter and our global spectral index values for NGC~6215, NGC~6221, NGC~3263, and NGC~3256C may also arise from the interaction state of these galaxies. NGC~6215 and NGC~6221 are an interacting pair connected by an \HI\ gas bridge \citep{koribalski-2004-IAU217, Koribalski-2004}. The position of the galaxies in Fig.\ref{fig:qvssp}, a steep $\alpha$ value and a low $q_{\rm W3PAH}$ value, suggests that NGC~6215 lies in regions associated with aged starbursts \citep{Bressan-2002}. While NGC~6221 is complex, because its global spectral index is steep, indicating ageing cosmic rays, but with a high fraction of dust and an intense star formation activity.

NGC~3263 is an edge-on galaxy with a tidal tail that extends to the east of it; \cite{English-2010} proposed that this galaxy interacts with NGC~3256C, along with NGC~3256B. The tidal shocks accompanying these interactions can increase the population of cosmic rays and thus enhance the non-thermal radio emission \citep{Prodanovic-2013}. The values of the spectral index and $q_{\rm W3PAH}$ parameter, see Fig.\ref{fig:qvssp}, suggest that NGC~3256B and NGC~3256 galaxies may be undergoing active or early-stage star formation episodes, possibly in NGC~3256B, triggered by group-scale interactions, while NGC~3263 galaxy lies in regions associated with aged starbursts \citep{Bressan-2002}. 

\cite{Donevski_2015} found that most of the galaxies in their sample classified as merger stage 3 have $q$ values between 1.8 and 2.6 with $\alpha$ values steeper than $-0.6$.  We find a similar trend. The galaxies in that regime are those with clear signs of interaction, classified by us as merger stages 2, 2.5 and 3. Fig.~\ref{fig:qvsMS} shows that galaxies with known \HI\ bridges or tails subclassified by us as merger stage 2.5 show $q_{\rm W3PAH}$ values close to 2.3, whereas the galaxies classified as stage 2.0 display a wide range of $q_{\rm W3PAH}$ values.

\subsection{Resolved properties of the brightest galaxies in the sample}

In this section, we examine the resolved radio continuum and W3PAH properties of the brightest galaxies of our sample. Four are members of the NGC~3256/3263 group, and the other two are of the NGC~6221 pair of galaxies. These groups are in intermediate and initial evolutionary stages, respectively. 

The radio continuum distribution along these galaxies is similar to what is observed in W3PAH and optical images. Radio continuum and W3PAH are tracing the inner structures of these galaxies well. The derived resolved spectral index map shows some structure in the spectral index distribution within the galaxies despite being created using the in-band spectral method; see Fig.~\ref{fig:spectral_index}. Moreover, the $q_{\rm W3PAH}$ maps reveal the presence of low-luminosity AGN due to the radio excess detection in the cores of NGC~6221 and NGC~3256. In the following subsections, we discuss these galaxies individually.

\subsubsection{NGC~3256}

This galaxy is a merging system; two extended tidal tails observed optically and in \HI\ revealed by \citet{English_2003}. Because two different nuclei are present \citep{Norris_1995}, we give the merger stage classification 3. Despite being very luminous in the far-infrared regime, it is not a far-infrared ultra-luminous source. Its starburst nature is indicated by data over a wide wavelength range. The presence of hot young stars in UV data \citep{Kinney_1993}, cosmic-ray electrons accelerated by supernovae, whose progenitors were massive stars in radio continuum data \citep{Norris_1995}, and OB stars ionising \HII\ regions in infrared observation \citep{Graham_1984}. Therefore, the galaxy's position in the WISE colour-colour diagram is expected to be on the right side, with a value of $W2 - W3$ larger than 3.5, which is what we find. Moreover, the value of $W1 - W2$  for this galaxy approaches the limit of AGN classification; see Fig.~\ref{fig:class}. 
Because of the presence of hot, young, massive stars and cosmic-ray electrons, the global spectral index is expected to be dominated by non-thermal emission. We find a global $\alpha^{1021}_{1345} \sim -0.9$ confirming that. However, the spectral index map also reveals a region with values around $-0.6 \pm 0.1$, suggesting a star-forming region.

We cannot distinguish both nuclei in our radio continuum map; see Fig.~\ref{fig:6221}. 
The star formation rate of this galaxy is the highest of our galaxy sample, $SFR_{\rm 20cm}$ [$>$0.1\Msun\,yr$^{-1}$]=98.15 and $SFR_{\rm W3PAH}$ [\Msun\,yr$^{-1}$]=34.

The $q_{\rm W3PAH}$ global value is 2.3; however, the $q_{\rm W3PAH}$ map reaches values lower than 1.5, which are predominant in regions where the star formation process occurs. Besides, there is a clear radio excess where the nuclei are, which points to the existence of AGN activity, which agrees with \cite{Ohyama_2015}. Those authors found evidence of a heavily absorbed low-luminosity AGN in the NGC~3256 southern nucleus. Furthermore, recent JWST observations by \citet{Bohn-2024} have revealed a prominent, collimated warm molecular outflow originating exclusively from the southern nucleus, extending up to $\sim 0.7$~kpc, with intrinsic velocities reaching $\sim$1000 \rm km s$^{-1}$. This outflow is spatially and kinematically aligned with previously known cold molecular and atomic outflows \citep[e.g.][]{Cazzoli-2016}, and is interpreted as being driven by the southern nucleus. In contrast, no warm molecular outflow is detected around the northern nucleus, suggesting significant differences in feedback processes between the two. These results reinforce the scenario in which the southern nucleus hosts a low-luminosity AGN, while the northern nucleus is dominated by starburst activity.

\subsubsection{NGC~3256B} 

This galaxy is classified as SB(s)bc galaxy. Its inclination is $\sim$73\degr\ based on the axis ratio (b/a) estimated by \cite{Lauberts_1989}. Because the brightness on one side of the galaxy is slightly fainter than the other, we give the merger stage classification 2. Its W3PAH luminosity is $\sim$10$^{9.5}$\Lsun. 

The star formation rate of this galaxy is $SFR_{\rm 20cm}$ [$>$0.1\Msun\,yr$^{-1}$] = 7.62 and $SFR_{\rm W3PAH}$ [\Msun\,yr$^{-1}$] = 1.4. These values agree with those expected for classical spiral galaxies. The WISE colour-colour diagram shows this galaxy is located in the star-forming region and somewhat above the star formation sequence track. We note that its modest $SFR_{\rm W3PAH}$ value is included in the range of SFRs associated with the star-forming category defined for this diagram in \citet{Jarrett_2019}. With respect to the higher $W1 - W2$ colour, there are sources of heating other than intense star formation that could contribute to enhancing the temperature of the dust. For example, for interacting galaxies, their dust can be heated by tidal shocks \citep{Donevski_2015} and by the radiation field of a larger companion galaxy
\citep{Nersesian-2020}.

Its global spectral index is $-0.8 \pm 0.2$, typical for normal star-forming galaxies. The spectral index map shows that the index is around $-$0.6 within the galaxy disk. In particular, it is observed that in the same disk, it reaches flatter values in the southeastern region, while in the north-west direction, where the spiral arm seems to be more open, it has steeper values. 

Although the $q_{\rm W3PAH}$ global value is 2.3, the $q_{\rm W3PAH}$ map shows that values around 1.5 are predominant in the central region of the galaxy and within the spiral arms, especially in the same region where the spectral index has the more flattened value. As it can be observed from the $q_{\rm W3PAH}$ map, Fig.~\ref{fig:qmaps}, the $q_{\rm W3PAH}$ values span from 1.0 to 3.0. This observation suggests that star formation is probably occurring in the southeastern region.

\subsubsection{NGC~3256C} 

This galaxy is classified as SB(rs)d. Its inclination is $\sim$42\degr\ based on the axis ratio (b/a) estimated by \cite{Lauberts_1989}. \cite{English-2010} proposed that the galaxy-galaxy interaction among NGC~3256C, NGC~3256B and NGC~3256 is the main process that produced the Vela Cloud complex. Because the brightness distribution of NGC~3256C is not symmetric, we give the merger stage classification 2. The arms towards the south-west side of the galaxy seem to be more open. Moreover, optical images show clumpy regions in the south-west. Its W3PAH luminosity is $\sim$10$^{9}$\Lsun. 

The SFR estimated using radio and W3PAH agrees, especially considering the star formation of massive stars. The star formation rate of this galaxy is low, $SFR_{\rm 20cm}$ [$>$0.1\Msun\,yr$^{-1}$] = 4.74 and $SFR_{\rm W3PAH}$ [\Msun\,yr$^{-1}$] = 1.1. The WISE colour-colour diagram shows that it is located in the star-forming region, which is unexpected if this region were to indicate enhanced star-forming activity. However, a few galaxies in \citet{Jarrett_2019} also have this value and lower yet reside in this region of the diagram.

Its global spectral index is $-1.5 \pm 0.4$. The spectral index map shows that the index is around $-0.6$ in the region where dust and stars are more concentrated. 
Although the $q_{\rm W3PAH}$ global value is 2.3, the $q_{\rm W3PAH}$ map shows that values around 1.25 are predominant in the central region of the galaxy and within the spiral arms. As it can be observed from the $q_{\rm W3PAH}$ map, Fig.~\ref{fig:qmaps}, the $q_{\rm W3PAH}$ values span from 1.0 to 3.0. 

\subsubsection{NGC~3263} 

This galaxy is classified as SB(rs)cd. It is an almost edge-on galaxy with a tail that extends to the east of it. \cite{English-2010} found a possible \HI\ bridge, 67~kpc long, that connects NGC~3263 with ESO\,263-G044. NGC~3263's arm extends in the opposite direction to the Vela Cloud, confirming that the Vela Cloud is superfluous for producing the tidal features of NGC~3263. Because of these signs of galaxy-galaxy interaction, we give the merger stage classification number 2.5. Its W3PAH luminosity is $\sim$10$^{10}$\Lsun. In the WISE colour-colour diagram, it is located close to the limit of being classified as a star-forming galaxy. Its global spectral index is $-1.2 \pm 0.1$.
The spectral index map shows that the index is around $-0.6$ in the core region.
The star formation rate of this galaxy is high, $SFR_{\rm 20cm}[>$0.1\Msun\,yr$^{-1}]$ = 30.5 and $SFR_{\rm W3PAH}$ [\Msun\,yr$^{-1}$] = 5.3. 
Although the global $q_{\rm W3PAH}$, value is 2.3, as shown in the $q_{\rm W3PAH}$ map (Fig. \ref{fig:qmaps}), the values span from $0.5$ to $3.0$. The map reveals slightly lower values within the stellar disk, possibly due to enhanced radio emission linked to environmental interaction. Interestingly, at the tip of the tail, there is a clear radio deficiency relative to the W3PAH emission. This region also shows a flatter spectral index of around $-0.2$, suggesting a larger thermal contribution. Although the uncertainties in this area are relatively high, both the spectral and $q_{\rm W3PAH}$ maps point toward the possibility of triggered star formation in this part of the galaxy. Further insight could be gained from higher-resolution data and a multi-wavelength analysis.

\subsubsection{NGC~6221} 

The galaxy NGC~6221 is classified as a barred spiral of type SB(s)c. It has an optical diameter of $3.5\arcmin \times 2.5\arcmin$ (inclination $\sim$43\degr) with a large amount of dust visible in both spiral arms as well as along the bar. The northern arm sharply turns ($>$90\degr) and continues for another 2\arcmin\ towards the south-east. A clumpy radio continuum region (north of the nucleus) agrees with a blue \HII\ region complex, suggesting recent star formation, possibly induced by the interaction. We give the merger stage classification 2.5. Its W3PAH luminosity is $\sim$10$^{10}$\Lsun. The WISE colour-colour diagram shows that it is located in the star-forming region. Its global spectral index is $-1.4 \pm 0.1$. 
The spectral index map shows that the index is around $-0.6$ in the core region, in the hammer-like structure, and a clump at $\alpha,\delta$(J2000) = $16^{\rm h}\,52^{\rm m}\,45^{\rm s}$, -59\degr\,12\arcmin\,00\arcsec. 

The star formation rate of this galaxy is high, $SFR_{\rm 20cm}[>$0.1\Msun\,yr$^{-1}]$ = 16.3 and $SFR_{\rm W3PAH}$ [\Msun\,yr$^{-1}$] = 5.1. Although the global value of $q_{\rm W3PAH}$ is 2.2, the $q_{\rm W3PAH}$ map shows low values, such as 0.6 in the galaxy core. This is a clear radio excess where the nuclei are, which points to the existence of AGN activity. Besides, NGC~6221 has been shown to host significant molecular outflows, as reported by \cite{Stone-2016}. The same authors identified NGC~6221 as a Type II AGN (dusty, obscured torus), which could affect the radio-to-IR connection in this system. The hammer-like structure and the clump region show values between 1.0 and 1.6, indicating a good correlation between W3PAH and radio continuum emission.

\subsubsection{NGC~6215} 

This galaxy is classified as SA(s)c. Its inclination is $\sim$38\degr. The galaxy has a smaller optical diameter of $2.1\arcmin \times 1.8\arcmin$. We give the merger stage classification 2.5. Its W3PAH luminosity is $\sim$10$^{9.5}$\Lsun. The WISE colour-colour diagram shows that it is located in the star-forming region. Its global spectral index is $-1.2 \pm 0.1$. The spectral index map shows that the index is around $-0.6$ in the core region, where the galaxy structure seems compressed.
The star formation rate of this galaxy is high, $SFR_{\rm 20cm}$ [$>$0.1\Msun\,yr$^{-1}$]=13.5 and $SFR_{\rm W3PAH}$ [\Msun\,yr$^{-1}$]=2.1. The $q_{\rm W3PAH}$ map does not show a considerable gradient. The lower value is in the core. But it does show that most of the star-forming activity occurs where the gas is more compressed due to the interaction with NGC~6221.\\

\section{Summary}

We obtained high-resolution radio continuum observations of three groups of galaxies at different evolutionary stages using the MeerKAT radio telescope with a mean angular resolution of $8\arcsec \times 6.7\arcsec$ and $\sim$5.6~$\mu $Jy\,beam$^{-1}$ rms. The radio continuum emission is detected from the galaxy pair NGC~6215/6221 and three background field galaxies, ten galaxies in the NGC~3256/3263 group, and four galaxies in the NGC~2434 group, as well as three background field galaxies. 

We complemented our data with WISE infrared images and measured the integrated WISE W1, W2 and W3 band flux densities for our galaxy sample using {\sc Photutils} \citep{Bradley-2016,Bradley-2020}. This allowed us to construct a WISE colour-colour diagram, perform SED fitting, and estimate a relation between the SFR and the stellar mass. The galaxies lying in the star-forming main sequence are distributed appropriately for the evolutionary stage of their group. For example, the galaxies residing in the intermediate evolutionary group stage (NGC~3256/3263) span all three categories, from spheroidals through intermediate disks to star-forming disks. One galaxy, NGC~3256, resides in a part of the colour-colour diagram towards the AGN region. Galaxies in the most evolved galaxy group (NGC~2434) reside in the spheroidal and intermediate disk regimes, and the galaxy pair in the group in the earliest evolution stage (NGC~6221) are both located in the star-forming region.

We estimated the contribution from the PAH following the method by \cite{Cluver_2017} and assessed the version of the infrared-radio correlation known as the W3PAH-radio relation.  Most galaxies in our sample follow that relation, except for ESO\,059-G012, which is a member of the most evolved group in the sample. The galaxy has already converted much of its gas into stars, and the infrared excess relative to the radio continuum emission is due to the heating of dust by the old stellar population.

For our sample, we found a $q_{\rm W3PAH}$ mean value of $2.5 \pm 0.1$. Even though it is offset from the empirical relation \citep[$q = 2.3$,][]{Yun-2001}, within uncertaintes, it agrees with the result presented in \cite{Shao-2018}, considering the entire sample, and \cite{Grundy-2023}, calculated at the same frequency for the galaxies in the Eridanus group. We observed that the $q_{\rm W3PAH}$ parameter remains independent of the radio continuum and W3PAH luminosities. 

The $q_{\rm W3PAH}$ maps of two galaxies, NGC~3256 and NGC~6221, highlight the presence of low-luminosity AGN. This agrees with what was proposed by \cite{Ohyama_2015} and \cite{Koribalski-2004} for each galaxy, respectively. Additionally, the $q_{\rm W3PAH}$ maps of the galaxies with clear signs of interactions, such as tails, bridges or classified as a merger, are the maps with a wide range of $q_{\rm W3PAH}$ values. In particular, the galaxy NGC~6215 shows a gradient towards the bridge that connects this galaxy with NGC~6221.

Since star formation is one of the dominant physical processes in the formation and evolution of galaxies, we examine their star formation rate characteristics in groups
at three different evolutionary stages. For this modest, preliminary investigation, we find:

\begin{itemize}

\item Representing an early stage of evolution, the galaxies in the NGC~6221 group are morphologically interacting rather than merging. 
We find, according to the radio continuum fluxes, that both galaxies have a high star formation rate of 13.5 and 16.3\Msun\,yr$^{-1}$.  Their spectral indices indicate they are dominated by synchrotron emission as expected from work by \citep{Prodanovic-2013}.  
\item Representing the intermediate stage of group evolution is the NGC~3256/NGC~3263 Group. Previous work showed that morphologically the galaxies include minor interactions (e.g. NGC~3263) and a major merger (NGC~3256), along with disturbed dwarf galaxies. From the radio fluxes, 4 out of 9 galaxies have high SFR above 5\Msun\,yr$^{-1}$, two have a modest SFR, while the remaining three have a low SFR below 0.3\Msun\,yr$^{-1}$.  
\item Representing a later evolutionary stage is the NGC~2434 Group, which has three disturbed spiral galaxies and one elliptical. The non-ellipticals have radio fluxes that indicate SFRs ranging from 13.4 to 0.008\Msun\,yr$^{-1}$.

\end{itemize}

\begin{acknowledgements}
We are grateful to the anonymous referee for a critical reading of the manuscript and for very useful suggestions. We acknowledge the effort of Ian Heywood and Dane Kleiner in processing the data and thank Jing Yeung for his contribution. J.S. thanks Alberto Noriega-Crespo for helpful discussions on WISE image units, and Ivan Lopez and Rodrigo Haack for checking the photometry procedure. P.K.H. gratefully acknowledges the Fundação de Amparo à Pesquisa do Estado de São Paulo (FAPESP) for the support grant 2023/14272-4. This paper is based on observations obtained with the MeerKAT telescope, where the South African Radio Astronomy Observatory, a facility of the National Research Foundation, an agency of the Department of Science and Innovation, operates. 
\end{acknowledgements}

\section*{Data Availability}

The MeerKAT data used here are available through the SARAO Data Archive\footnote{\url{https://www.sarao.ac.za/}}. The radio continuum cutouts images of the galaxies are available in electronic form at the CDS via anonymous ftp to cdsarc.u-strasbg.fr (130.79.128.5) or via http://cdsweb.u-strasbg.fr/cgi-bin/qcat?J/A+A/. Besides, all images used in this publication will be made available upon reasonable request to the lead author.\\
 

\bibliographystyle{aa}
\bibliography{star-forming-gx.bib} 



\begin{appendix}

\section{Galaxy groups}\label{appen:galaxy_groups}

\subsection{The NGC~6221 galaxy group}

The NGC~6221 galaxy group consists of about 5 galaxies. There are at least three dwarf galaxies \citep{Koribalski-2004} and another two form a spiral galaxy pair NGC~6221 (\vhel\ = 1499\kms) and NGC~6215 (\vhel\ = 1564\kms), where \vhel\ is the galaxy's velocity in the heliocentric frame.
The pair separation is $\sim$100~kpc at the adopted group distance of $D$ = 18~Mpc. Their relative proximity, the peculiar appearance of NGC~6221 at different wavelengths such as optical \citep{Pence-1984}, \Ha\ \citep{Vega-1998}, and the presence of an extended \HI\ bridge connecting the pair \citep{Koribalski-2004} indicate that these galaxies are interacting. In addition, NGC~6221 is likely interacting with its three low-surface brightness dwarf companions \citep{koribalski-1996}. 

NGC~6221 is a barred spiral galaxy. The bar identified in optical and infrared images \citep{1987ess..book.....L,1994cag..book.....S} lies at a position angle ($PA$) of 118\degr\ \citep{Pence-1984} and has a length of 5.2~kpc. From its ends, two spiral arms extend symmetrically. However, the northern arm has a peculiar extension, which is relatively faint in infrared images. NGC~6215 is generally classified as a non-barred spiral galaxy and has an optical diameter 40\% smaller than NGC~6221.

\subsection{The NGC~3256/3263 galaxy group}

The NGC~3256/3263 galaxy group consists of three sub-groups centred on (from north to south) NGC~3256 (\vhel\ = 2836\kms), NGC~3263 (\vhel\ = 2935\kms) and NGC~3261 (\vhel\ = 2512\kms). The NGC~3256 and NGC~3263 galaxy groups are difficult to distinguish from each other spatially and kinematically as their systemic velocities differ by less than 200\kms. Different authors have applied group-finding algorithms to determine the members of these groups \citep[e.g.,][]{fouque-1992,Garcia-1993,lipari-2000}. The latest publication of them \citep{kourkchi-2017} considers ten galaxies to be members of a single large group spanning a few degrees at a distance of 38~Mpc ($\sim$1\degr\ = 600~kpc).

All members of the NGC~3256/3263 group show varying degrees of tidal interactions and complex kinematic behaviour \citep{lipari-2000,English-2010}. In particular, the group contains the gas-rich merging galaxy NGC~3256, surrounded by numerous \HI\ fragments. Additionally, there are distorted spiral galaxies, such as the edge-on galaxy NGC~3263 with a tidal tail, NGC~3256B and NGC~3256C, and the face-on galaxy NGC~3261 with three dwarf companions in its surroundings.

The most interesting feature within the group is the galaxy-sized \HI\ complex (the "Vela Cloud") that was discovered by \cite{English-2010} using the Australia Telescope Compact Array (ATCA) and the 64-m Parkes telescope. The Vela Cloud is located just west of NGC~3263 and south of the merger NGC~3256, spans at least $9' \times 16'$ (100~kpc $\times$ 175~kpc, at the adopted distance of 38~Mpc) and has an \HI\ mass of 3--5 $\times 10^9$\Msun\  \citep{English-2010}. In addition, a possible \HI\ bridge between NGC~3263 and ESO\,263-G044, 67~kpc long, appears to connect both galaxies \citep{English-2010}. The \HI\ column density of the bridge is 3--6 $\times$ 10$^{19}$~atoms\,cm$^{-2}$, which is similar to the Vela Cloud diffuse components. 

\subsection{The NGC~2434 galaxy group}

The NGC~2434 group, named after the central gas-poor elliptical galaxy (\vhel\ = 1367\kms), contains four large galaxies and several dwarf galaxies. The most prominent group member is the one-armed spiral NGC~2442 (\vhel\ = 1466\kms), close to which \cite{Ryder-2001} discovered a massive ($\sim$10$^9$\Msun) \HI\ cloud catalogued as HIPASS J0731--69 \citep[see also][]{Koribalski-2004-BGC}. NGC~2434 and NGC~2442 are separated by only 17\arcmin\ or 85~kpc, at the adopted distance of $D$ = 17~Mpc. The \HI\ cloud is located between NGC~2442 and the galaxy pair NGC~2397 (\vhel\ = 1355\kms) / NGC~2397B (\vhel\ = 1376\kms). Close inspection of the HIPASS data cube revealed \HI\ emission along a partial ring, spanning about $40' \times 20'$ (200 kpc $\times$ 100 kpc) connecting to the one-armed spiral NGC~2442 \citep{Koribalski-2020}. High-resolution ATCA \HI\ data reveal numerous high-density \HI\ clouds along the ring structure \citep{RK-2004}, but no stellar counterparts have yet been detected. 

\section{MeerKAT observing parameters.}

\begin{table}[h!]
\centering
\caption{MeerKAT observing parameters.}
\begin{adjustbox}{max width=0.5\textwidth}
\begin{tabular}{@{}lccc@{}}
\toprule
 Group & NGC~6221 & NGC~3256/3263 &  NGC~2434 \\
\midrule
RA (h m s, J2000) & 16:52:30.2 & 10:29:13.1 &   07:31:40.2 \\
DEC ($^{\circ}\,'\,''$, J2000)
& --59:12:01.2 & --44:07:21.2 & --69:01:37.2 \\
observing dates & 21-05-2021 & 27-05-2021 & 04-05-2021 \\ 
 &  & 05-01-2022 &  \\  
field of view (deg$^2$) & & 1.89 &  \\
time on source (hr) & 4.92 &  4.64, 4.86 & 4.92 \\
number of antennas  &  63  &  59, 62 &  59 \\
phase calibrator & J1726--5529  & J0825--5010  & J0906--6829  \\
phase calibrator flux (Jy) & 5.53 & 5.73 & 1.88 \\
centre frequency (MHz) &  \multicolumn{3}{c}{1284} \\
bandwidth (MHz) &  \multicolumn{3}{c}{856}  \\
median rms ($\mu$Jy\,beam$^{-1}$) & 7 & 3  & 7 \\
synthesized beam size & 7.4\arcsec $\times$ 7.1\arcsec & 7.6\arcsec $\times$ 6.7\arcsec & 8.7\arcsec $\times$ 6.4\arcsec  \\
\bottomrule
\end{tabular}
\label{tab:obs}
\end{adjustbox}
\end{table}

\onecolumn

\section{MeerKAT radio continuum data} \label{sec:RC}

\begin{figure}[ht!]
\centering
\includegraphics[width=0.25\textwidth]{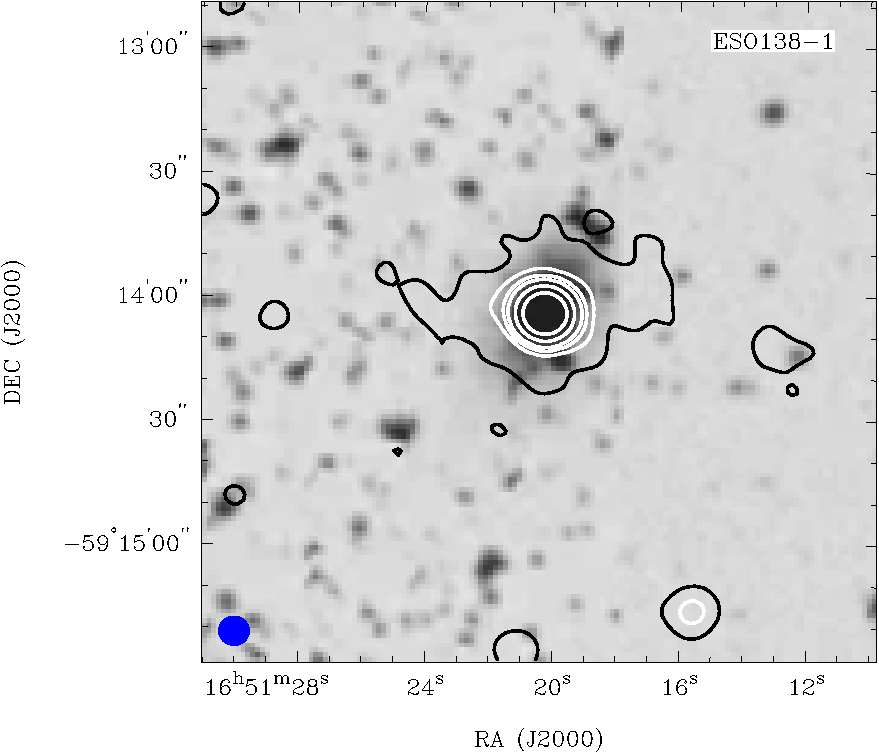}
\includegraphics[width=0.25\textwidth]{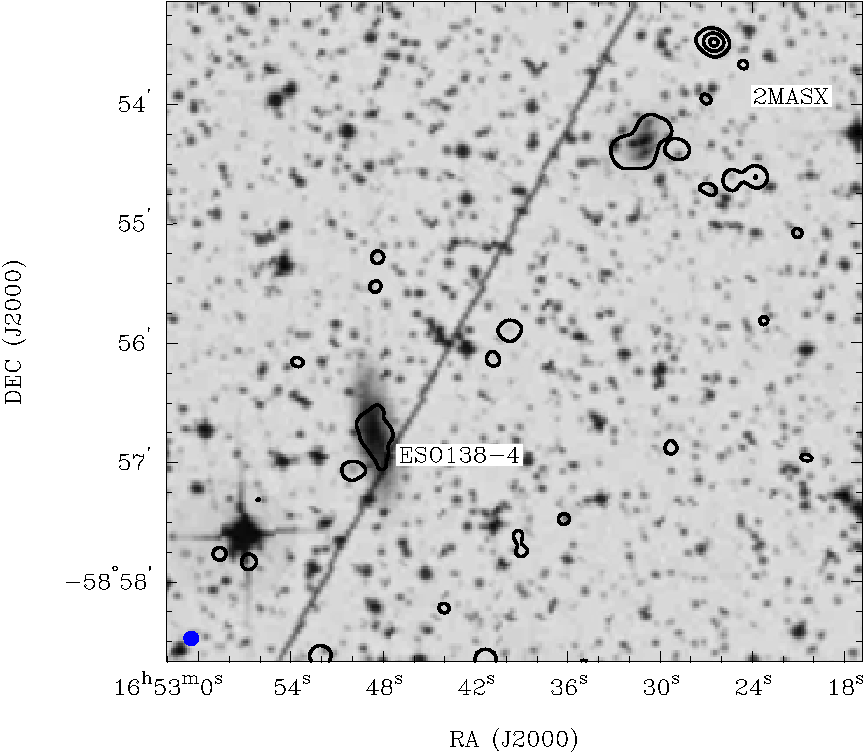} \includegraphics[width=0.25\textwidth]{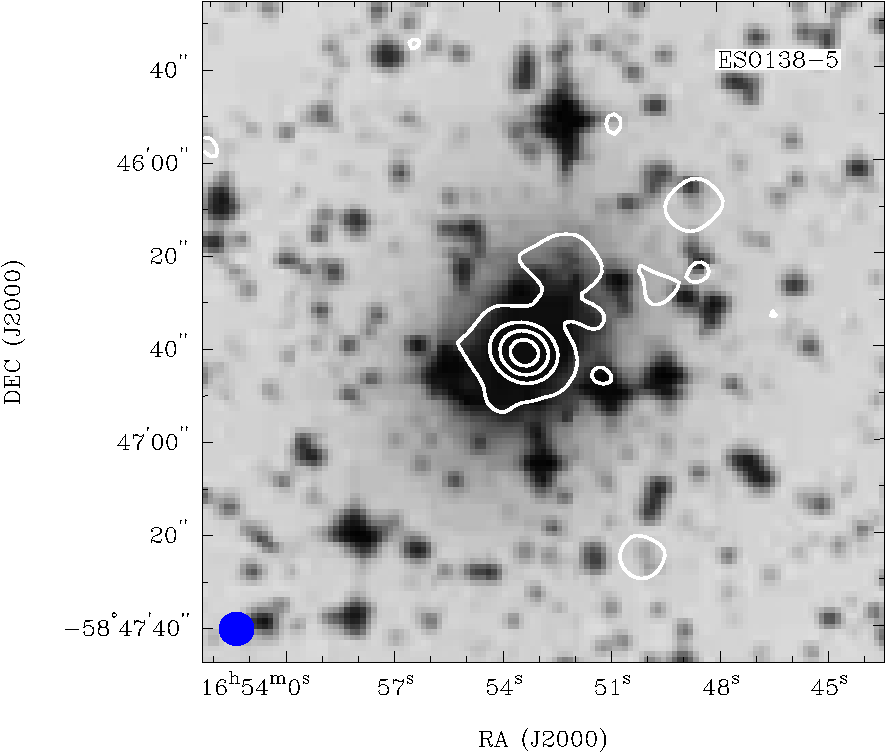}\\
\includegraphics[width=0.25\textwidth]{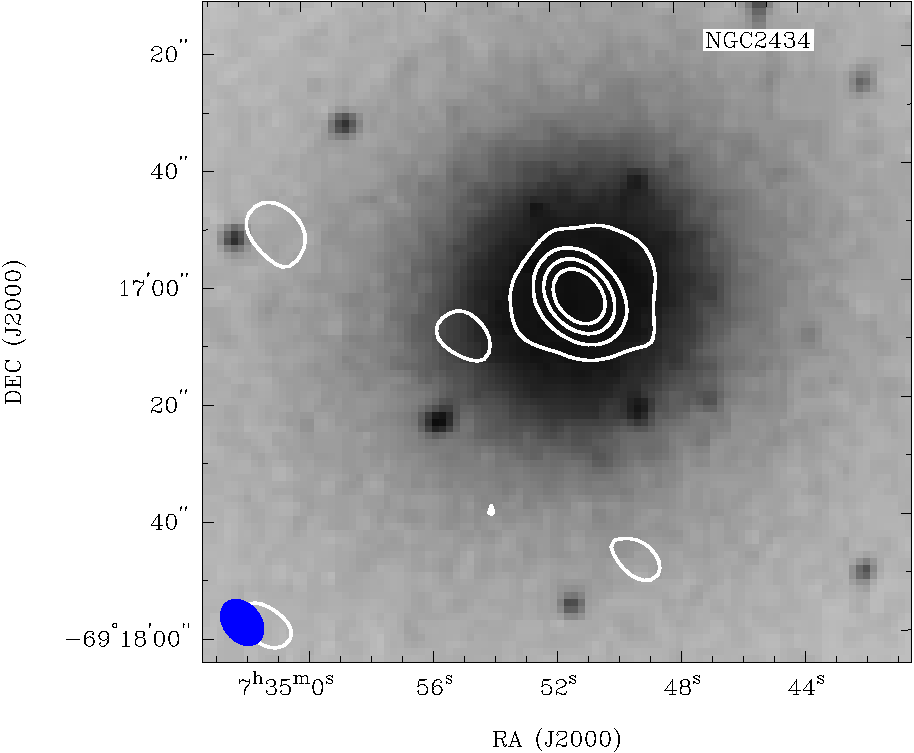}
\includegraphics[width=0.25\textwidth]{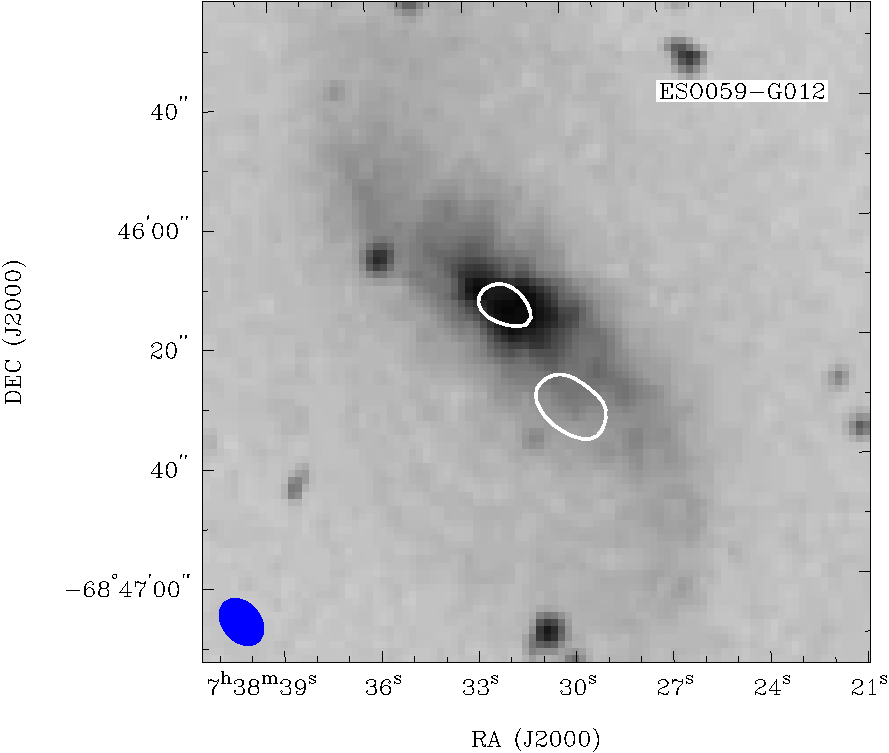}
\includegraphics[width=0.26\textwidth]{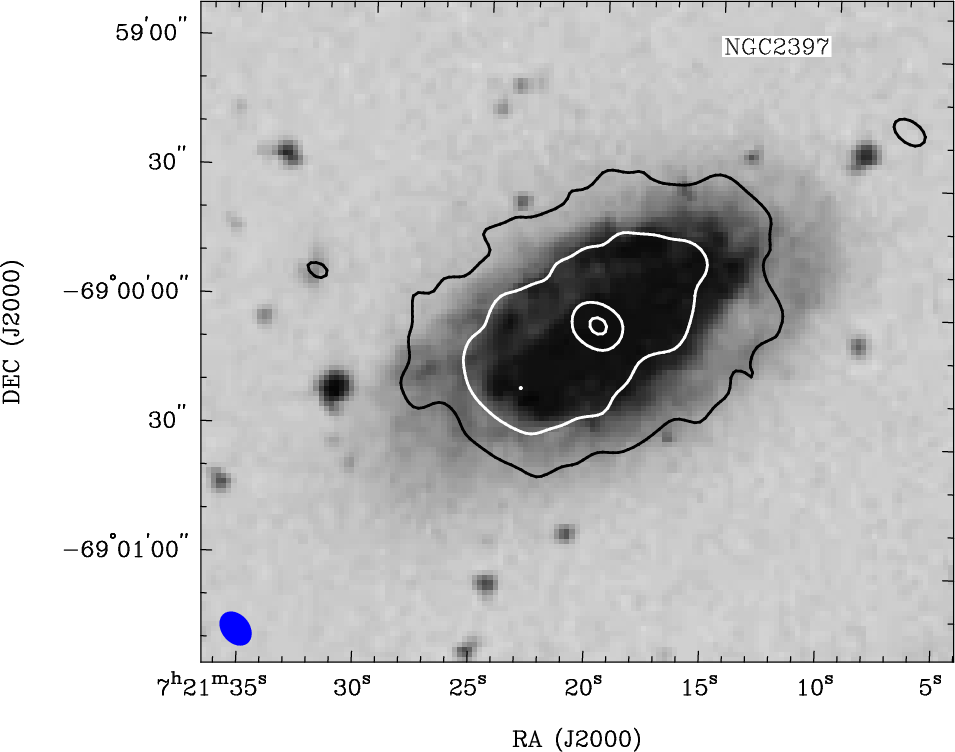}\\
\includegraphics[width=0.25\textwidth]{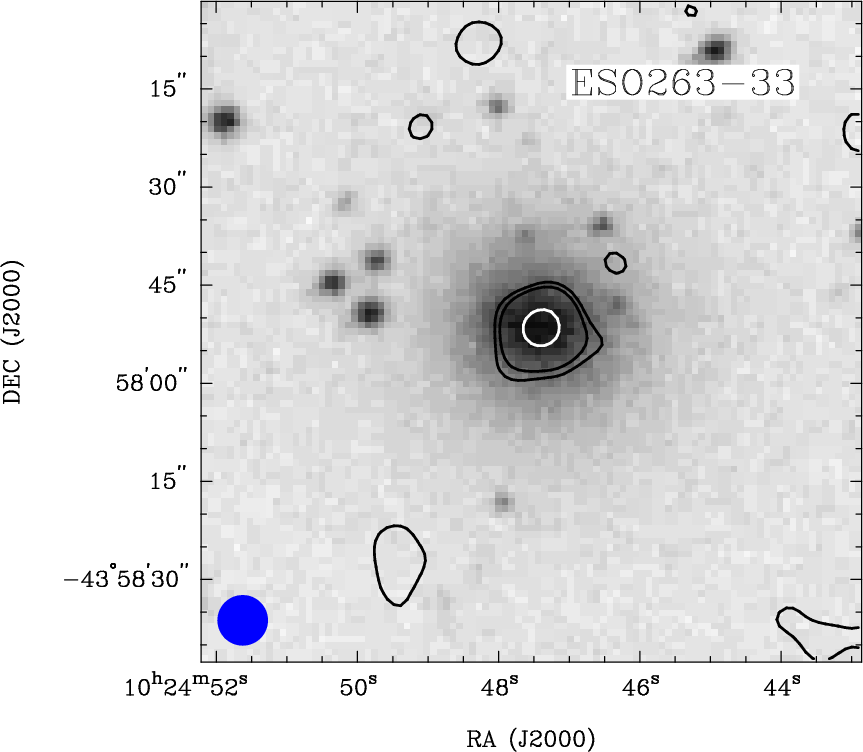}
\includegraphics[width=0.25\textwidth]{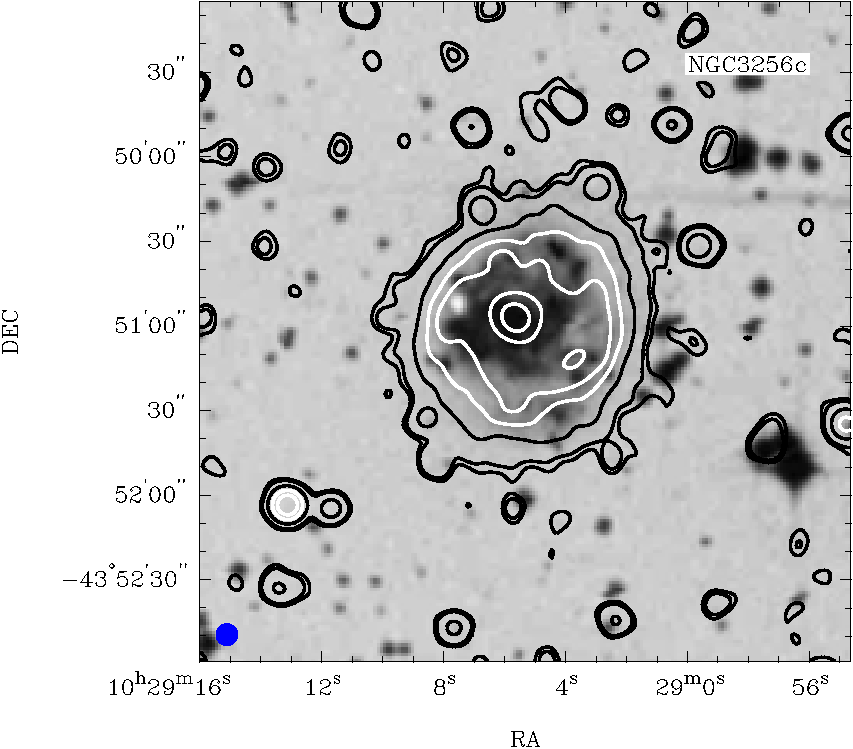}
\includegraphics[width=0.27\textwidth]{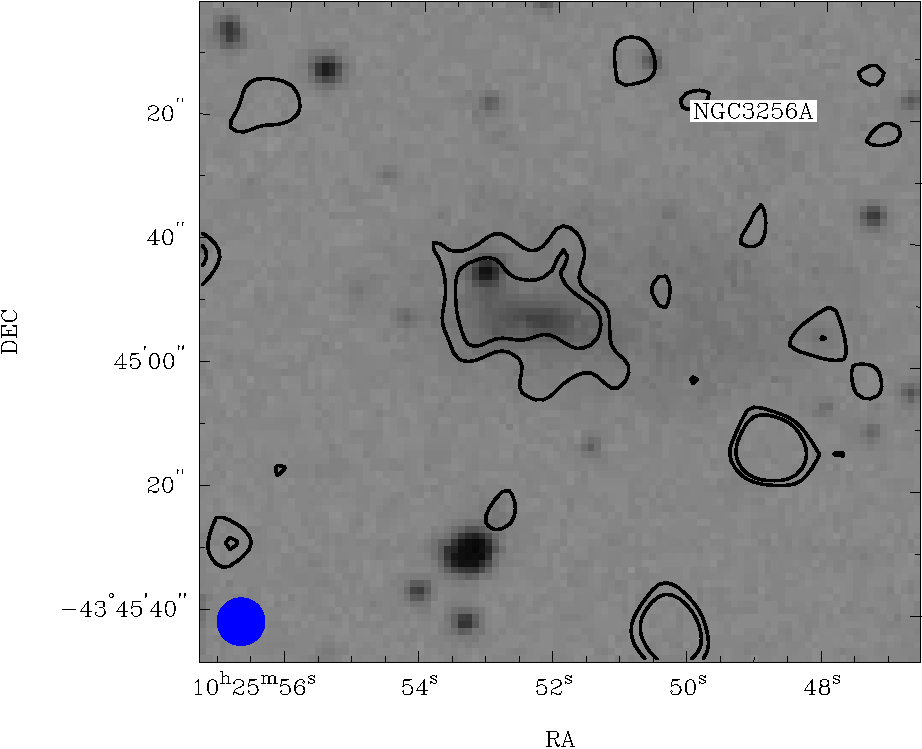}\\
\includegraphics[width=0.27\textwidth]{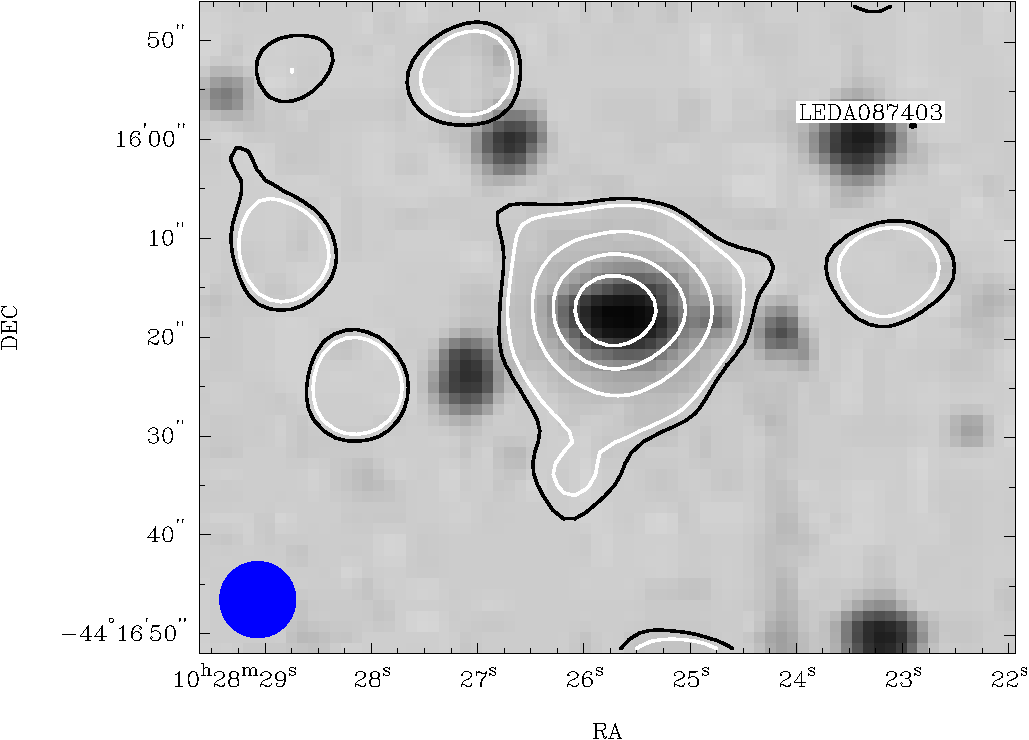}
\includegraphics[width=0.23\textwidth]{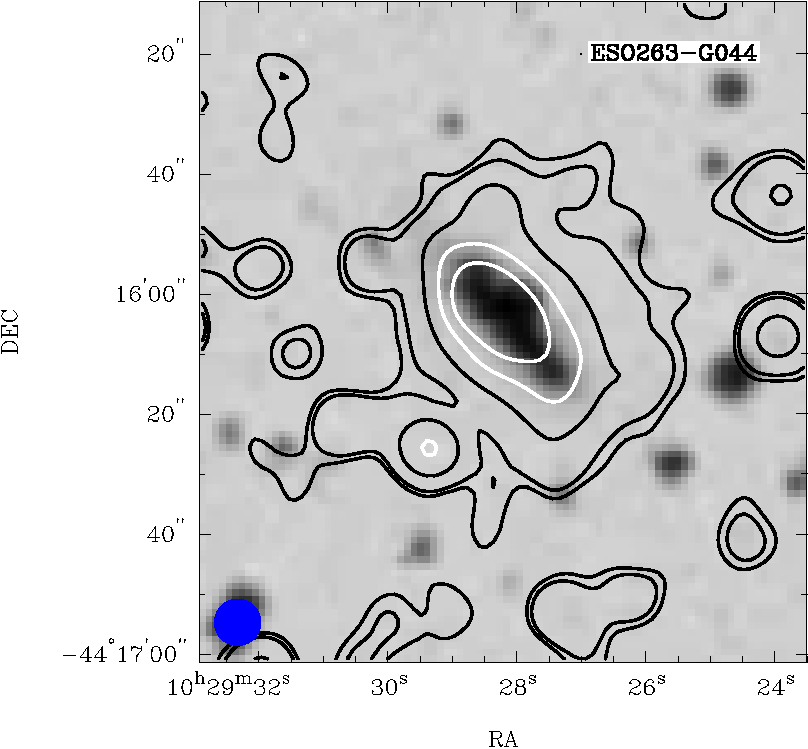}
\includegraphics[width=0.27\textwidth]{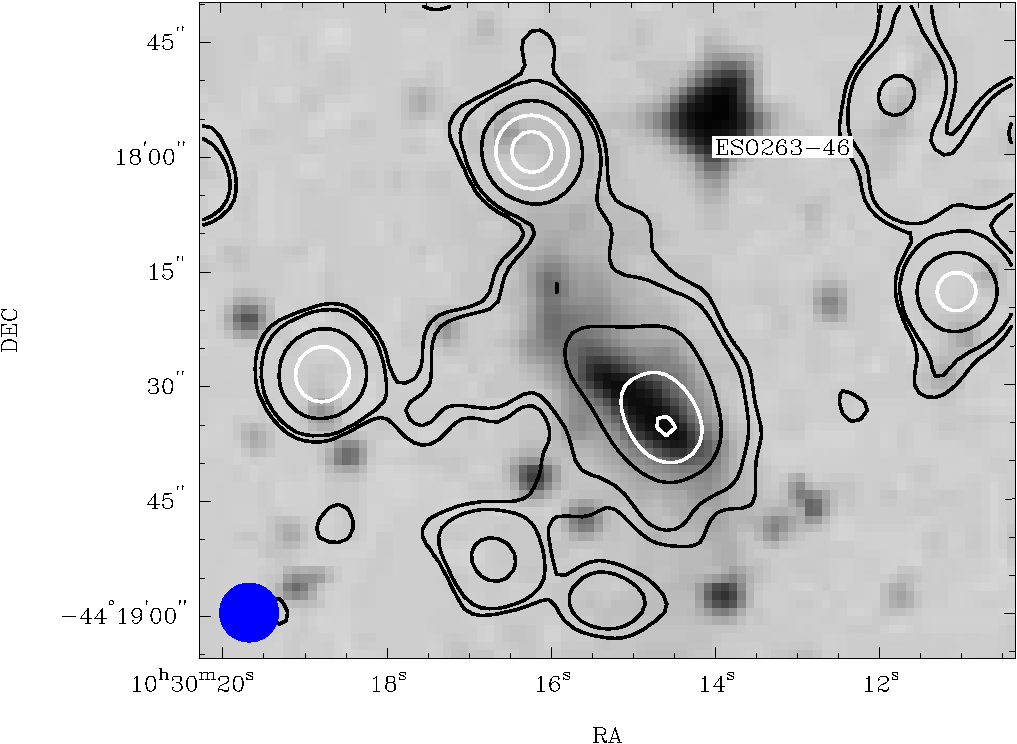}\\
\includegraphics[width=0.25\textwidth]{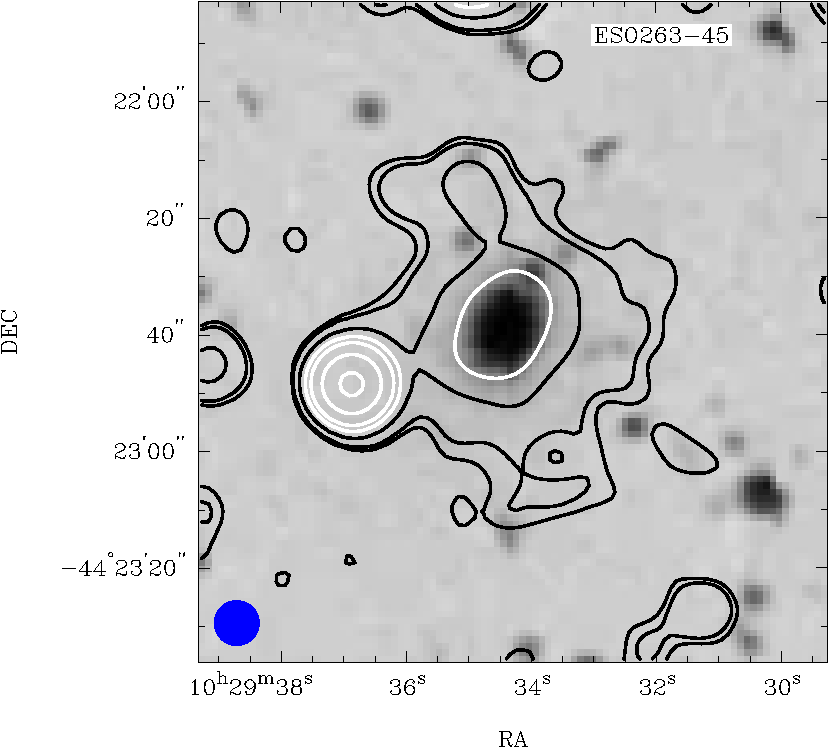}
\includegraphics[width=0.28\textwidth]{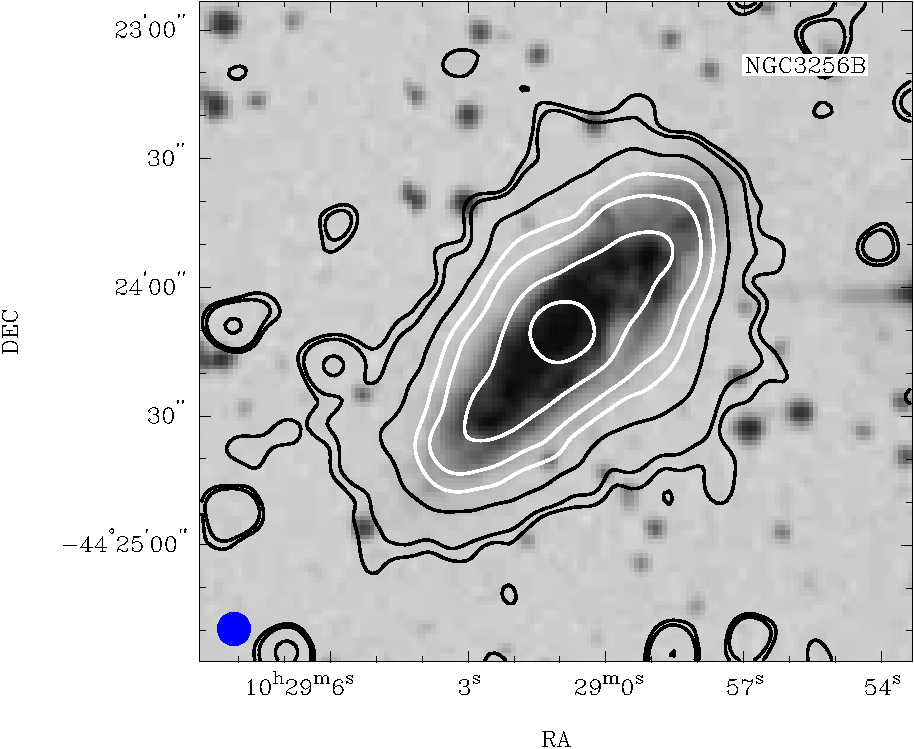}
\includegraphics[width=0.24\textwidth]{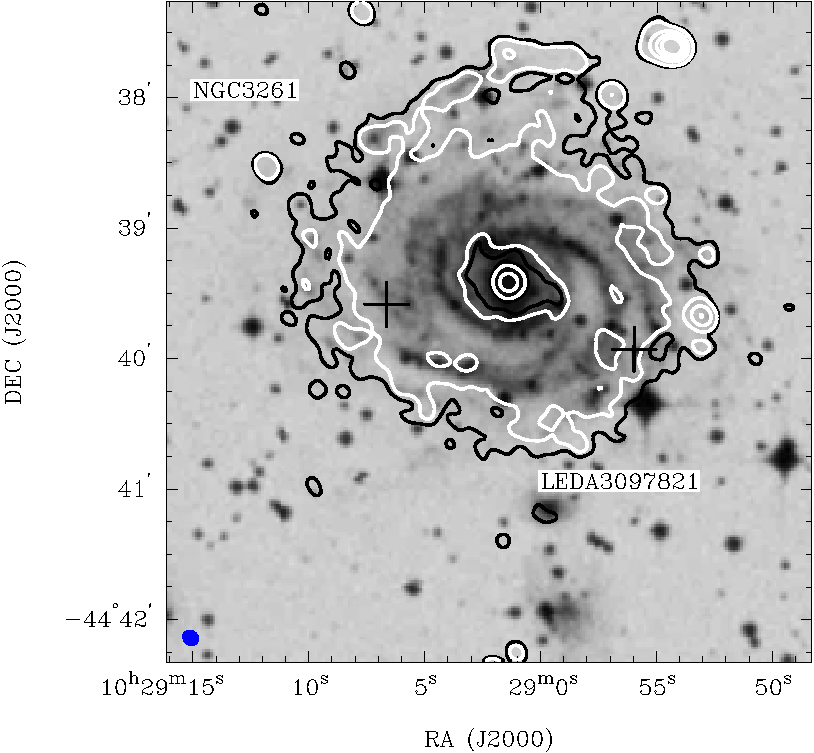}

\caption{High-resolution MeerKAT 20-cm radio continuum of the galaxies overlaid onto DSS R-band optical images.
First row: ESO\,138-1, ESO\,138-4, and ESO\,138-5 (background galaxies, NGC~6221 group), the contour levels are 0.03, 0.2, 0.5, 1, 3 and 8~mJy\,beam$^{-1}$. Second row: NGC~2434, ESO\,59-G0012, and NGC~2397 (NGC~2442 group), the contour levels are 0.027, 0.09, and 0.27~mJy\,beam$^{-1}$. From the third to the fifth rows: ESO263-44, NGC~3256C, NGC~3256A, NGC~3263, NGC~3256, LEDA~087403, ESO\,263-44, ESO\,263-46, ESO\,263-45 (background galaxy), NGC~3256B and NGC~3261 (NGC~3256/3263 group). The contour levels are 0.027, 0.045, 0.18, 0.45, 0.9, 2.7, and 7.2~mJy\,beam$^{-1}$. Contours are shown in black or white for display purposes. The synthesised beam is displayed at the bottom-left corner of each panel.}
\label{fig:continuo}
\end{figure}

\section{WISE photometry} \label{sec:WISE}

\begin{figure}[!ht]
\centering
\includegraphics[width=0.76\textwidth]{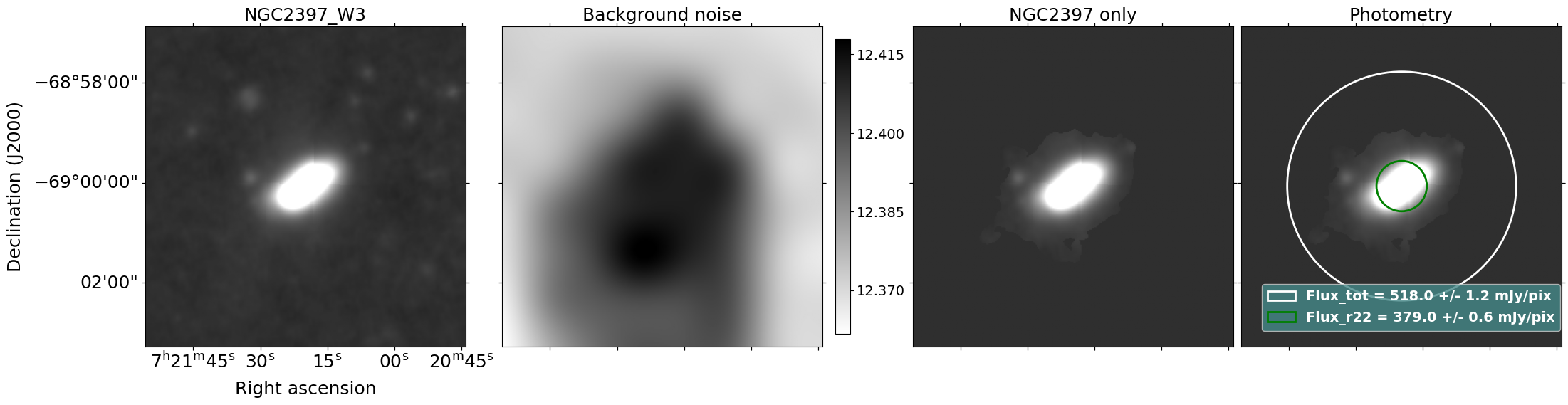} \\
\includegraphics[width=0.76\textwidth]{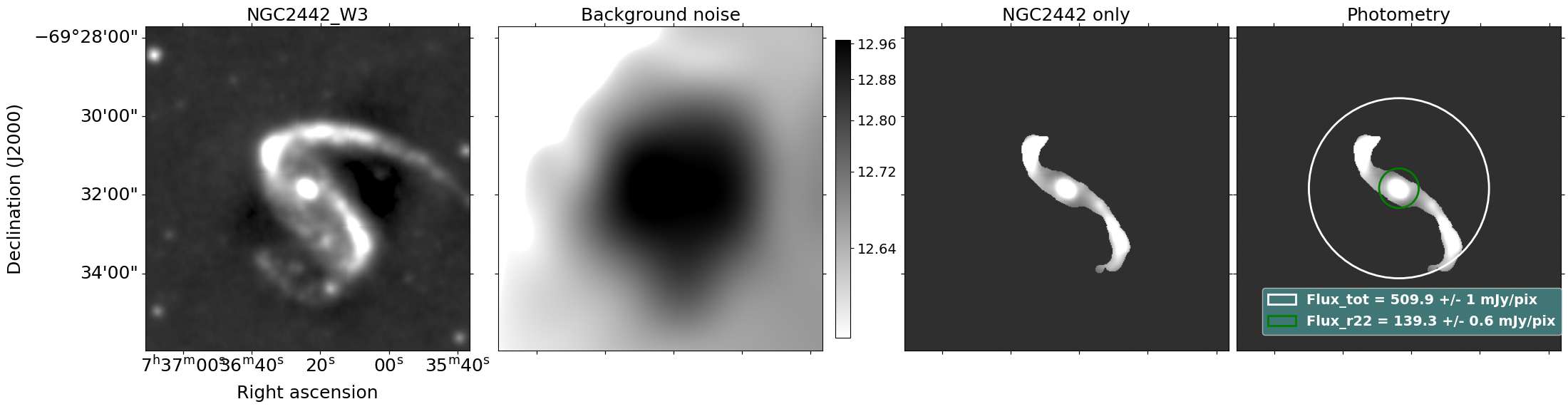}\\
\includegraphics[width=0.76\textwidth]{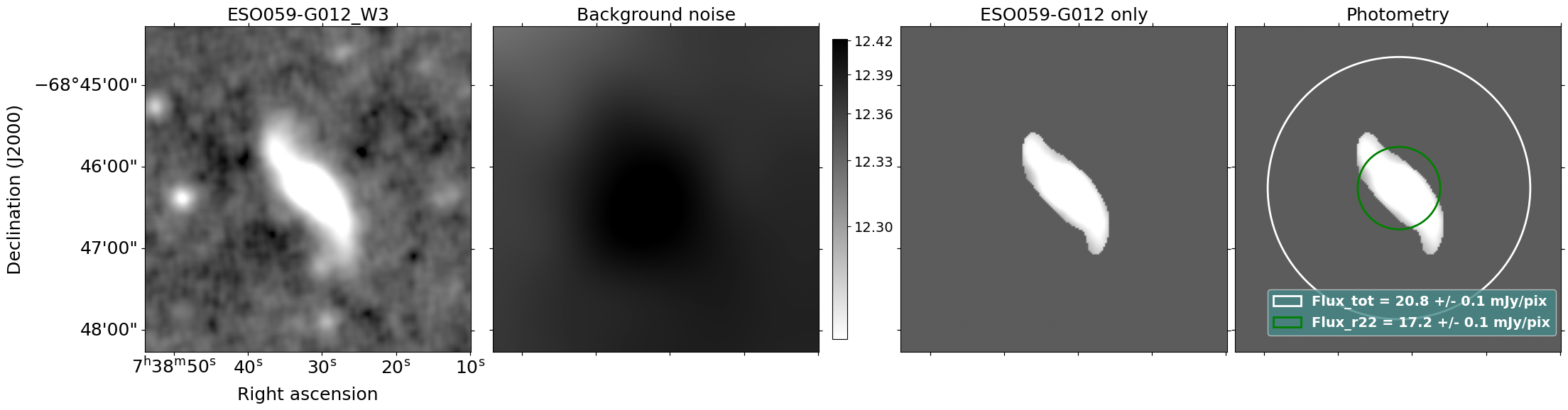}\\
\includegraphics[width=0.76\textwidth]{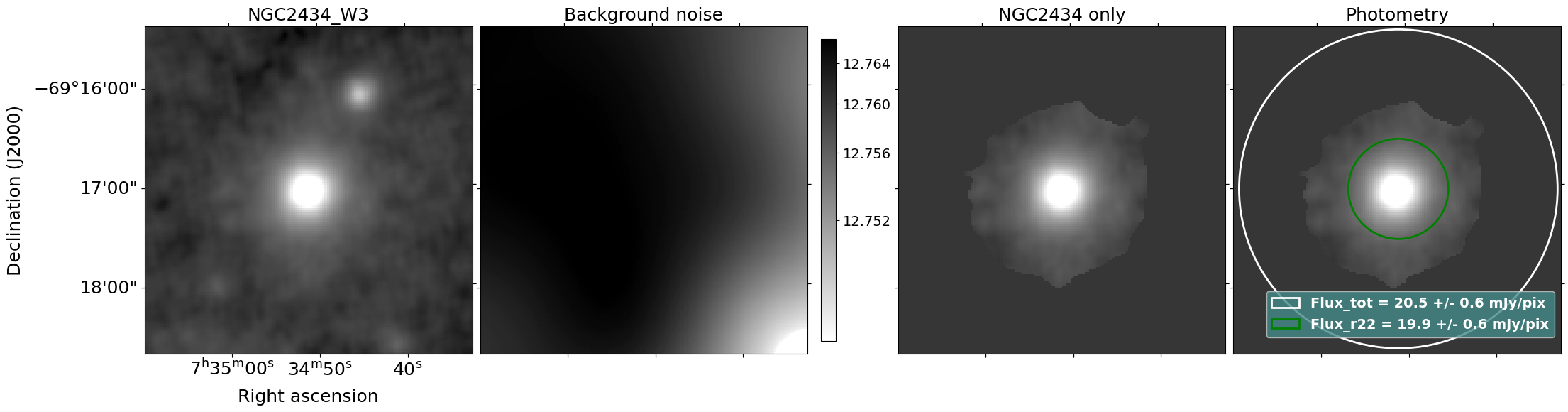}\\
\includegraphics[width=0.76\textwidth]{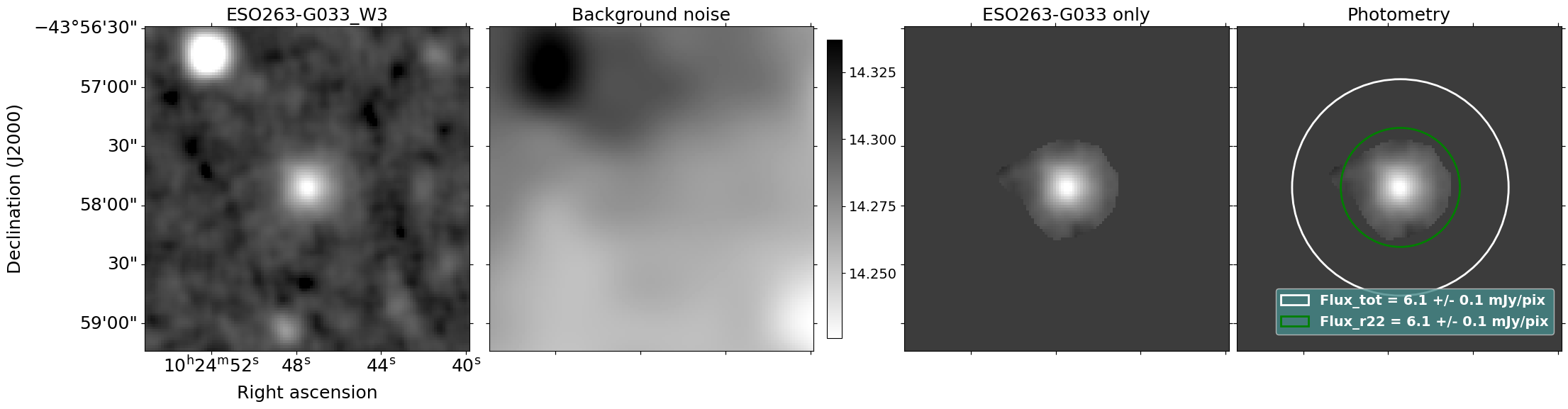}\\
\includegraphics[width=0.76\textwidth]{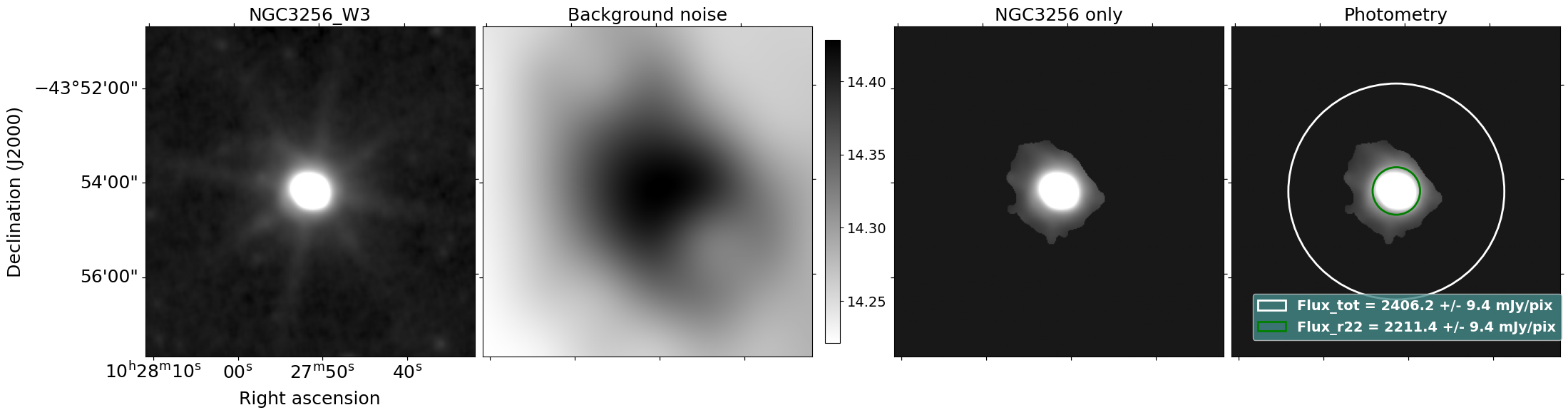}\\
\caption{Source identification and photometry using {\sc Photutils} over the WISE W3-band images. From left to right: original W3 image, background model, background-subtracted galaxy image, and final photometry. The white circle indicates the aperture used to measure the total flux density. The green circle corresponds to the 22\arcsec\ isophote employed for comparison with catalogue values.}
\label{fig:3_wise}
\end{figure}

\begin{figure}[!ht]
\centering
\includegraphics[width=0.76\textwidth]{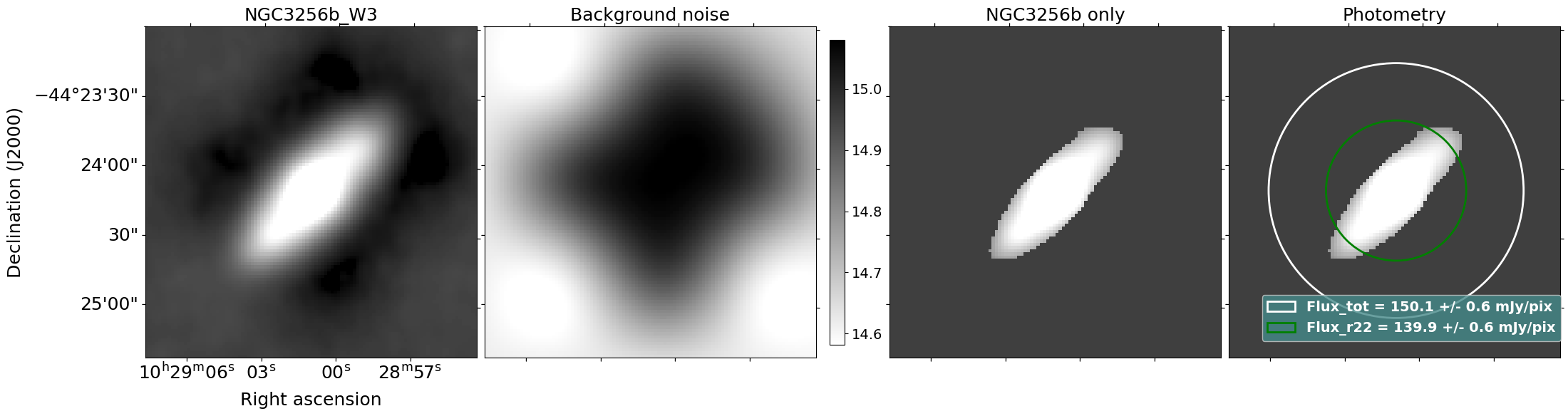}\\
\includegraphics[width=0.76\textwidth]{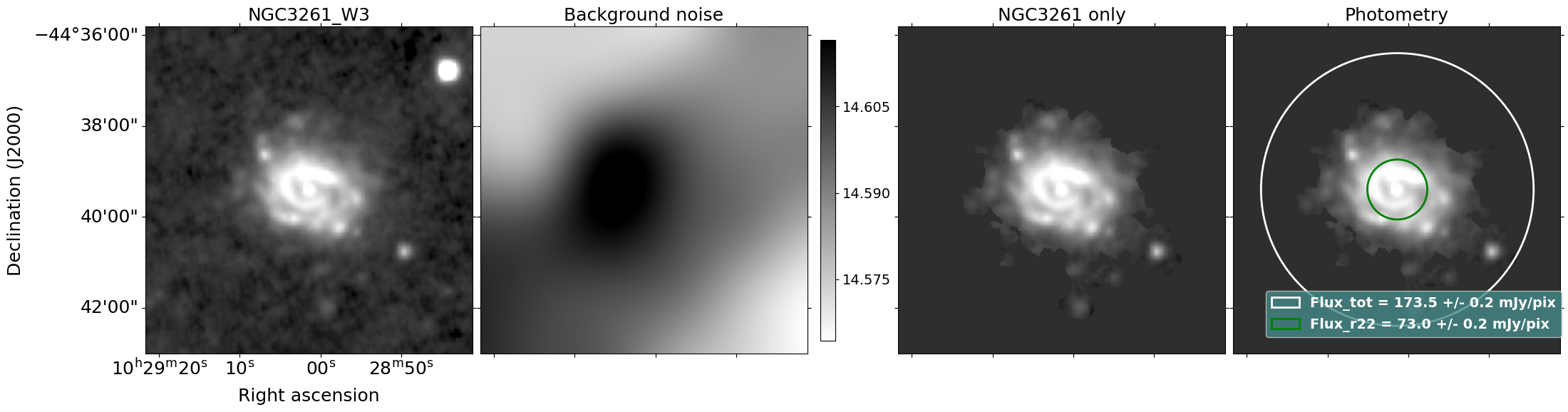}\\
\includegraphics[width=0.76\textwidth]{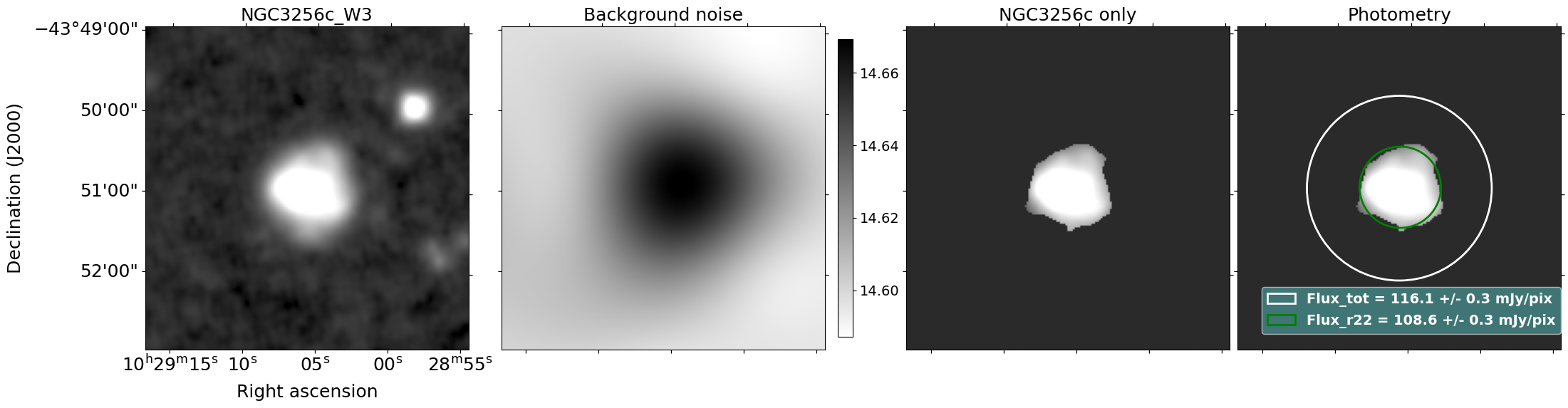}\\
\includegraphics[width=0.76\textwidth]{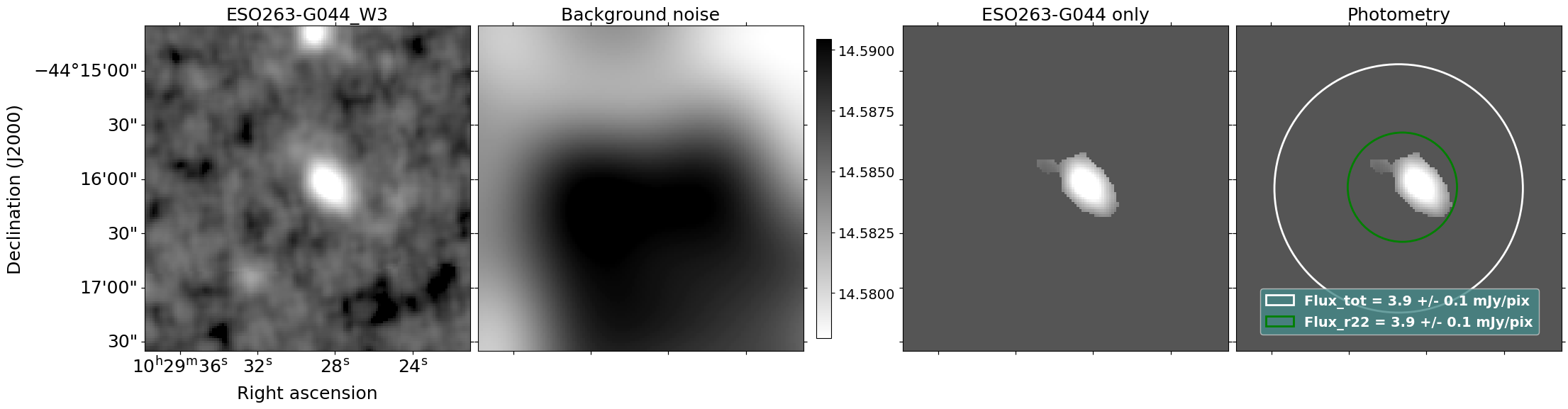}\\
\includegraphics[width=0.76\textwidth]{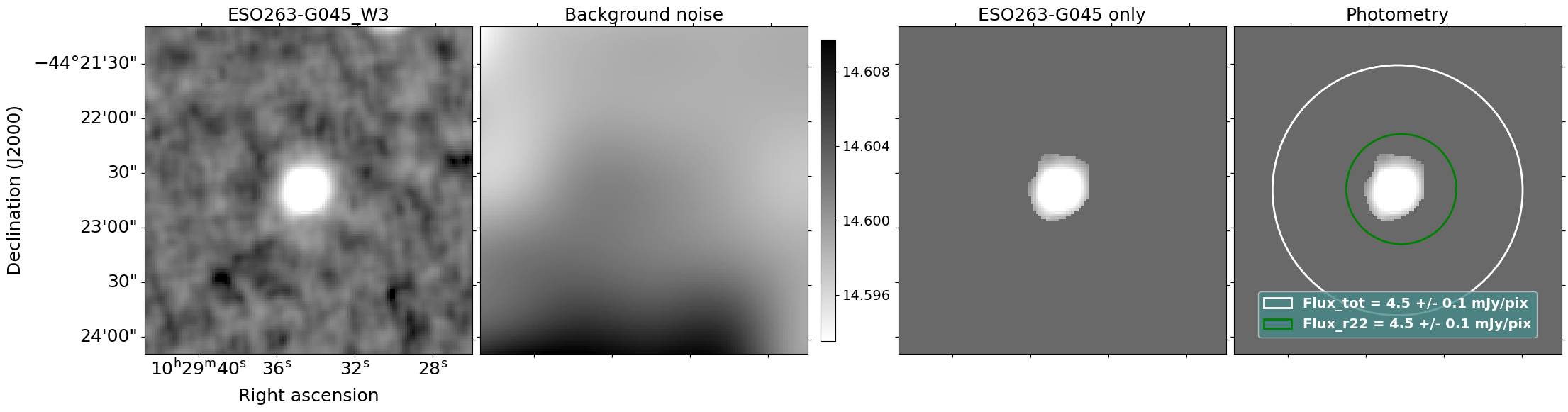}\\
\includegraphics[width=0.76\textwidth]{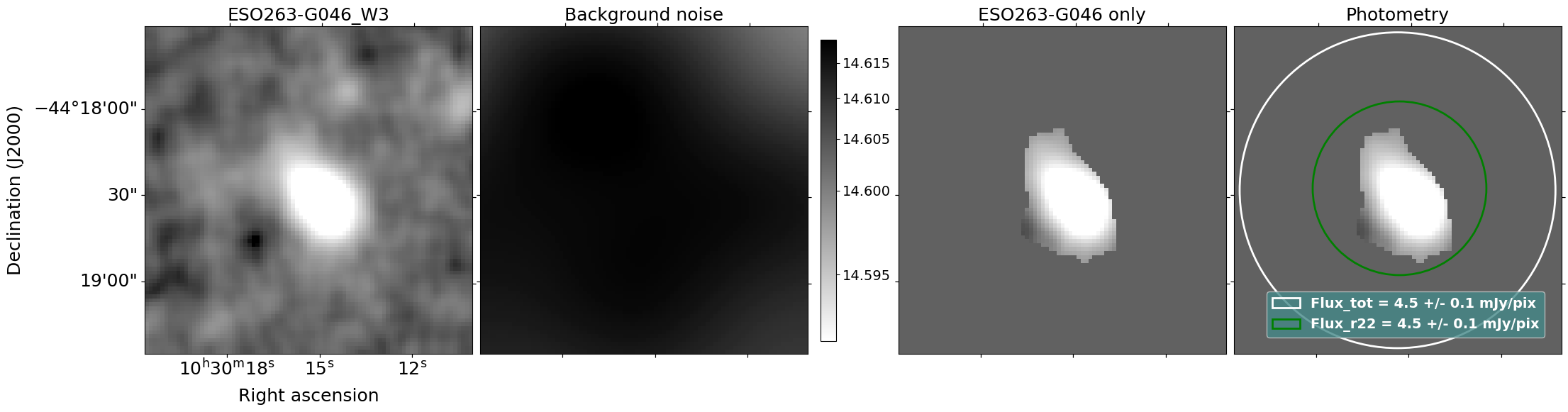}\\
\caption{continued.}
\label{fig:2_wise}
\end{figure}

\twocolumn

\begin{strip}
\centering
\includegraphics[width=0.76\textwidth]{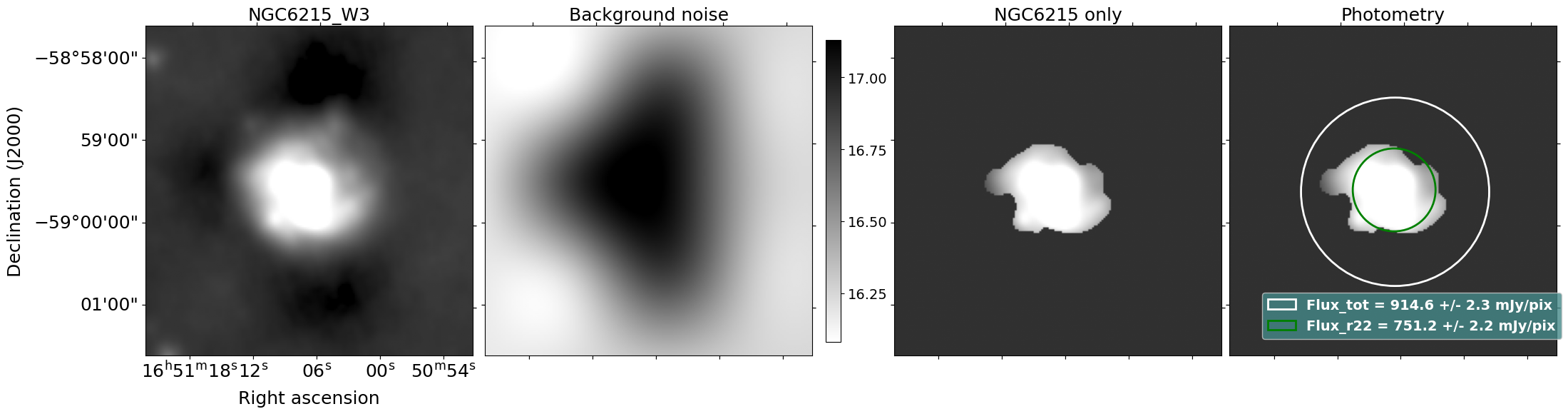}\\
\includegraphics[width=0.76\textwidth]{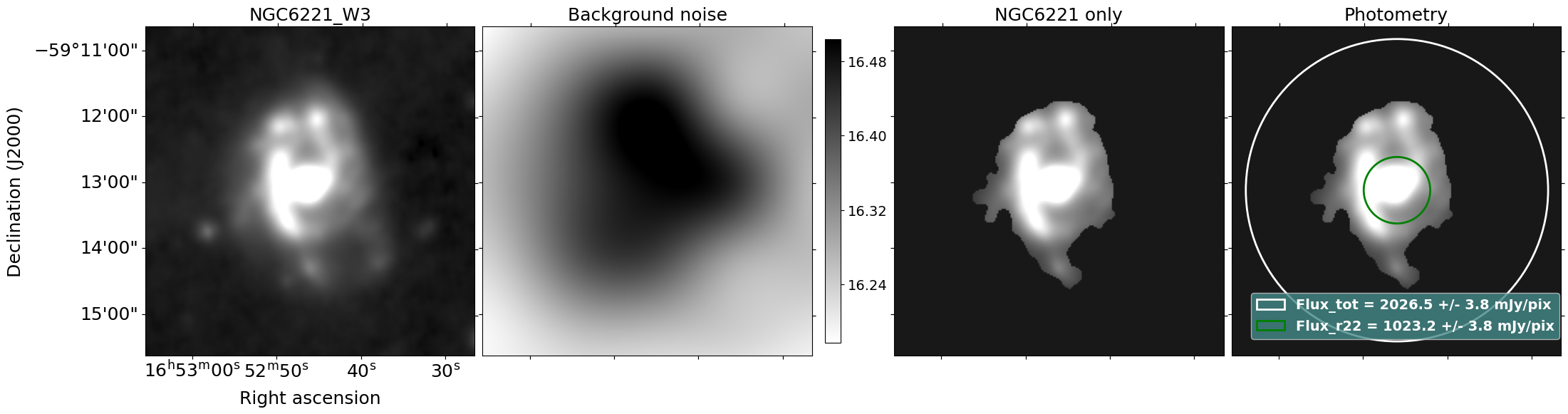}\\
{\small\raggedright \textbf{Fig. C.3.} continued.\par}
\label{fig:1_wise}
\end{strip}

\subsection*{}
\vspace{5em}

\begin{figure}
\centering 
\includegraphics[width=1\columnwidth]{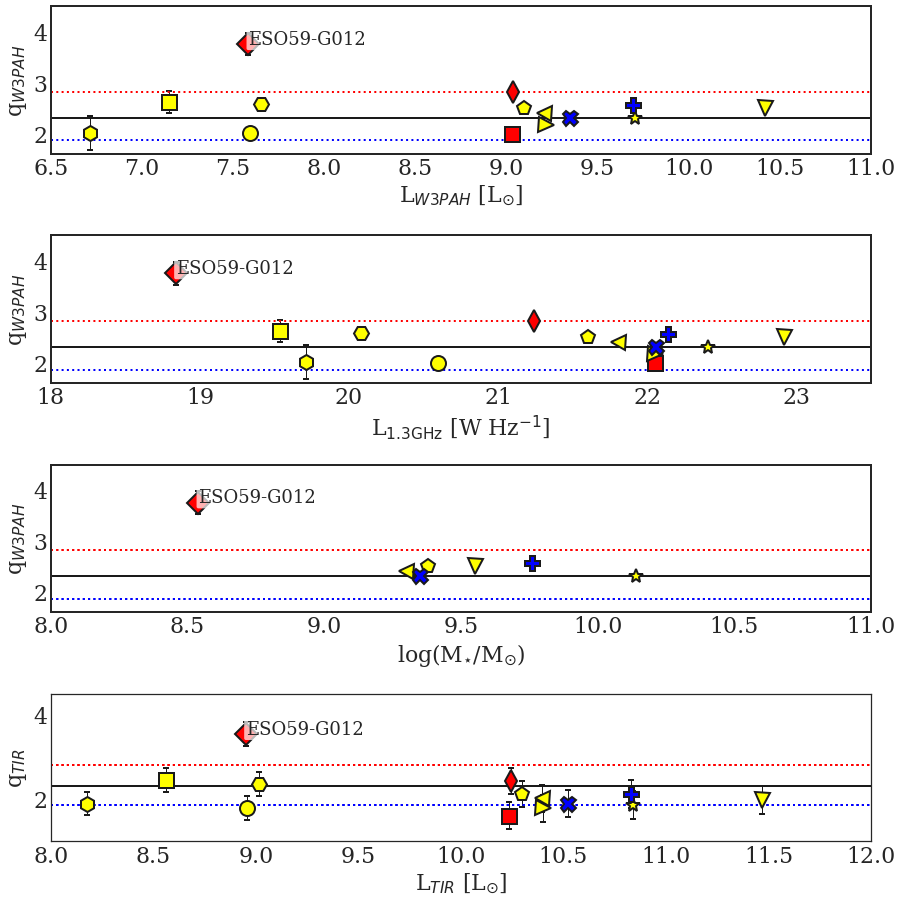}
\caption{The $q_{\rm W3PAH}$ value as a function of W3PAH, 1.3~GHz luminosity, stellar mass, and $q_{\rm TIR}$ vs TIR luminosity. Error bars are displayed for each object unless smaller than the symbol. Except for the errors in the mass determination, which are not shown. The solid line is the same as in \citet{Yun-2001} $q = 2.34$.
The blue and red dotted lines show the value corresponding to three times larger radio and W3PAH flux density than expected.} 
\label{fig:qlum}
\end{figure}

\begin{figure}[!ht]
   \centering  
   \includegraphics[width=0.67\columnwidth]{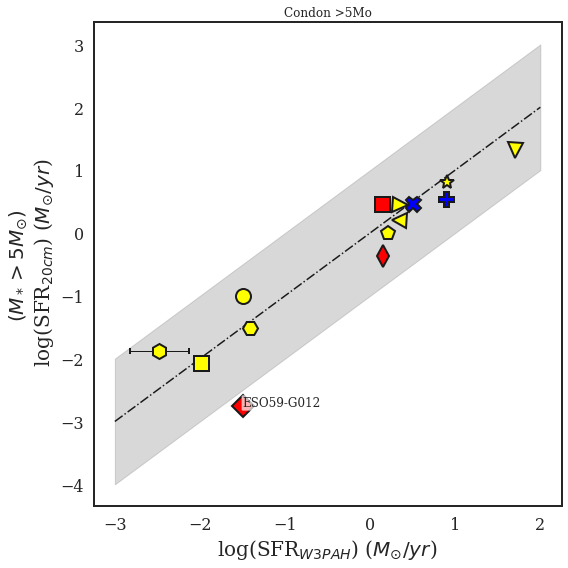}\\
\includegraphics[width=0.67\columnwidth]{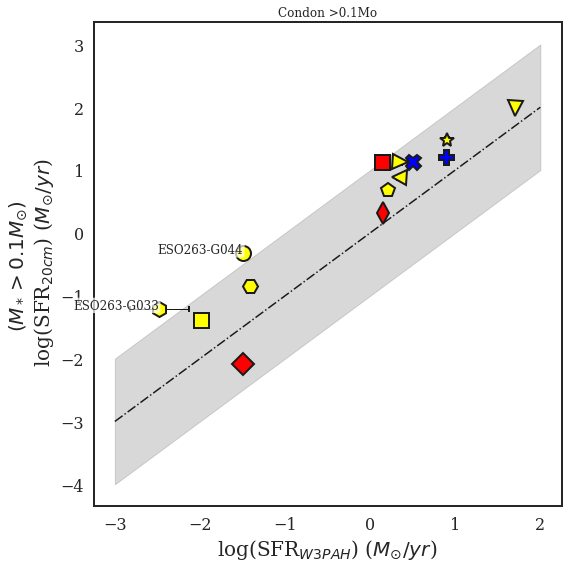}
   \caption{Comparison of galaxy star formation rates based on 20-cm radio continuum and W3PAH flux densities. The top panel shows the SFR using Eq.~\ref{eq:SFR_cont}, which accounts for stars with masses $>$5\Msun, while the bottom panel accounts for stars with masses $>$0.1\Msun. The linear function plotted represents the 1:1 relation; the shaded region indicates the 10\% scatter.}
   \label{fig:SFR}
\end{figure}

\begin{figure}[!ht]
\centering
    \includegraphics[width=0.8\columnwidth]{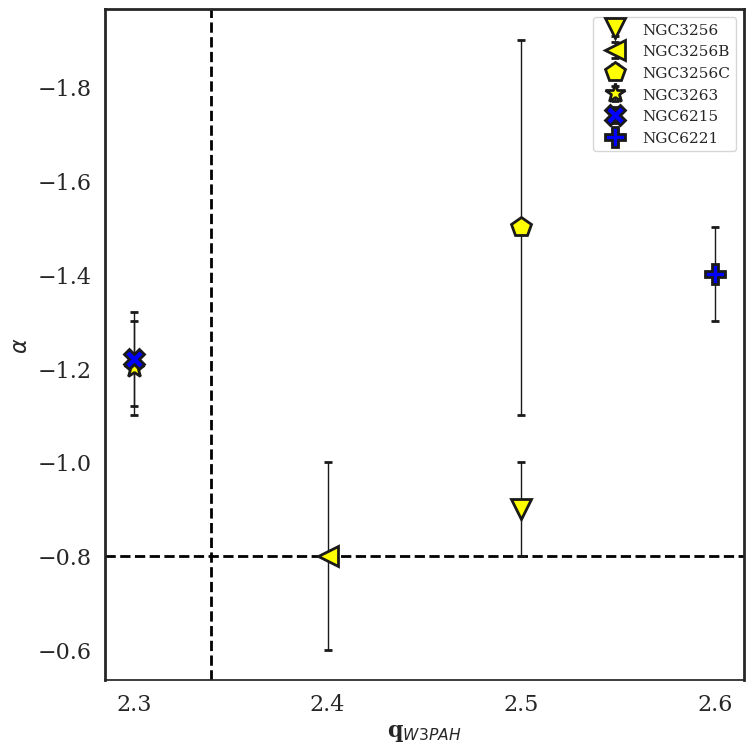}
\caption{Comparison of the $q_{\rm W3PAH}$ parameter with global spectral index. Error bars, in the spectral index determination, are displayed for each object. The dotted line at $q = 2.34$ corresponds to the value found by \citet{Yun-2001}, and the dotted line at $\alpha = -0.8$ marks the classical value, which is adopted to show the predominance of non-thermal emission.}
\label{fig:qvssp}
\end{figure}

\begin{figure}[!ht]
    \centering
    \includegraphics[width=0.8\columnwidth]{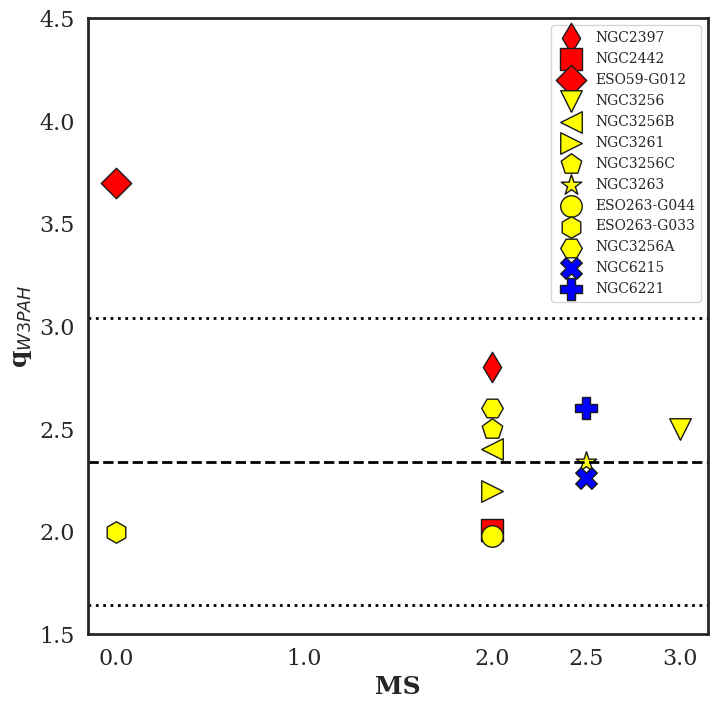}
\caption{Comparison of $q_{\rm W3PAH}$ parameters and merger stage. The dashed line is the same as in \citet[][$q = 2.34$]{Yun-2001}, and the dotted lines show the value corresponding to five times larger radio or FIR flux density than expected.}
    \label{fig:qvsMS}
\end{figure}

\FloatBarrier

\section{SED fitting and data acquisition}

\label{appen.SED_modeling}

\begin{figure*}[htpb]
\centering
\includegraphics[width=0.825\textwidth, trim={0 0 0 0}, clip]{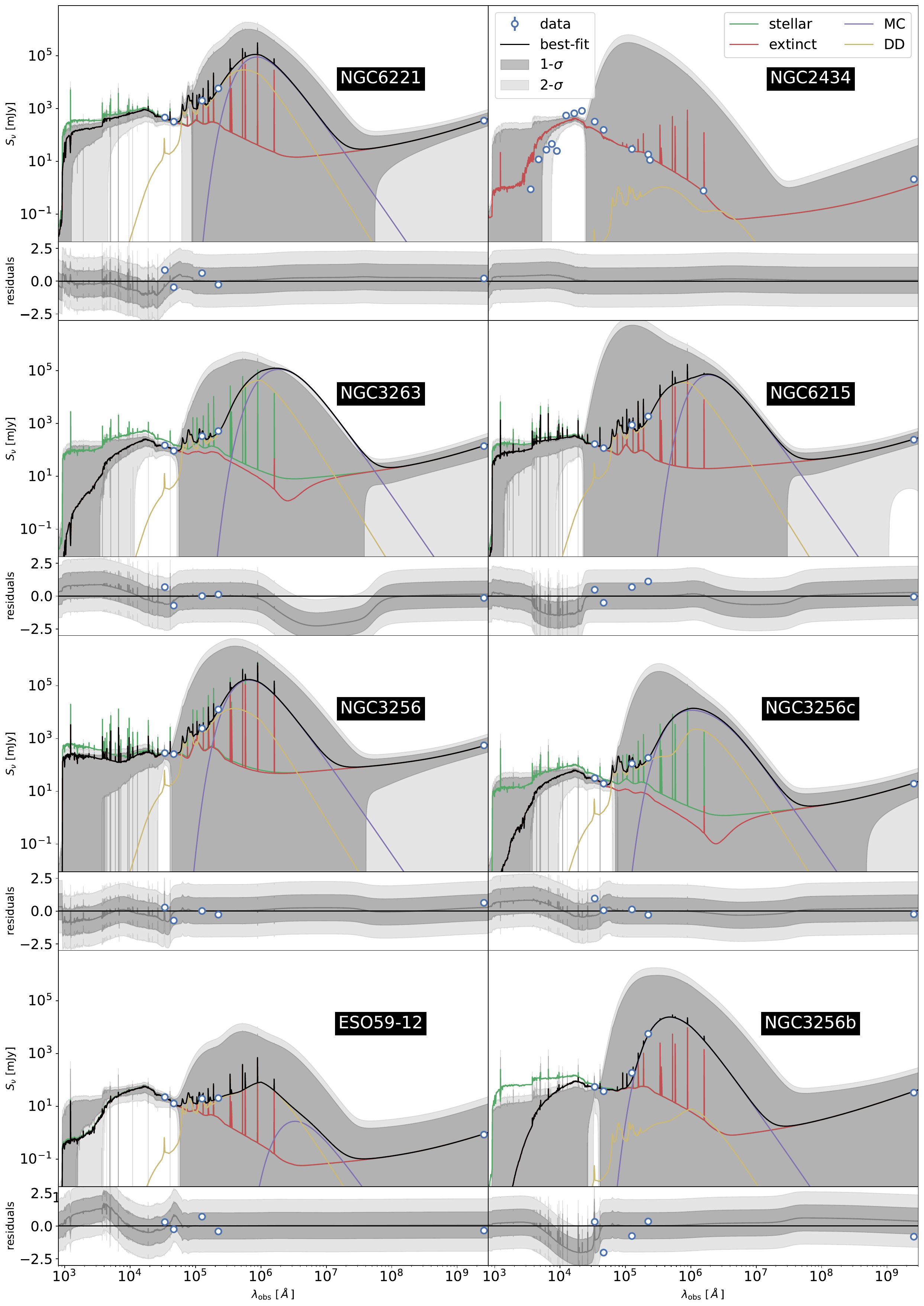}
\caption{SED models obtained with the GalaPy software for the studied galaxies, as explained in Appendix~\ref{appen.SED_modeling}, with available enough photometric observations in the literature. In the main panels, the abscissas and ordinates correspond to the observed wavelength in \AA ngströms and the flux density in mJy. Points correspond to the photometric measurements obtained from the WISE archive and Meerkat continuum levels (this work), while solid lines correspond to the unattenuated stellar emission (green), molecular cloud component (MC, purple), stellar emission considering extinction (extinct, red), and diffuse dust (DD, yellow). The bottom sub-panels highlight the residuals with 1 and 2$\sigma$ shaded areas, which are also labelled in the main panels.}
\label{fig:SEDs}
\end{figure*}

We performed the spectral energy distribution (SED) fitting of galaxies within our sample for which sufficient photometric data were available in the WISE archive, obtained via the SkyView service\footnote{\url{https://skyview.gsfc.nasa.gov/current/cgi/titlepage.pl}}. This modelling was conducted using GalaPy \citep{Ronconi2024}, an open-source software package implemented in Python/C$^{++}$ capable of modelling emission from the X-ray to the radio regimes. The underlying stellar populations were assumed to follow a Chabrier initial mass function \citep{Chabrier2003}, within a $\Lambda$CDM cosmological framework \citep{Planck2020}.
GalaPy offers a selection of distinct star formation history models. For the present study, we employed the In-Situ model, which has demonstrated efficacy in predicting the emission characteristics of galaxies across a range of types and redshift \citep{Ronconi2024,Humire2025}. This model facilitates a self-consistent tracking of the evolution of gas, dust, and metallicity in conjunction with star formation processes, thereby ensuring internal coherence among the derived physical parameters. In terms of dust treatment, GalaPy incorporates a two-component model that circumvents the assumption of a pre-defined attenuation curve, instead deriving it from structural parameters. These two components represent: (1) the molecular cloud (MC) phase, associated with young stellar populations, and (2) a diffuse dust (DD) medium that further attenuates the stellar emission. Therefore, the emission from both components is modelled via two independent modified gray bodies, as evidenced by the infrared bump in Fig.~\ref{fig:SEDs}.

Among the various free parameters available in GalaPy, we fixed specific, physically motivated values and ranges to ensure meaningful results and optimise computational efficiency. Specifically, the redshift for each source was fixed to the value reported in the SIMBAD Astronomical Database {\footnote{\url{https://simbad.cds.unistra.fr/simbad/}}}. The number of molecular clouds, \( N_\text{MC} \), was constrained to a 10$^{2.5-3.5}$ range, reflecting a typical population of isolated molecular clouds (MCs) within individual galaxies. Furthermore, the dusty and molecular radii of these clouds were constrained to a characteristic range of 10--1000~pc. A comprehensive compilation of the initial conditions adopted for our SED models is provided in Table~\ref{tab.gmcs_GalaPy_constant_parameters}, while the resulting diffuse dust ($T_{\rm{DD}}$) and molecular cloud ($T_{\rm{MC}}$) temperatures are presented in Table~\ref{tab.gmcs_GalaPy_constant_parameters}, whose lower and upper uncertainties correspond to the 16th and 84th percentiles, respectively, with the main value corresponding to the 50th percentile (median) of the derived parameter.

\begin{table}[htbp]
\centering

\caption{Initial conditions for GalaPy input values or ranges.}
\begin{adjustbox}{max width=0.5\textwidth}

\begin{tabular}{@{}lll@{}}
\toprule
Parameter & Value/Range & Brief description \\
\midrule
redshift & from SIMBAD   & see text\\
age      & ([7,11], log)             & age of the MC \\
$\tau_{\bigstar}$ &  ([4,11], log)       & characteristic timescale (yrs.)\\
ism.tau\_esc  & ([4,11], log)        & Stars' escape time from MC (yrs.)\\
$\psi_{\rm{max}}$  & ([0, 4], log)    & Maximum SFR ($M_{\rm{\odot}}$~yr$^{-1}$) \\
ism.R\_MC     &  ([0.0, 3], log)    & MC radius (pc) \\  
sfh.tau\_quench & 1e+20             & Star formation quenching (yrs.)\\
ism.f\_MC & ([0.0, 1.0], lin)         & MCs' fraction into the ISM\\
ism.norm\_MC & 100.0                & MCs' normalization factor\\
ism.N\_MC & 1.0                     & number of MCs\\
ism.dMClow & 1.3                    & Extinction index $<$100$\mu$\\
ism.dMCupp & 1.6                    & Extinction index $\gtrsim$100$\mu$\\
ism.norm\_DD & 1.0                  & Diffuse dust norm. factor\\
ism.Rdust & ([0.0, 3.0], log)         & Radius of the diffuse dust (DD, pc)\\
$f_{\rm{PAH}}$ & ([0.0, 1.0], lin)        & DD fraction radiated by PAH\\
ism.dDDlow & 0.7                    & DD extinction index $<$100$\mu$\\
ism.dDDupp & 2.0                    & DD extinction index $\gtrsim$100$\mu$\\
syn.alpha\_syn & 0.75               & Spectral index\\
syn.nu\_self\_syn & 0.2             & Self-absorption frequency (GHz)\\
f\_cal & ([-5.0, 0.0], log)           & Calibration uncertainty\\
\bottomrule
\end{tabular}
\end{adjustbox}
\tablefoot{The terms ``log'' and ``lin'' next to the ranges indicate logarithmic (log10) and linear values, respectively.}
\label{tab.gmcs_GalaPy_constant_parameters}
\end{table}

\begin{landscape}
\begin{table}
\centering
\caption{Stellar- and dust-related parameters for all SED results derived from GalaPy. The initial conditions are provided in Table~\ref{tab.gmcs_GalaPy_constant_parameters}.}
\setlength{\tabcolsep}{2pt} 
\begin{adjustbox}{max width=\textheight}
\begin{tabular}{lllllllllllrrrrr}
\toprule
Source             & SFR               & $T_{\rm{MC}}$     & $T_{\rm{DD}}$     & $M_{\rm{\bigstar}}$  & $M_{\rm{dust}}$   & $Z_{\bigstar}$      & $L_{\rm{bol}}$     & Age       & $\psi_{\rm{max}}$    & $\tau_{\bigstar}$    & $f_{\rm{PAH}}$    & $\chi^{2}_{\rm{red}}$  \\
& [$M_{\odot}$ yr$^{-1}$] & [K]       & [K]       & [log$_{10}$ ($M_{\odot}$)] & [log$_{10}$ ($M_{\odot}$)] &                      & [log$_{10}$ ($L_{\odot}$)] & [log$_{10}$ (yrs)]  & [log$_{10}$ ($M_{\odot}$ yr$^{-1}$)] & [log$_{10}$ (yrs)]  & [\%]               &                    \\
 (1)& (2) & (3) & (4) & (5) & (6) & (7) & (8) & (9)& (10)& (11)  & (12)  \\ \midrule

ESO59-12 & 0.01$_{-0.01}^{+0.01}$ & 18.94$_{-12.36}^{+20.01}$ & 40.52$_{-14.77}^{+22.07}$ & 8.541$_{-0.199}^{+0.323}$ & 3.494$_{-1.015}^{+0.714}$ & 0.022$_{-0.003}^{+0.003}$ & 9.413$_{-0.099}^{+0.357}$ & 8.48$_{-0.40}^{+0.52}$ & 1.20$_{-0.34}^{+0.28}$  & 7.15$_{-0.51}^{+0.58}$  & 2.01$_{-0.38}^{+0.58}$  & -0.56$_{-0.23}^{+0.25}$ \\
NGC2434 & 0.026$_{-0.011}^{+0.016}$ & 10.140$_{-10.140}^{+10.548}$ & 28.269$_{-6.291}^{+17.698}$ & 10.299$_{-0.430}^{+0.123}$ & 5.442$_{-0.631}^{+0.256}$ & 0.026$_{-0.001}^{+0.003}$ & 10.259$_{-0.045}^{+0.199}$ & 9.98$_{-0.60}^{+0.12}$ & 8.64$_{-0.62}^{+0.12}$ & 1.62$_{-0.11}^{+0.22}$ & 2.77$_{-0.47}^{+0.17}$ &  $-0.08_{-0.07}^{+0.05}$ \\
NGC3256  & 42.1$_{-20.3}^{+41.7}$ & 32.58$_{-21.07}^{+21.74}$ & 67.59$_{-20.78}^{+29.12}$ & 9.559$_{-0.413}^{+0.775}$ & 7.441$_{-0.952}^{+0.965}$ & 0.028$_{-0.025}^{+0.012}$ & 11.993$_{-0.311}^{+0.056}$ & 7.88$_{-0.69}^{+0.96}$ & 2.64$_{-0.40}^{+0.47}$  & 7.28$_{-0.90}^{+2.00}$  & 2.60$_{-0.52}^{+0.27}$  & -0.40$_{-0.25}^{+0.21}$ \\
NGC3256b & 4.8$_{-2.9}^{+5.5}$   & 24.31$_{-24.31}^{+35.57}$ & 72.11$_{-36.53}^{+62.07}$ & 9.309$_{-1.168}^{+0.793}$ & 6.699$_{-1.008}^{+1.264}$ & 0.013$_{-0.012}^{+0.020}$ & 11.033$_{-0.214}^{+0.157}$ & 8.30$_{-0.98}^{+1.24}$ & 1.86$_{-0.66}^{+0.75}$  & 8.43$_{-1.54}^{+1.54}$  & 1.59$_{-0.98}^{+0.74}$  & -0.20$_{-0.15}^{+0.12}$ \\
NGC3256c & 1.4$_{-0.9}^{+0.6}$   & 24.80$_{-12.24}^{+14.15}$ & 36.40$_{-10.93}^{+31.23}$ & 9.388$_{-0.321}^{+0.321}$ & 6.757$_{-1.831}^{+0.752}$ & 0.019$_{-0.006}^{+0.012}$ & 10.581$_{-0.118}^{+0.223}$ & 9.02$_{-1.11}^{+0.61}$ & 1.07$_{-0.45}^{+1.09}$  & 8.34$_{-1.64}^{+0.94}$  & 2.59$_{-0.63}^{+0.30}$  & -0.55$_{-0.24}^{+0.28}$ \\
NGC3263  & 12.1$_{-8.7}^{+4.7}$  & 24.96$_{-11.03}^{+14.49}$ & 47.02$_{-8.95}^{+27.74}$  & 10.142$_{-0.442}^{+0.462}$ & 7.703$_{-2.240}^{+0.973}$ & 0.029$_{-0.011}^{+0.012}$ & 11.516$_{-0.155}^{+0.092}$ & 8.83$_{-1.05}^{+0.81}$ & 2.07$_{-0.44}^{+0.82}$  & 8.10$_{-1.73}^{+1.19}$  & 2.81$_{-0.42}^{+0.14}$  & -0.53$_{-0.21}^{+0.28}$ \\
NGC6215  & 6.5$_{-2.3}^{+3.6}$   & 26.40$_{-13.04}^{+18.87}$ & 46.53$_{-14.73}^{+15.37}$ & 9.350$_{-0.839}^{+0.394}$ & 7.268$_{-1.189}^{+0.716}$ & 0.017$_{-0.014}^{+0.011}$ & 11.082$_{-0.144}^{+0.121}$ & 8.62$_{-1.30}^{+0.62}$ & 1.53$_{-0.37}^{+0.61}$  & 8.24$_{-1.47}^{+1.62}$  & 2.67$_{-0.34}^{+0.23}$  & -0.48$_{-0.23}^{+0.24}$ \\
NGC6221  & 5.7$_{-1.3}^{+1.5}$   & 21.86$_{-21.86}^{+15.17}$ & 58.68$_{-12.33}^{+11.01}$ & 9.782$_{-0.220}^{+0.365}$ & 6.969$_{-0.512}^{+0.728}$ & 0.029$_{-0.006}^{+0.003}$ & 11.149$_{-0.061}^{+0.051}$ & 8.60$_{-0.34}^{+0.61}$ & 1.99$_{-0.36}^{+0.21}$  & 7.71$_{-0.44}^{+0.84}$  & 2.53$_{-0.17}^{+0.17}$  & -0.84$_{-0.24}^{+0.39}$ \\
\bottomrule
\end{tabular}
\end{adjustbox}
\tablefoot{{\small Columns description from left to right: (1) Galaxy name, (2) star formation rate (SFR), (2) molecular cloud (MC; $T_{\rm{MC}}$) and (4) diffuse dust (DD; $T_{\rm{DD}}$) temperatures, obtained from gray body fittings, (5) stellar and (6) dust masses ($M_{\rm{\bigstar}}$, $M_{\rm{dust}}$; respectively), (7) stellar metallicity ($Z_{\bigstar}$), (8) bolometric luminosity ($L_{\rm{bol}}$), (9) stellar age (Age), (10) maximum SFR ($\psi_{\rm{max}}$) that experienced the galaxy at a certain time ($\tau_{\bigstar}$), known as the characteristic timescale, (11) DD fraction radiated by polyaromatic hydrocarbons (PAH), (12) reduced $\chi^{2}$.}}
\label{tab:SED-results}
\end{table}
\end{landscape}

\label{LastPage}

\end{appendix}
\end{document}